# Comprehensive Time-Series Regression Models Using Gretl—U.S. GDP and Government Consumption Expenditures & Gross Investment from 1980 to 2013[1]

Juehui Shi
Department of Operations Management and Strategy
University at Buffalo, The State University of New York (SUNY)

**Abstract**: Using Gretl, I apply ARMA, Vector ARMA, VAR, state-space model with a Kalman filter, transfer-function and intervention models, unit root tests, cointegration test, volatility models (ARCH, GARCH, ARCH-M, GARCH-M, Taylor-Schwert GARCH, GJR, TARCH, NARCH, APARCH, EGARCH) to analyze quarterly time series of GDP and Government Consumption Expenditures & Gross Investment (GCEGI) from 1980 to 2013. The article is organized in three sections: (I) Definition; (II) Regression Models; (III) Discussion [Summary of Major Findings and Their Managerial Implications, Comparison of Empirical Results, Contributions to Literature, Limitations and Future Research, Gretl Scripts]. Additionally, I discovered a unique interaction between GDP and GCEGI in both the short-run and the long-run and provided policy makers with some suggestions. For example in the short run, GDP responded positively and very significantly (0.00248) to GCEGI, while GCEGI reacted positively but not too significantly (0.08051) to GDP. In the long run, current GDP responded negatively and permanently (0.09229) to a shock in past GCEGI, while current GCEGI reacted negatively yet temporarily (0.29821) to a shock in past GDP. Therefore, policy makers should not adjust current GCEGI based merely on the condition of current and past GDP. Although increasing GCEGI does help GDP in the short-term, significantly abrupt increase in GCEGI might not be good to the long-term health of GDP. Instead, a balanced, sustainable, and economically viable solution is recommended, so that the short-term benefits to the current economy from increasing GCEGI often largely secured by the long-term loan outweigh or at least equal to the negative effect to the future economy from the long-term debt incurred by the loan. Finally, I found that non-normally distributed volatility models generally perform better than normally distributed ones. More specifically, TARCH-GED performs the best in the group of non-normally distributed, while GARCH-M does the best in the group of normally distributed.

---

Work is completed under the supervision of my advisor Dr. Winston T. Lin.

## SECTION I: DEFINTION

### *(1) Nonstationarity*

Phenomena: In order to use statistical tests for an ARMA model selection, we need to estimate the underlying stochastic process, which can be derived by the mean, variance, and covariance of the sample data. However, these quantities are only meaningful if they are independent of time. If that is the case, the series is stationary. If not, then it is nonstationary. The most commonly considered is covariance stationarity with those following conditions: $E(y_t) = m; E[(y_t - m)^2] = \gamma_0 < \infty;$ $E[(y_t - m)(y_{t-i} - m)] = \gamma_i$, where t=1,…,T and i=…,-2, -1, 0, 1, 2, … The nonstationary series do not meet all or either one of the conditions.

Methodology: Univariate ARIMA for example $ARIMA(1,1,1)$: $\Delta y_t = \mu + \varphi \Delta y_{t-1} + \varepsilon_t - \theta \varepsilon_{t-1}$, where the nonstationary time series $y_t$ is differenced once to obtain stationary series $\Delta y_t (y_t - y_{t-1})$. Sometimes, it is necessary to differentiate the data more than once to obtain a stationary series. The process can be also denoted as $ARIMA(p,d,q) = \varphi_p(L)\Delta^d y_t = \mu + \theta_q(L)\varepsilon_t$, where d is the order of the differencing operator $\Delta^d = (1-L)^d$. For example, if d=1, $\Delta_{y_t}^1 = (1-L)^1 y_t = y_t - y_{t-1}$. $ARIMA(1,1,1) = \varphi_1(L)\Delta^1 y_t = \mu + \theta_1(L)\varepsilon_t$. Additionally, the most widely used test for non-stationarity is the Augmented Dickey-Fuller (ADF) unit root test. If the tested data has unit root, then the series is non-stationary.

Supply chain implication: Often, we observe some kind of trending behavior within wholesaler or retailer's sales data, for example the sales in the end of the sample are higher (lower) than the initial sales. Since the stationary data requires the mean level is independent of time (which implies that the average sales level at the beginning of the sample period is equal to the average sales level at the end of the sample), it is normal for supply chain time series to be nonstationary.

### *(2) Seasonality*

Phenomena: Seasonality exists when there is a highly fluctuating pattern such as seasonal sales or advertising pattern in the retailing business. Seasonal processes can be identified from the ACF and PACF functions, similarly to the nonseasonal ARIMA processes, except that the patterns occur at lags s, 2s, 3s, … instead of 1, 2, 3, … In addition, seasonal unit root tests have been developed to detect the order of seasonal integration (Franses 1998 and Hylleberg 1992).

Methodology: Univariate ARIMA with (i) purely deterministic seasonal processes, (ii) stationary seasonal processes, and (iii) integrated seasonal processes (Maddala and Kim 1998, p. 363). For example (i) can be model as $y_t = \mu + \sum_{s=1}^{S-1} b_s d_{st} + \varepsilon_t$ where S is the maximum number of seasons (12 for monthly data, 4 for quarterly data), $d_{st}$ is a dummy variable taking the value one in season s and zero otherwise, and $b_s$ is a parameter. (ii) ARMA type if seasonal fluctuations in sales levels or random shocks die out over time in a seasonal way. (iii) integrated, when nonstationary seasonal patterns exist. It can be models as $ARIMA(P, D, Q)_s$, where s is the seasonal lag. For instance, $ARIMA(1, 0, 1)_{12}$: $y_t = \mu + \varphi y_{t-12} + \varepsilon_t - \theta\varepsilon_{t-12}$.

Supply chain implication: Retailer could have much higher advertising expenditures and sales figures in Spring than in Summer. In fact, many time series in supply chain display such seasonal patterns caused by managerial decisions, weather conditions, events, holidays, etc. Additionally, manufacturer could face similar fluctuation for raw material shortage due to adverse weather impact or unfavorable market condition, which also inevitably affect wholesaler's procurement and sales.

***(3) Changing volatility***

Phenomena: Volatility exists when the variance of the dependent variable changes over time. The volatility can also change over time. For instance the U.S. stock returns index (NASDAQ) experiences a relatively sedate period from 1992 to 1996. Then, stock returns become much more volatile until early 2004. Volatility increases again at the end of 2009.

Methodology: ARCH, GARCH, ARCH-M, EGARCH and other ARCH variants or VAR and impulse response functions.

Supply chain implication: Volatility widely exists in the supply chain where manufacturers require to procure precious metal (silver, gold, platinum) as raw materials whose price are usually volatile over time, due to uncertain supply and demand.

***(4) Dynamics***

Phenomena: Dynamics exists when the time series exhibit a deterministic trend which implies that the level is not constant, but can be perfectly predicted if the underlying deterministic function is known. The linear time trend is the most commonly used function for such purpose:

$y_t = \mu + \beta t + \varepsilon_t$, t=1, … , T, where the long run behavior of $y_t$ is perfectly determined by the series' individual growth path $\beta t$. In the long run the series always returns to its individual growth path $\beta t$. Therefore, such series is often called Trend Stationary (TS), because it is stationary around a trend.

Methodology: We can incorporate the deterministic trend into ARMA model. For example: $dynamic\ AR(1)\text{:}\ y_t = \mu + \beta t + \varphi y_{t-1} + \varepsilon_t, or\ ARMA(1,1)\text{:}\ y_t = \mu + \beta t + \varphi y_{t-1} - \theta \varepsilon_{t-1} + \varepsilon_t$.

Supply chain implication: Wholesaler's demand could be dynamic but predictable with a pattern, so is manufacturer.

***(5) Randomness***

Phenomena: Randomness exists when the time series exhibit a stochastic trend which implies that the variation is systematic but hardly predictable, because every temporary deviation may change the long-run performance of the series. Such phenomenon is also called "shock persistence". A simple example of such series is named as the Random Walk (RW) process: $y_t = \beta + y_{t-1} + \varepsilon_t$.

Methodology: ARMA with a stochastic trend: $y_t = \beta + \sum_{i=1}^{p} \varphi_p y_{t-p} + \sum_{i=1}^{q} \theta_q \varepsilon_{t-q} + \varepsilon_t$. A simple case of $AR(1)$: $y_t = \beta + \varphi_1 y_{t-1} + \varepsilon_t, or\ ARMA(1,1)\text{:}\ y_t = \beta + \varphi_1 y_{t-1} - \theta_1 \varepsilon_{t-1} + \varepsilon_t$.

Supply chain implication: Retailer's demand is unpredictable, but rather stochastic due to the nature of retailing business.

***(4) + (5) Dynamics + Randomness***

Phenomena: A deterministic trend and a stochastic trend can coexist in a very complex situation, for example: $y_t = \mu + \beta t + \sum_{i=1}^{t} \varepsilon_{t-i} + \varepsilon_t$, where $\mu + \beta t$ is the deterministic trend, $\sum_{i=1}^{t} \varepsilon_{t-i}$ is the stochastic trend, and $\varepsilon_t$ is the noise term.

Methodology: ARMA with a dynamic and stochastic trend: $y_t = \mu + \beta t + \sum_{i=1}^{p} \varphi_p y_{t-p} + \sum_{i=1}^{q} \theta_q \varepsilon_{t-q} + \varepsilon_t$. A simple case of $AR(1)$: $y_t = \mu + \beta t + \varphi_1 y_{t-1} + \varepsilon_t, or\ ARMA(1,1)\text{:}\ y_t = \mu + \beta t + \varphi_1 y_{t-1} - \theta_1 \varepsilon_{t-1} + \varepsilon_t$.

Supply chain implication: Consumer's demand is changing all the time, sometime predictable but other time not, therefore the combination of both dynamics and stochastic works better.

***(6) Nonnormality***

Phenomena: Nonnormality exists if residuals of the time series are not normally distributed. We have to utilize GARCH or its variants integrating nonnormal distributions such as student's t or generalized error to explain the endogenous variables.

Methodology: ARCH, ARCH-M, GARCH, GARCH-M, EGARCH and other GARCH variants with nonnormal residual distributions such as t, GED, skewed t, skewed-GED.

Supply chain implication: Error distributions within time series, especially of volatile data, are nonnormal. Sales or price series within the supply chain can often be quite heteroscedastic with high volatility over time.

***(7) Nonlinearity***

Phenomena: Sometimes, the observed series are not only influenced by its own lagged past value but also by other determinant exogenous variables. For example, one of the main fields of interest in marketing is the determination of the effect of marketing actions (advertising campaign, pricing) on sales fluctuation.

Methodology: ARMAX, transfer function and intervention model.

Supply chain implication: Since retailer side of the supply chain frequently involves marketing decisions such as pricing and advertising expenditures, sales time series can be effectively modeled as ARMAX by incorporating pricing and advertising as exogenous determining variables. Such model provides the researchers with important insights of how pricing and advertising decisions contribute to demand changes within the supply chain. Moreover, pricing also heavily influences demand on the side of both manufacturer and wholesaler. We can even further study the joint cause and effect of pricing-demand from downstream (manufacturer→wholesaler→retailer) to upstream (retailer→wholesaler→manufacturer).

## SECTION II: REGRESSION MODELS

Data source: $GDP$[1], Government Consumption Expenditures & Gross Investment[2]

Annotations: 1 = GDPC1Q, 2 = GCEC1Q

### *(1) The Box-Jenkins Univariate ARIMA Approach:*

Note: Maximum lag for the correlogram (ACF and PACF) is set to 20. I denote observations GDPC1Q as $Y_{1t}$ and GCEC1Q as $Y_{2t}$.

Data collected:

a) Use for model construction (Sample Range): 1980 Q1-2006 Q1, Billions of Chained 2005 Dollars, Quarterly, Seasonally Adjusted Annual Rate.
b) Use for forecast performance assessment (Forecast Range): 2006 Q2-2013 Q1

Sample size for each observation: 105.

### I. Identification

In this section, I draw the graphs of ACF and PACF for all the observed variables, including GDPC1Q and GCEC1Q, which help me to initially identify fitted models.

Autocorrelation function for GDPC1Q

| LAG | ACF | | PACF | | Q-stat. | [p-value] |
|---|---|---|---|---|---|---|
| 1 | 0.9725 | *** | 0.9725 | *** | 102.1765 | [0.000] |
| 2 | 0.9445 | *** | -0.0250 | | 199.4772 | [0.000] |
| 3 | 0.9157 | *** | -0.0280 | | 291.8301 | [0.000] |
| 4 | 0.8873 | *** | -0.0062 | | 379.4130 | [0.000] |
| 5 | 0.8591 | *** | -0.0123 | | 462.3396 | [0.000] |
| 6 | 0.8308 | *** | -0.0187 | | 540.6633 | [0.000] |
| 7 | 0.8027 | *** | -0.0103 | | 614.5253 | [0.000] |
| 8 | 0.7739 | *** | -0.0280 | | 683.8957 | [0.000] |
| 9 | 0.7443 | *** | -0.0319 | | 748.7268 | [0.000] |
| 10 | 0.7145 | *** | -0.0187 | | 809.1079 | [0.000] |
| 11 | 0.6846 | *** | -0.0205 | | 865.1257 | [0.000] |
| 12 | 0.6555 | *** | -0.0032 | | 917.0296 | [0.000] |
| 13 | 0.6268 | *** | -0.0089 | | 965.0109 | [0.000] |
| 14 | 0.5986 | *** | -0.0109 | | 1009.2447 | [0.000] |
| 15 | 0.5702 | *** | -0.0194 | | 1049.8350 | [0.000] |
| 16 | 0.5423 | *** | -0.0093 | | 1086.9642 | [0.000] |
| 17 | 0.5148 | *** | -0.0108 | | 1120.8026 | [0.000] |
| 18 | 0.4880 | *** | -0.0058 | | 1151.5546 | [0.000] |
| 19 | 0.4610 | *** | -0.0213 | | 1179.3186 | [0.000] |
| 20 | 0.4332 | *** | -0.0352 | | 1204.1175 | [0.000] |

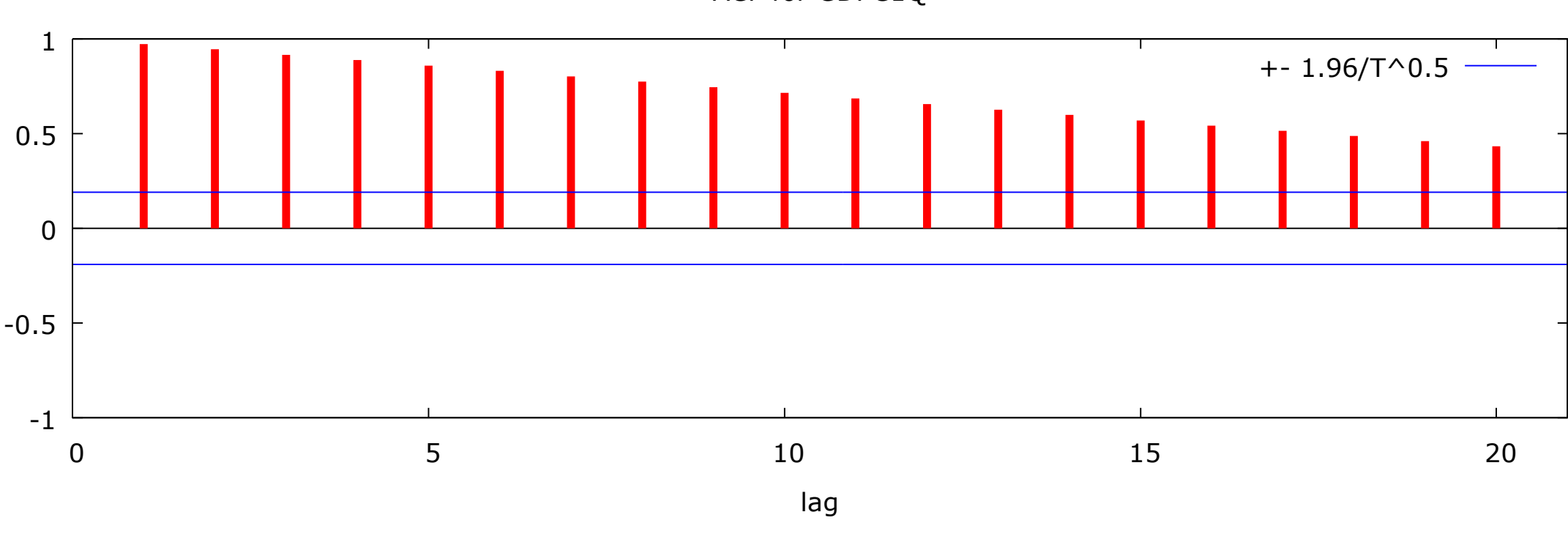

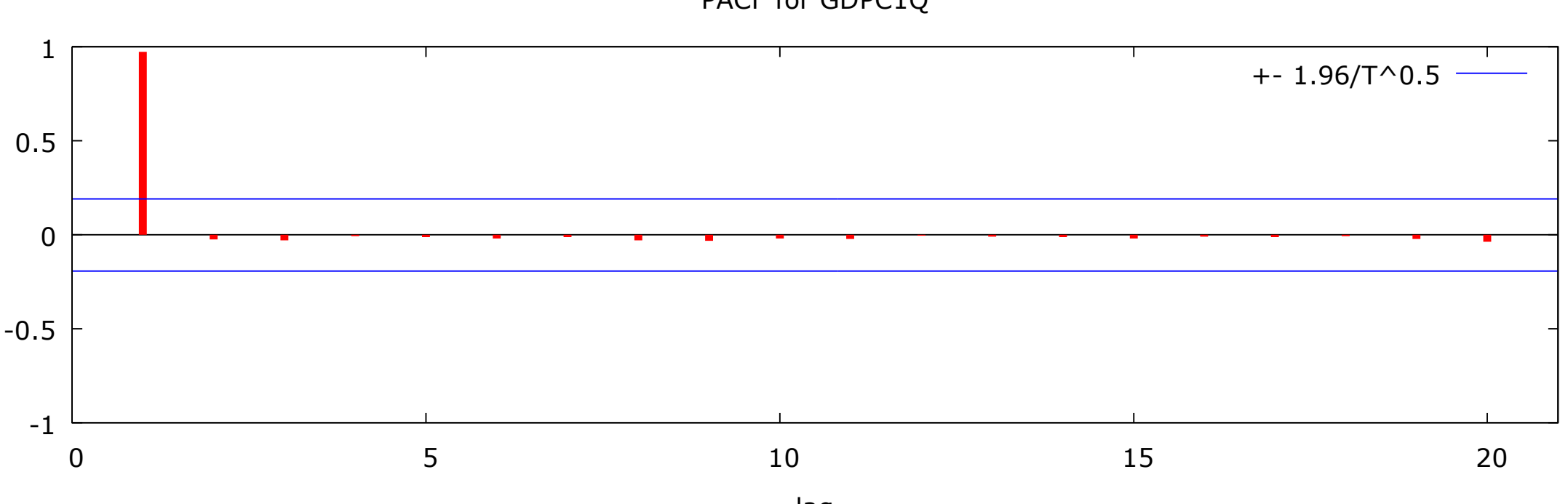

The ACF graph for GDPC1Q dies out slowly (exponentially decaying), with one spike in PACF that cuts off after lag 1. Data is nonstationary.Therefore, the first model for GDPC1Q ($Y_{1t}$) is initially identified as ARIM(1, 1, 0), ARIMA(0, 1, 1), or ARIMA (1, 1, 1).

Autocorrelation function for GCEC1Q

| LAG | ACF | | PACF | | Q-stat. | [p-value] |
|---|---|---|---|---|---|---|
| 1 | 0.9697 | *** | 0.9697 | *** | 101.5772 | [0.000] |
| 2 | 0.9399 | *** | -0.0063 | | 197.9379 | [0.000] |
| 3 | 0.9071 | *** | -0.0651 | | 288.5790 | [0.000] |
| 4 | 0.8738 | *** | -0.0281 | | 373.5108 | [0.000] |
| 5 | 0.8400 | *** | -0.0229 | | 452.7871 | [0.000] |
| 6 | 0.8053 | *** | -0.0339 | | 526.3749 | [0.000] |
| 7 | 0.7681 | *** | -0.0600 | | 594.0054 | [0.000] |
| 8 | 0.7310 | *** | -0.0185 | | 655.8961 | [0.000] |
| 9 | 0.6931 | *** | -0.0308 | | 712.1199 | [0.000] |
| 10 | 0.6549 | *** | -0.0286 | | 762.8375 | [0.000] |
| 11 | 0.6161 | *** | -0.0309 | | 808.2024 | [0.000] |
| 12 | 0.5781 | *** | -0.0083 | | 848.5792 | [0.000] |
| 13 | 0.5415 | *** | 0.0006 | | 884.3816 | [0.000] |
| 14 | 0.5043 | *** | -0.0325 | | 915.7805 | [0.000] |
| 15 | 0.4682 | *** | -0.0080 | | 943.1473 | [0.000] |
| 16 | 0.4308 | *** | -0.0461 | | 966.5724 | [0.000] |
| 17 | 0.3946 | *** | -0.0062 | | 986.4527 | [0.000] |
| 18 | 0.3611 | *** | 0.0199 | | 1003.2914 | [0.000] |
| 19 | 0.3294 | *** | 0.0052 | | 1017.4698 | [0.000] |
| 20 | 0.2986 | *** | -0.0133 | | 1029.2550 | [0.000] |

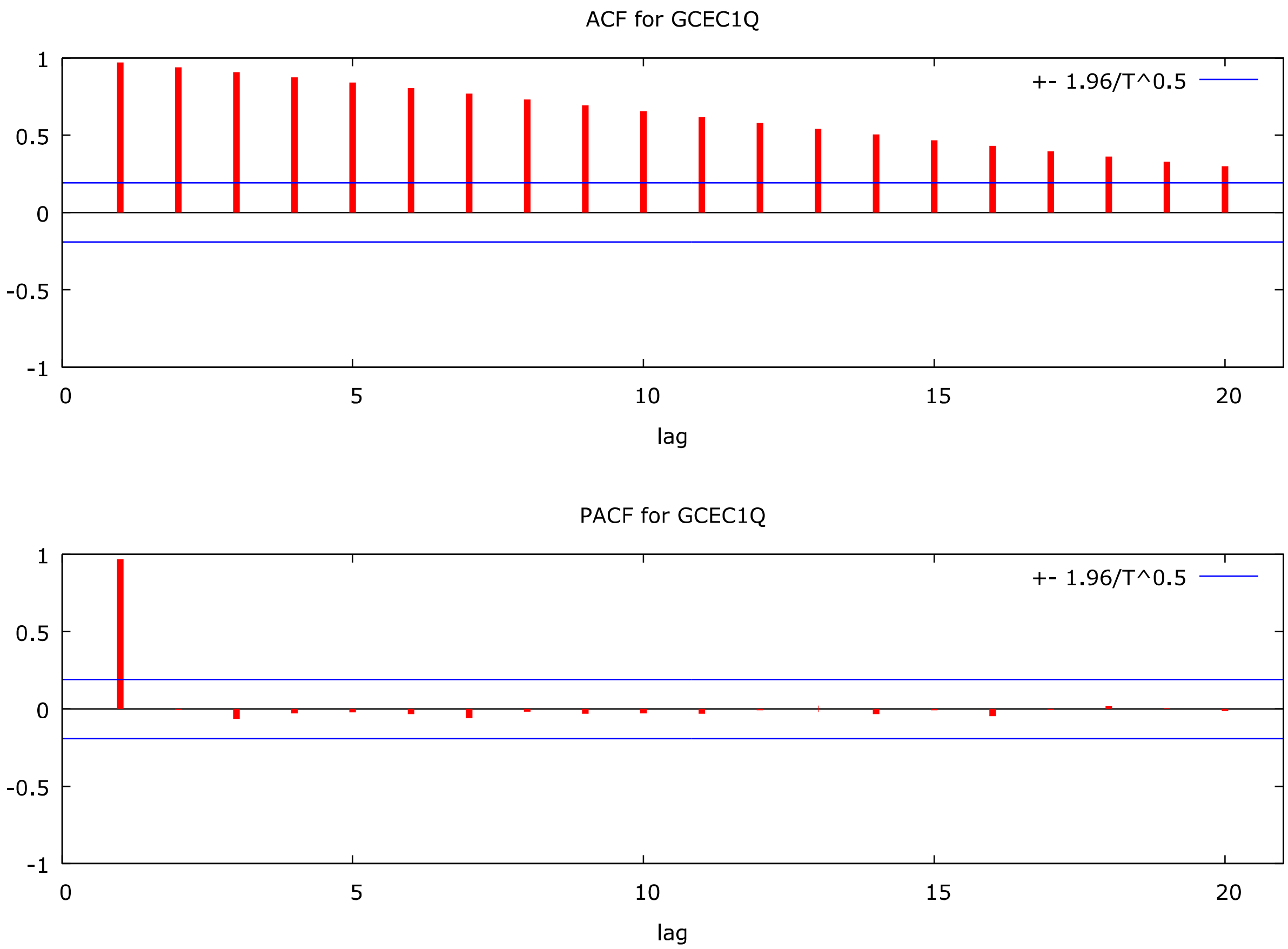

The ACF graph for GCEC1Q dies out slowly (exponentially decaying), with one spike in PACF that cuts off after lag 1. Data is nonstationary.Therefore, the second model for GCEC1Q ($Y_{2t}$) is initially identified as ARIMA(1, 1, 0), ARIMA(0, 1, 1), or ARIMA (1, 1, 1).

## II & III. Estimation and Diagnostic Checking

In this section, I fit the following ARIMA models into collected data (GDPC1Q and GCEC1Q), for example ARIMA (1, 0, 0); (1, 1, 0); (0, 0, 1); (0, 1, 1); (1, 0, 1); (1, 1, 1). I will rule out any ARIMA model, which either does not have significant parameters or pass the initial normality test with significantly low p-value that rejects the null of normal distribution. Additionally, the model must meet those calculation requirements, such as convergence criterion and finite initial value in the objective function. Finally, I might need to increase the AR or MA order to 2 if necessary to obtain normally-distributed residual.

If the tested model meets the preliminary requirements for significant parameters and residual normality distribution with successful calculation, the estimated models are suitable to be tested further against autocorrelation and heteroskedasticity.

For diagnostic checking, I will perform additional tests on all the models passing the normality test, for instance Ljung-Box Q test for autocorrelation and ARCH test for heteroskedasticity. The selection technique for the best fitted model is to have the best overall white noise, which is normally distributed, independently distributed (NO autocorrelation), and homoskedastic (NO ARCH effect).

Below is the summarized testing results:

| | Normality ($\chi^2$) Test | Autocorrelation (Ljung-Box Q) Test (lag order 4) | ARCH Effect (LM) Test (lag order 4) |
|---|---|---|---|
| Model 1<br>GDPC1Q ($Y_{1t}$) | | | |
| ARIMA (1, 1, 0) | 6.406 (0.04063) | 9.64775 (0.02181) | 11.4214 (0.0222148) |
| ARIMA (0, 0, 1) | 3.548 (0.16963) | 343.18 (4.471e-074) | 99.6012 (1.19602e-020) |
| ARIMA (0, 1, 1) | 6.943 (0.03107) | 14.2602 (0.002572) | 8.3121 (0.0807918) |
| ARIMA (1, 1, 1) | 4.976 (0.08308) | 3.64901 (0.1613) | 12.384 (0.0147128) |
| ARIMA (2, 1, 0) | 5.343 (0.06915) | 1.05835 (0.5891) | 15.8064 (0.00329019) |
| ARIMA (0, 1, 2) | 7.162 (0.02784) | 2.50263 (0.2861) | 13.6701 (0.00842606) |
| | | | |
| Model 2<br>GCEC1Q ($Y_{2t}$) | | | |
| ARIMA (0, 0, 1) | 0.791 (0.67336) | 317.396 (1.708e-068) | 97.9932 (2.63013e-020) |
| ARIMA (2, 1, 1) | 0.491 (0.78234) | 3.69189 (0.05468) | 3.0714 (0.545949) |
| ARIMA (1, 1, 2) | 0.411 (0.81410) | 2.13669 (0.1438) | 2.68573 (0.611714) |

P value for the testing statistics is in the parenthesis. 0.02 is used as significance level for all the hypothesis testing.
Null hypothesis for ARCH Effect (LM) test is that no ARCH effect is present.
Feasible models are highlighted in red, which need to be further screened for the best appropriate model.

Ideally, studied models should pass all three tests. But ARCH model can be used to correct the heteroskedasticity if the model has passed both normality and autocorrelation tests. ARIMA (1, 1, 0) is appropriate for model 1, because it has passed all the tests. The reason for not selecting ARIMA (2, 1, 0) or ARIMA (0, 1, 2) is that they did not pass the ARCH effect test, which also have more parameters than ARIMA (1, 1, 0). However, ARIMA (1, 1, 1) would be appropriate if I choose 0.01 as significance level for all the hypothesis testing. Therefore, I will estimate both ARIMA (1, 1, 0) and ARIMA (1, 1, 1) model, then decide which one I will utilize as model 1 based on their forecasting performance and overall the goodness of fit.

Additionally, I choose ARIMA (1, 1, 2) for model 2, because it has passed all the tests with the lowest test statistics (highest p value) for each test.

Below are the estimated results of ARIMA (1, 1, 0)/ARIMA (1, 1, 1) for model 1 GDPC1Q ($Y_{1t}$) and ARIMA (1, 1, 2) for model 2 GCEC1Q ($Y_{2t}$).

Model 1.1 GDPC1Q ($Y_{1t}$): ARIMA (1, 1, 0), using observations 1980:1-2006:1 (T = 105)
Dependent variable: (1-L) GDPC1Q
Standard errors based on Hessian

| | *Coefficient* | *Std. Error* | *z* | *p-value* | |
|---|---|---|---|---|---|
| const | 66.9945 | 7.93778 | 8.4400 | <0.00001 | *** |
| phi_1 | 0.327894 | 0.0931787 | 3.5190 | 0.00043 | *** |

| | | | |
|---|---|---|---|
| Mean dependent var | 66.78000 | S.D. dependent var | 58.36696 |
| Mean of innovations | 0.148774 | S.D. of innovations | 54.91789 |
| Log-likelihood | -569.6585 | Akaike criterion | 1145.317 |
| Schwarz criterion | 1153.279 | Hannan-Quinn | 1148.543 |

| | *Real* | *Imaginary* | *Modulus* | *Frequency* |
|---|---|---|---|---|
| AR | | | | |
| Root 1 | 3.0498 | 0.0000 | 3.0498 | 0.0000 |

**For Model 1.1:**

$$\Delta^{1d}Y_{1t} = c_1 + \varphi_1\Delta^{1d}Y_{1,t-1} + \varepsilon_{1t} \quad (1.1)$$

Where $c_1$ *is a constant*, $\varepsilon_{1t}$ *is white noise*

$$\Delta^{1d} Y_{1t} = Y_{1t} - Y_{1,t-1} \tag{1.2}$$

Combine (1.1) and (1.2), we have: $Y_{1t} - Y_{1,t-1} = c_1 + \varphi_1 \left(Y_{1,t-1} - Y_{1,t-2}\right) + \varepsilon_{1t}$

Replace the symbol with estimated parameters, we have:

$$Y_{1t} - Y_{1,t-1} = 66.9945 + 0.327894\left(Y_{1,t-1} - Y_{1,t-2}\right) + \varepsilon_{1t} \tag{1.3}$$

$$\therefore Y_{1t} = 66.9945 + 1.327894 Y_{1,t-1} - 0.327894 Y_{1,t-2} + \varepsilon_{1t} \tag{1.4}$$

More simply, let first difference of GDPC1Q be d_GDPC1Q; $Y_{d_GDPC1Q}$ be $\Delta^{1d} Y_1$, then:

$$\Delta^{1d} Y_{1t} = 66.9945 + 0.327894 \Delta^{1d} Y_{1,t-1} + \varepsilon_{1t} \tag{1.5}$$

Regression residuals (= observed - fitted GDPC1Q)

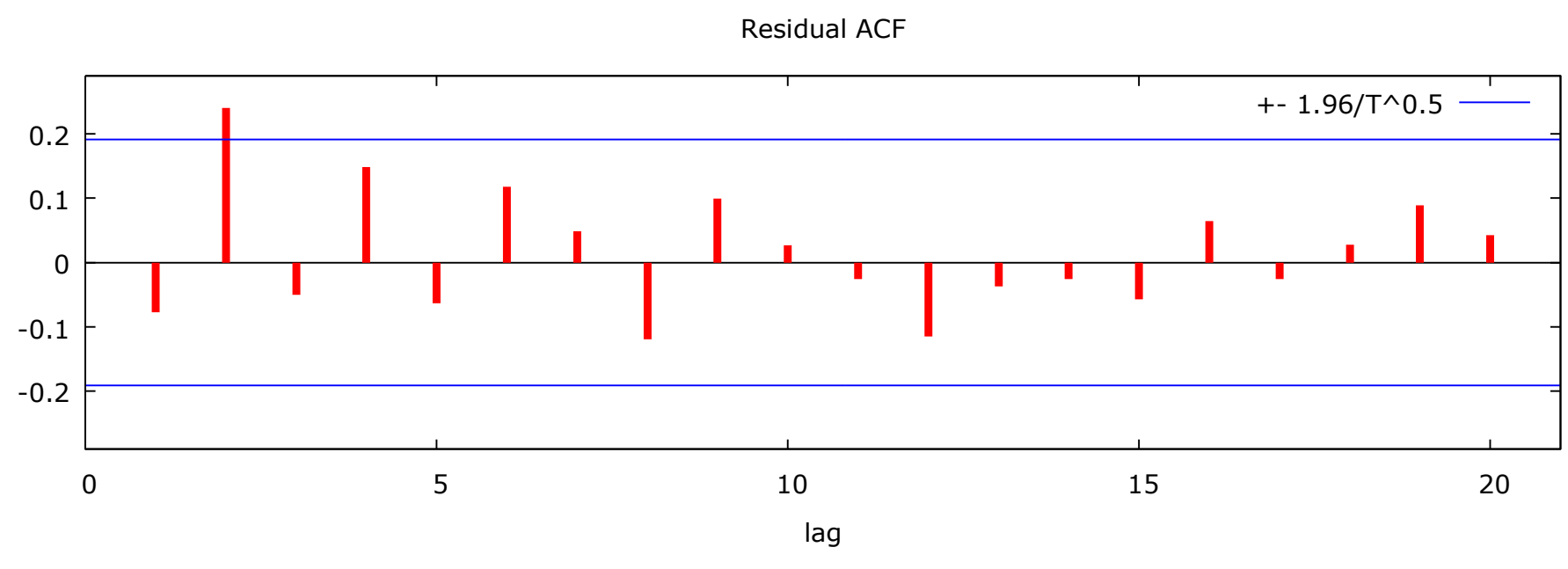

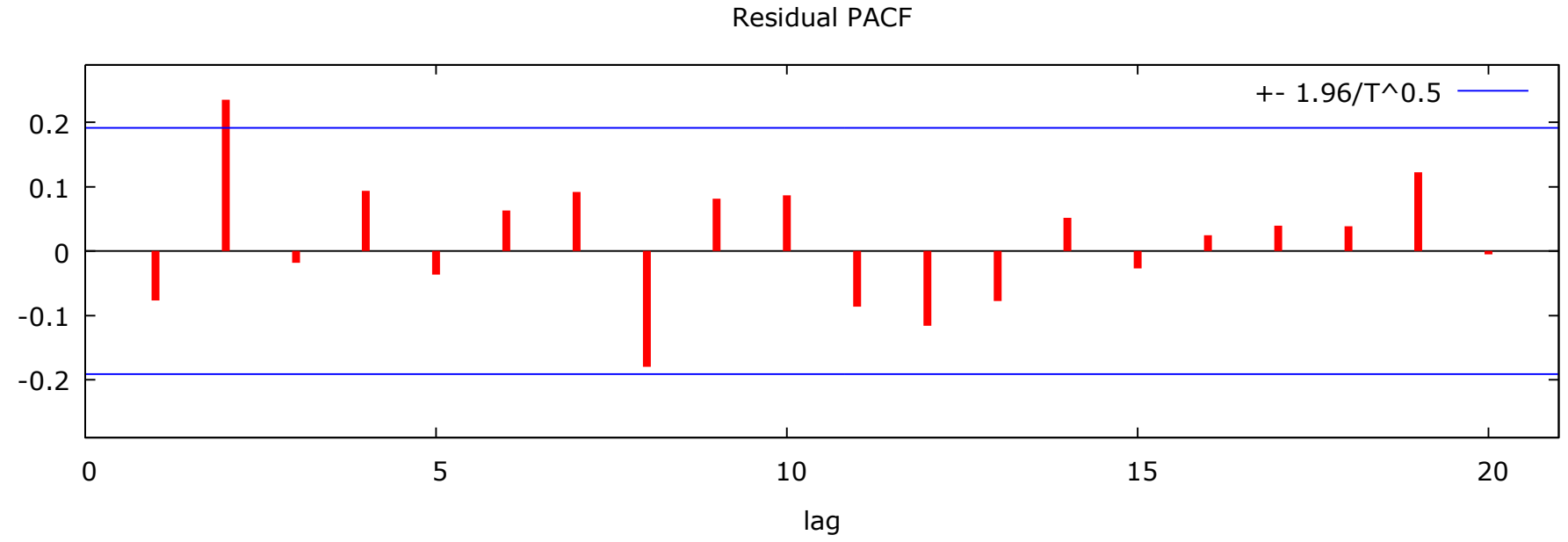

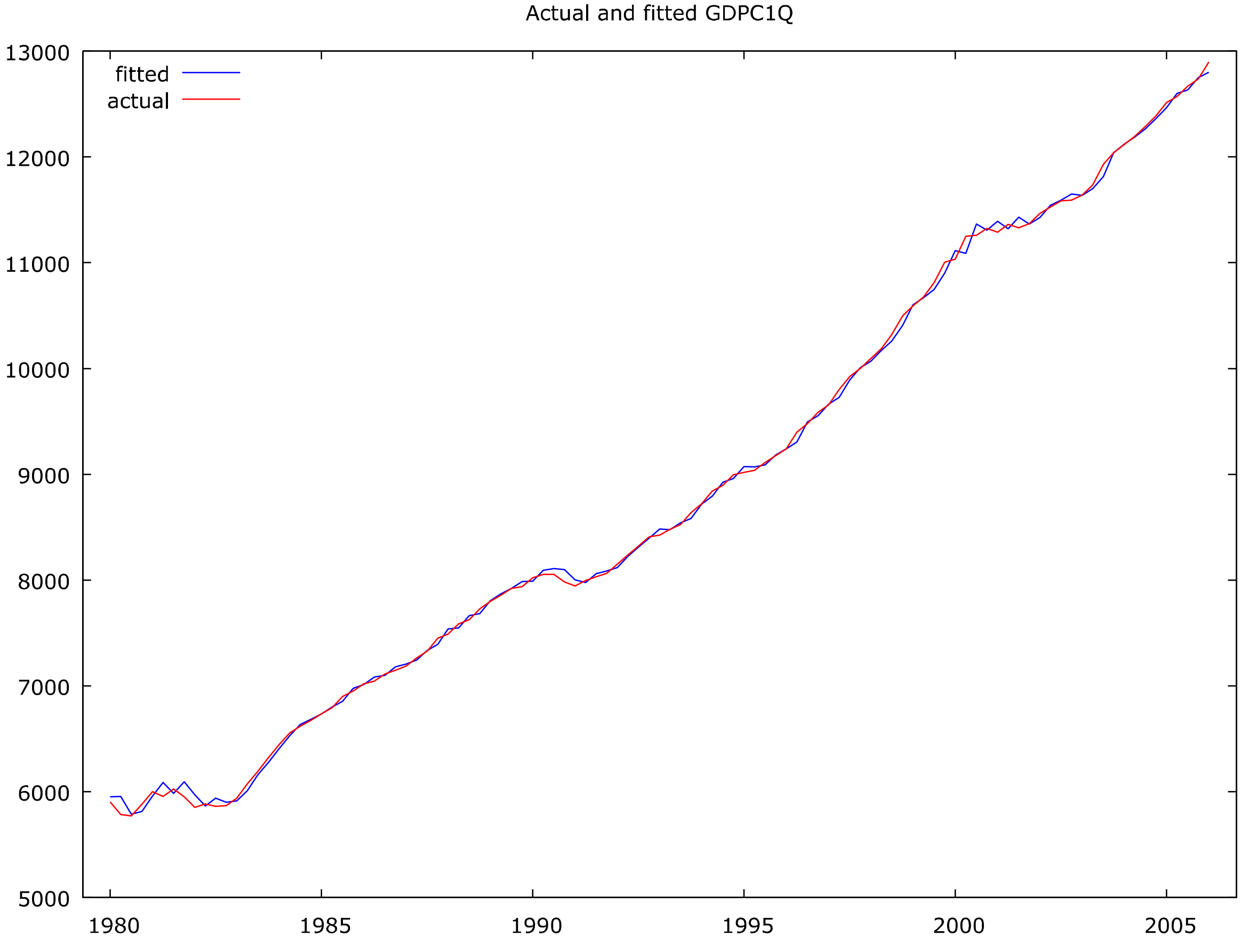

Model 1.1 estimation range: 1980:1 - 2006:1
Standard error of residuals = 54.9179

| | GDPC1Q | fitted | residual | |
|---|---|---|---|---|
| 1980:1 | 5903.40 | 5951.49 | -48.0923 | |
| 1980:2 | 5782.40 | 5954.62 | -172.223 | * |
| 1980:3 | 5771.70 | 5787.75 | -16.0508 | |
| 1980:4 | 5878.40 | 5813.22 | 65.1825 | |
| 1981:1 | 6000.60 | 5958.41 | 42.1878 | |
| 1981:2 | 5952.70 | 6085.69 | -132.995 | |
| 1981:3 | 6025.00 | 5982.02 | 42.9802 | |
| 1981:4 | 5950.00 | 6093.73 | -143.733 | * |
| 1982:1 | 5852.30 | 5970.43 | -118.134 | |
| 1982:2 | 5884.00 | 5865.29 | 18.7093 | |
| 1982:3 | 5861.40 | 5939.42 | -78.0202 | |
| 1982:4 | 5866.00 | 5899.02 | -33.0155 | |
| 1983:1 | 5938.90 | 5912.53 | 26.3657 | |
| 1983:2 | 6072.40 | 6007.83 | 64.5706 | |
| 1983:3 | 6192.20 | 6161.20 | 31.0002 | |
| 1983:4 | 6320.20 | 6276.51 | 43.6924 | |

| | | | |
|---|---|---|---|
| 1984:1 | 6442.80 | 6407.20 | 35.6036 |
| 1984:2 | 6554.00 | 6528.03 | 25.9743 |
| 1984:3 | 6617.70 | 6635.49 | -17.7877 |
| 1984:4 | 6671.60 | 6683.61 | -12.0128 |
| 1985:1 | 6734.50 | 6734.30 | 0.200579 |
| 1985:2 | 6791.50 | 6800.15 | -8.65047 |
| 1985:3 | 6897.60 | 6855.22 | 42.3841 |
| 1985:4 | 6950.00 | 6977.42 | -27.4155 |
| 1986:1 | 7016.80 | 7012.21 | 4.59242 |
| 1986:2 | 7045.00 | 7083.73 | -38.7293 |
| 1986:3 | 7112.90 | 7099.27 | 13.6275 |
| 1986:4 | 7147.30 | 7180.19 | -32.8899 |
| 1987:1 | 7186.90 | 7203.61 | -16.7055 |
| 1987:2 | 7263.30 | 7244.91 | 18.3895 |
| 1987:3 | 7326.30 | 7333.38 | -7.07703 |
| 1987:4 | 7451.70 | 7391.98 | 59.7167 |
| 1988:1 | 7490.20 | 7537.84 | -47.6438 |
| 1988:2 | 7586.40 | 7547.85 | 38.5501 |
| 1988:3 | 7625.60 | 7662.97 | -37.3693 |
| 1988:4 | 7727.40 | 7683.48 | 43.9206 |
| 1989:1 | 7799.90 | 7805.81 | -5.90554 |
| 1989:2 | 7858.30 | 7868.70 | -10.3982 |
| 1989:3 | 7920.60 | 7922.47 | -1.87494 |
| 1989:4 | 7937.90 | 7986.05 | -48.1537 |
| 1990:1 | 8020.80 | 7988.60 | 32.2015 |
| 1990:2 | 8052.70 | 8093.01 | -40.3083 |
| 1990:3 | 8052.60 | 8108.19 | -55.5858 |
| 1990:4 | 7982.00 | 8097.59 | -115.593 |
| 1991:1 | 7943.40 | 8003.88 | -60.4766 |
| 1991:2 | 7997.00 | 7975.77 | 21.2308 |
| 1991:3 | 8030.70 | 8059.60 | -28.9011 |
| 1991:4 | 8062.20 | 8086.78 | -24.5760 |
| 1992:1 | 8150.70 | 8117.55 | 33.1454 |
| 1992:2 | 8237.30 | 8224.74 | 12.5555 |
| 1992:3 | 8322.30 | 8310.72 | 11.5785 |
| 1992:4 | 8409.80 | 8395.20 | 14.6031 |
| 1993:1 | 8425.30 | 8483.52 | -58.2167 |
| 1993:2 | 8479.20 | 8475.41 | 3.79170 |
| 1993:3 | 8523.80 | 8541.90 | -18.0994 |
| 1993:4 | 8636.40 | 8583.45 | 52.9500 |
| 1994:1 | 8720.50 | 8718.35 | 2.15321 |
| 1994:2 | 8839.80 | 8793.10 | 46.6982 |
| 1994:3 | 8896.70 | 8923.94 | -27.2437 |
| 1994:4 | 8995.50 | 8960.38 | 35.1169 |
| 1995:1 | 9017.60 | 9072.92 | -55.3219 |
| 1995:2 | 9037.00 | 9069.87 | -32.8724 |

| | | | | |
|---|---|---|---|---|
| 1995:3 | 9112.90 | 9088.39 | 24.5129 | |
| 1995:4 | 9176.40 | 9182.81 | -6.41309 | |
| 1996:1 | 9239.30 | 9242.25 | -2.94720 | |
| 1996:2 | 9399.00 | 9304.95 | 94.0495 | |
| 1996:3 | 9480.80 | 9496.39 | -15.5906 | |
| 1996:4 | 9584.30 | 9552.65 | 31.6523 | |
| 1997:1 | 9658.00 | 9663.26 | -5.26296 | |
| 1997:2 | 9801.20 | 9727.19 | 74.0083 | |
| 1997:3 | 9924.20 | 9893.18 | 31.0197 | |
| 1997:4 | 10000.3 | 10009.6 | -9.25688 | |
| 1998:1 | 10094.8 | 10070.3 | 24.5213 | |
| 1998:2 | 10185.6 | 10170.8 | 14.7881 | |
| 1998:3 | 10320.0 | 10260.4 | 59.6013 | |
| 1998:4 | 10498.6 | 10409.1 | 89.5051 | |
| 1999:1 | 10592.1 | 10602.2 | -10.0878 | |
| 1999:2 | 10674.9 | 10667.8 | 7.11598 | |
| 1999:3 | 10810.7 | 10747.1 | 63.6244 | |
| 1999:4 | 11004.8 | 10900.3 | 104.546 | |
| 2000:1 | 11033.6 | 11113.5 | -79.8701 | |
| 2000:2 | 11248.8 | 11088.1 | 160.731 | * |
| 2000:3 | 11258.3 | 11364.4 | -106.089 | |
| 2000:4 | 11325.0 | 11306.4 | 18.5591 | |
| 2001:1 | 11287.8 | 11391.9 | -104.096 | |
| 2001:2 | 11361.7 | 11320.6 | 41.0717 | |
| 2001:3 | 11330.4 | 11431.0 | -100.557 | |
| 2001:4 | 11370.0 | 11365.2 | 4.83713 | |
| 2002:1 | 11467.1 | 11428.0 | 39.0895 | |
| 2002:2 | 11528.1 | 11544.0 | -15.8644 | |
| 2002:3 | 11586.6 | 11593.1 | -6.52747 | |
| 2002:4 | 11590.6 | 11650.8 | -60.2077 | |
| 2003:1 | 11638.9 | 11636.9 | 1.96248 | |
| 2003:2 | 11737.5 | 11699.8 | 37.7368 | |
| 2003:3 | 11930.7 | 11814.9 | 115.844 | |
| 2003:4 | 12038.6 | 12039.1 | -0.475032 | |
| 2004:1 | 12117.9 | 12119.0 | -1.10569 | |
| 2004:2 | 12195.9 | 12188.9 | 6.97208 | |
| 2004:3 | 12286.7 | 12266.5 | 20.1983 | |
| 2004:4 | 12387.2 | 12361.5 | 25.7013 | |
| 2005:1 | 12515.0 | 12465.2 | 49.8207 | |
| 2005:2 | 12570.7 | 12601.9 | -31.2308 | |
| 2005:3 | 12670.5 | 12634.0 | 36.5104 | |
| 2005:4 | 12735.6 | 12748.2 | -12.6497 | |
| 2006:1 | 12896.4 | 12802.0 | 94.4282 | |

Note: * denotes a residual in excess of 2.5 standard errors

Model 1.2 GDPC1Q ($Y_{1t}$): ARIMA (1, 1, 1), using observations 1980:1-2006:1 (T = 105)
Dependent variable: (1-L) GDPC1Q
Standard errors based on Hessian

| | *Coefficient* | *Std. Error* | *z* | *p-value* | |
|---|---|---|---|---|---|
| Const | 66.2629 | 11.1287 | 5.9542 | <0.00001 | *** |
| phi_1 | 0.759918 | 0.139983 | 5.4286 | <0.00001 | *** |
| theta_1 | -0.479305 | 0.180154 | -2.6605 | 0.00780 | *** |

| | | | |
|---|---|---|---|
| Mean dependent var | 66.78000 | S.D. dependent var | 58.36696 |
| Mean of innovations | 0.894814 | S.D. of innovations | 53.61904 |
| Log-likelihood | -567.1968 | Akaike criterion | 1142.394 |
| Schwarz criterion | 1153.010 | Hannan-Quinn | 1146.695 |

| | | *Real* | *Imaginary* | *Modulus* | *Frequency* |
|---|---|---|---|---|---|
| AR | | | | | |
| | Root 1 | 1.3159 | 0.0000 | 1.3159 | 0.0000 |
| MA | | | | | |
| | Root 1 | 2.0864 | 0.0000 | 2.0864 | 0.0000 |

**For Model 1.2:**

$$\Delta^{1d}Y_{1t} = c_1 + \varphi_1\Delta^{1d}Y_{1,t-1} + \varepsilon_{1t} + \theta_1\varepsilon_{1,t-1} \tag{1.6}$$

Where $c_1$ *is a constant*, $\varepsilon_{1t}$ *is white noise*

$$\Delta^{1d}Y_{1t} = Y_{1t} - Y_{1,t-1} \tag{1.7}$$

Combine (1.6) and (1.7), we have: $Y_{1t} - Y_{1,t-1} = c_1 + \varphi_1(Y_{1,t-1} - Y_{1,t-2}) + \varepsilon_{1t} + \theta_1\varepsilon_{1,t-1}$

Replace the symbol with estimated parameters, we have:

$$Y_{1t} - Y_{1,t-1} = 66.2629 + 0.759918(Y_{1,t-1} - Y_{1,t-2}) + \varepsilon_{1t} - 0.479305\varepsilon_{1,t-1} \tag{1.8}$$

$$\therefore Y_{1t} = 66.2629 + 1.759918Y_{1,t-1} - 0.759918Y_{1,t-2} + \varepsilon_{1t} - 0.479305\varepsilon_{1,t-1} \tag{1.9}$$

More simply, let first difference of GDPC1Q be d_GDPC1Q; $Y_{d_GDPC1Q}$ be $\Delta^{1d}Y_1$, then:

$$\Delta^{1d}Y_{1t} = 66.2629 + 0.759918\Delta^{1d}Y_{1,t-1} + \varepsilon_{1t} - 0.479305\varepsilon_{1,t-1} \tag{1.10}$$

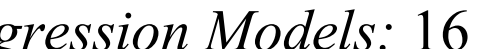

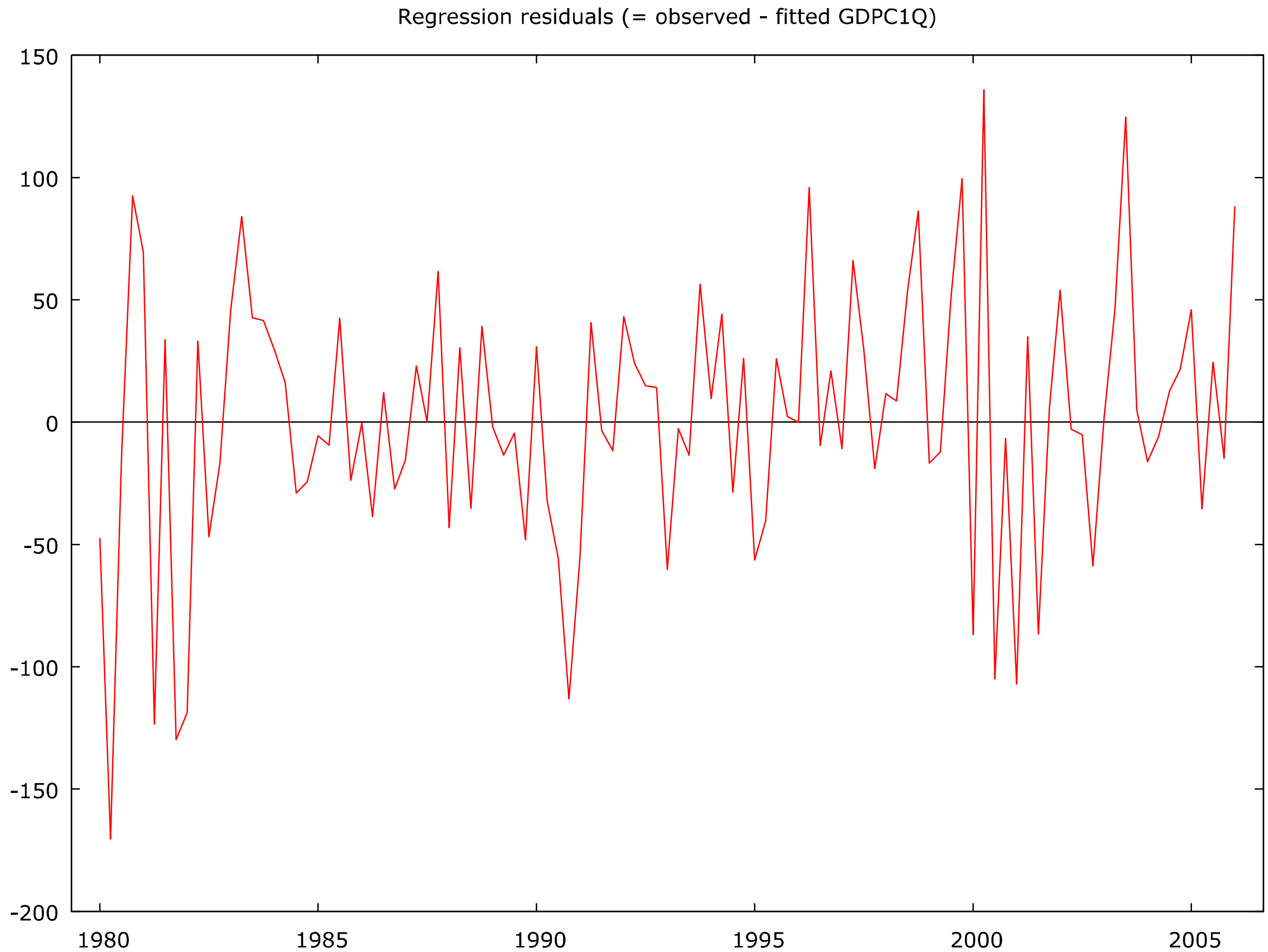

Regression residuals (= observed - fitted GDPC1Q)

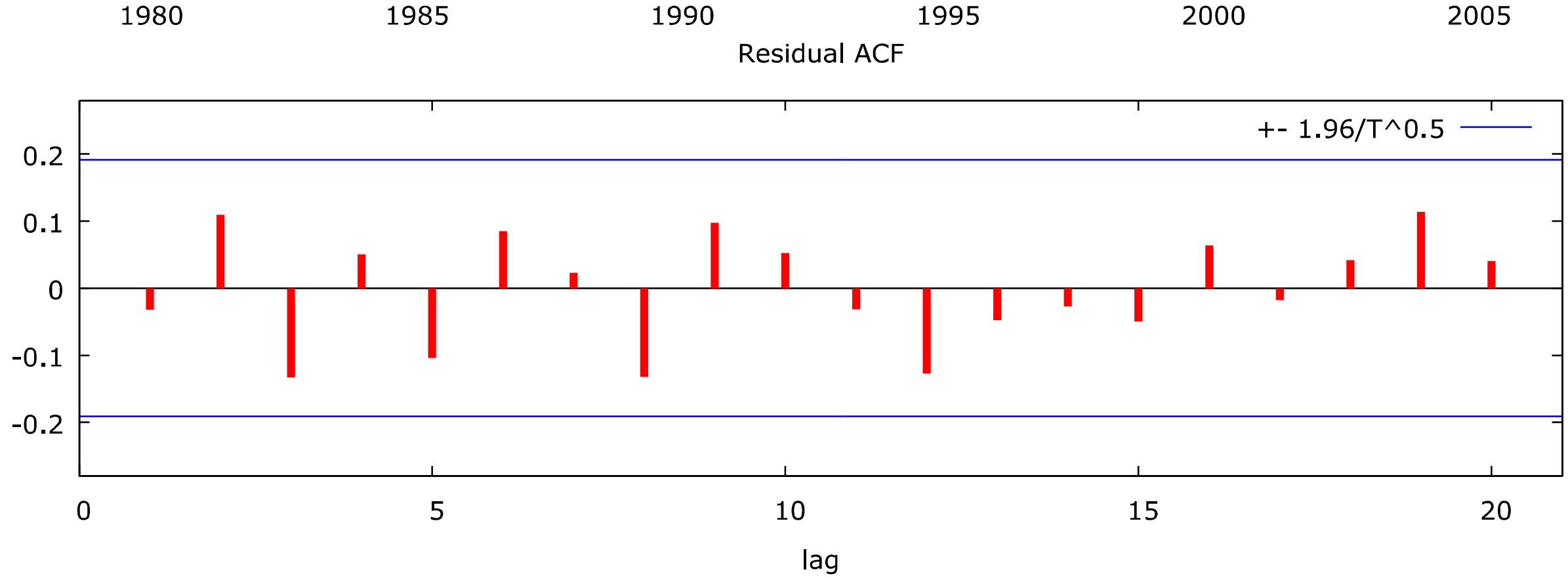

Residual ACF

+- 1.96/T^0.5

lag

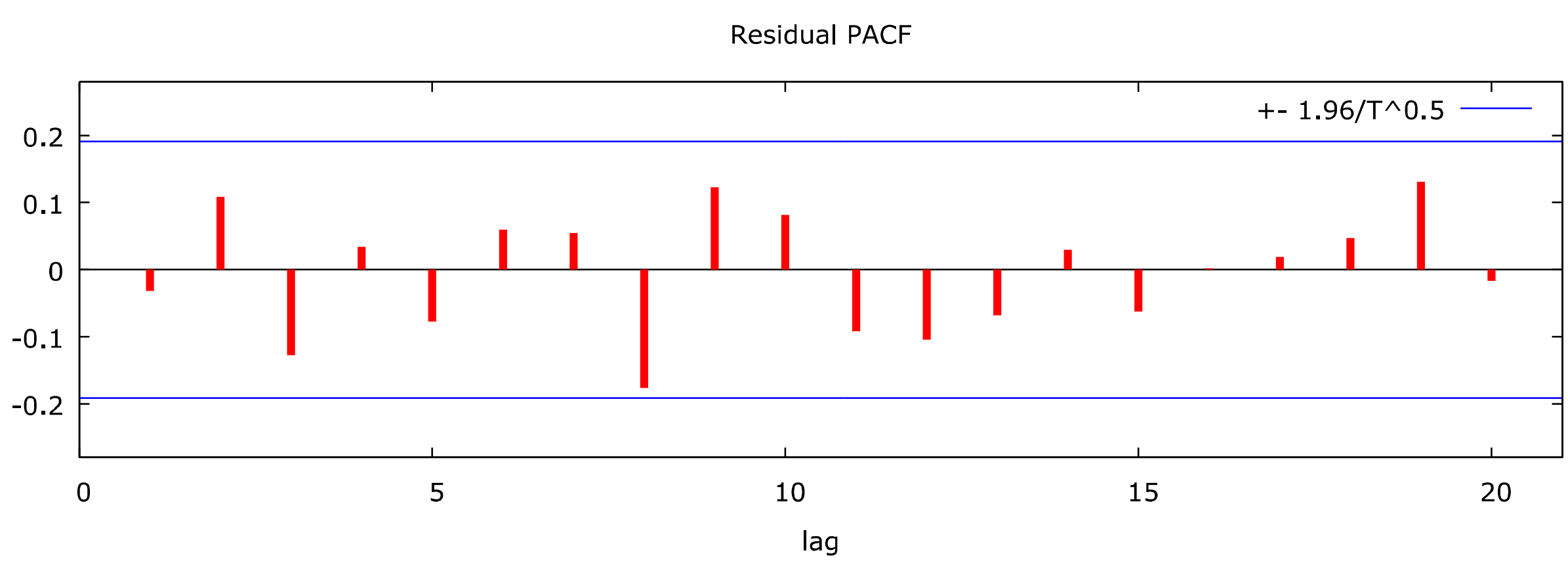

Residual PACF

+- 1.96/T^0.5

lag

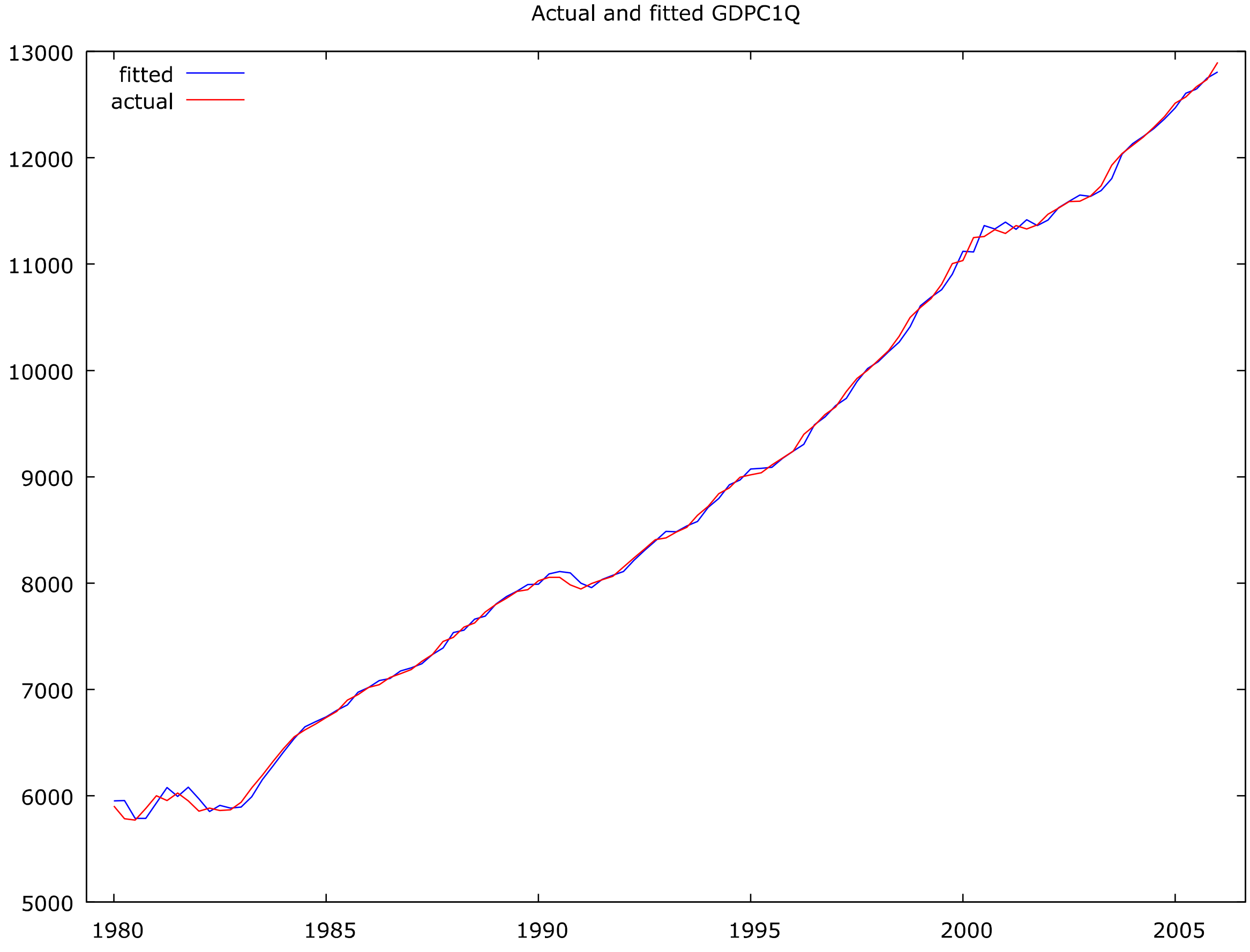

Model 1.2 estimation range: 1980:1 - 2006:1
Standard error of residuals = 53.619

| | GDPC1Q | fitted | residual | |
|---|---|---|---|---|
| 1980:1 | 5903.40 | 5950.76 | -47.3624 | |
| 1980:2 | 5782.40 | 5952.81 | -170.406 | * |
| 1980:3 | 5771.70 | 5785.19 | -13.4899 | |
| 1980:4 | 5878.40 | 5785.89 | 92.5081 | |
| 1981:1 | 6000.60 | 5931.13 | 69.4672 | |
| 1981:2 | 5952.70 | 6076.09 | -123.388 | |
| 1981:3 | 6025.00 | 5991.34 | 33.6567 | |
| 1981:4 | 5950.00 | 6079.72 | -129.719 | |
| 1982:1 | 5852.30 | 5971.09 | -118.789 | |
| 1982:2 | 5884.00 | 5850.90 | 33.0992 | |
| 1982:3 | 5861.40 | 5908.13 | -46.7333 | |
| 1982:4 | 5866.00 | 5882.53 | -16.5338 | |
| 1983:1 | 5938.90 | 5893.33 | 45.5711 | |
| 1983:2 | 6072.40 | 5988.36 | 84.0360 | |
| 1983:3 | 6192.20 | 6149.48 | 42.7214 | |
| 1983:4 | 6320.20 | 6278.67 | 41.5300 | |

| | | | |
|---|---|---|---|
| 1984:1 | 6442.80 | 6413.47 | 29.3277 |
| 1984:2 | 6554.00 | 6537.82 | 16.1826 |
| 1984:3 | 6617.70 | 6646.65 | -28.9549 |
| 1984:4 | 6671.60 | 6695.89 | -24.2934 |
| 1985:1 | 6734.50 | 6740.11 | -5.61192 |
| 1985:2 | 6791.50 | 6800.90 | -9.39708 |
| 1985:3 | 6897.60 | 6855.23 | 42.3722 |
| 1985:4 | 6950.00 | 6973.83 | -23.8265 |
| 1986:1 | 7016.80 | 7017.15 | -0.348276 |
| 1986:2 | 7045.00 | 7083.64 | -38.6379 |
| 1986:3 | 7112.90 | 7100.86 | 12.0426 |
| 1986:4 | 7147.30 | 7174.63 | -27.3348 |
| 1987:1 | 7186.90 | 7202.45 | -15.5513 |
| 1987:2 | 7263.30 | 7240.36 | 22.9450 |
| 1987:3 | 7326.30 | 7326.27 | 0.0314818 |
| 1987:4 | 7451.70 | 7390.07 | 61.6318 |
| 1988:1 | 7490.20 | 7533.36 | -43.1617 |
| 1988:2 | 7586.40 | 7556.05 | 30.3471 |
| 1988:3 | 7625.60 | 7660.87 | -35.2670 |
| 1988:4 | 7727.40 | 7688.20 | 39.1992 |
| 1989:1 | 7799.90 | 7801.88 | -1.97971 |
| 1989:2 | 7858.30 | 7871.85 | -13.5513 |
| 1989:3 | 7920.60 | 7925.08 | -4.48285 |
| 1989:4 | 7937.90 | 7986.00 | -48.1000 |
| 1990:1 | 8020.80 | 7990.01 | 30.7904 |
| 1990:2 | 8052.70 | 8084.95 | -32.2476 |
| 1990:3 | 8052.60 | 8108.31 | -55.7063 |
| 1990:4 | 7982.00 | 8095.13 | -113.133 |
| 1991:1 | 7943.40 | 7998.48 | -55.0834 |
| 1991:2 | 7997.00 | 7956.38 | 40.6226 |
| 1991:3 | 8030.70 | 8034.17 | -3.46948 |
| 1991:4 | 8062.20 | 8073.88 | -11.6806 |
| 1992:1 | 8150.70 | 8107.64 | 43.0555 |
| 1992:2 | 8237.30 | 8213.22 | 24.0756 |
| 1992:3 | 8322.30 | 8307.48 | 14.8222 |
| 1992:4 | 8409.80 | 8395.70 | 14.1029 |
| 1993:1 | 8425.30 | 8485.44 | -60.1416 |
| 1993:2 | 8479.20 | 8481.81 | -2.61333 |
| 1993:3 | 8523.80 | 8537.32 | -13.5206 |
| 1993:4 | 8636.40 | 8580.08 | 56.3187 |
| 1994:1 | 8720.50 | 8710.88 | 9.61867 |
| 1994:2 | 8839.80 | 8795.71 | 44.0928 |
| 1994:3 | 8896.70 | 8925.23 | -28.5327 |
| 1994:4 | 8995.50 | 8969.52 | 25.9764 |
| 1995:1 | 9017.60 | 9074.04 | -56.4377 |
| 1995:2 | 9037.00 | 9077.35 | -40.3535 |

| | | | | |
|---|---|---|---|---|
| 1995:3 | 9112.90 | 9086.99 | 25.9075 | |
| 1995:4 | 9176.40 | 9174.07 | 2.33140 | |
| 1996:1 | 9239.30 | 9239.45 | -0.145765 | |
| 1996:2 | 9399.00 | 9303.08 | 95.9229 | |
| 1996:3 | 9480.80 | 9490.29 | -9.49098 | |
| 1996:4 | 9584.30 | 9563.42 | 20.8813 | |
| 1997:1 | 9658.00 | 9668.85 | -10.8514 | |
| 1997:2 | 9801.20 | 9735.12 | 66.0845 | |
| 1997:3 | 9924.20 | 9894.25 | 29.9460 | |
| 1997:4 | 10000.3 | 10019.2 | -18.9250 | |
| 1998:1 | 10094.8 | 10083.1 | 11.6910 | |
| 1998:2 | 10185.6 | 10176.9 | 8.68292 | |
| 1998:3 | 10320.0 | 10266.3 | 53.6528 | |
| 1998:4 | 10498.6 | 10412.3 | 86.2747 | |
| 1999:1 | 10592.1 | 10608.9 | -16.7778 | |
| 1999:2 | 10674.9 | 10687.1 | -12.2024 | |
| 1999:3 | 10810.7 | 10759.6 | 51.1218 | |
| 1999:4 | 11004.8 | 10905.3 | 99.4977 | |
| 2000:1 | 11033.6 | 11120.5 | -86.9187 | |
| 2000:2 | 11248.8 | 11113.1 | 135.745 | * |
| 2000:3 | 11258.3 | 11363.2 | -104.879 | |
| 2000:4 | 11325.0 | 11331.7 | -6.69674 | |
| 2001:1 | 11287.8 | 11394.8 | -107.005 | |
| 2001:2 | 11361.7 | 11326.7 | 34.9726 | |
| 2001:3 | 11330.4 | 11417.0 | -86.6038 | |
| 2001:4 | 11370.0 | 11364.0 | 5.96732 | |
| 2002:1 | 11467.1 | 11413.1 | 53.9589 | |
| 2002:2 | 11528.1 | 11530.9 | -2.83368 | |
| 2002:3 | 11586.6 | 11591.7 | -5.12162 | |
| 2002:4 | 11590.6 | 11649.4 | -58.8184 | |
| 2003:1 | 11638.9 | 11637.7 | 1.15991 | |
| 2003:2 | 11737.5 | 11691.0 | 46.5435 | |
| 2003:3 | 11930.7 | 11806.0 | 124.672 | |
| 2003:4 | 12038.6 | 12033.7 | 4.93147 | |
| 2004:1 | 12117.9 | 12134.1 | -16.2398 | |
| 2004:2 | 12195.9 | 12201.9 | -5.95371 | |
| 2004:3 | 12286.7 | 12273.9 | 12.7644 | |
| 2004:4 | 12387.2 | 12365.5 | 21.7091 | |
| 2005:1 | 12515.0 | 12469.1 | 45.9251 | |
| 2005:2 | 12570.7 | 12606.0 | -35.3138 | |
| 2005:3 | 12670.5 | 12645.9 | 24.6381 | |
| 2005:4 | 12735.6 | 12750.4 | -14.8390 | |
| 2006:1 | 12896.4 | 12808.1 | 88.3085 | |

Note: * denotes a residual in excess of 2.5 standard errors

Model 1.3 GCEC1Q ($Y_{2t}$): ARIMA (1, 1, 2), using observations 1980:1-2006:1 (T = 105)

Dependent variable: (1-L) GCEC1Q
Standard errors based on Hessian

| | *Coefficient* | *Std. Error* | *z* | *p-value* | |
|---|---|---|---|---|---|
| const | 9.7675 | 2.54067 | 3.8445 | 0.00012 | *** |
| phi_1 | 0.811342 | 0.12216 | 6.6416 | <0.00001 | *** |
| theta_1 | -0.943283 | 0.141859 | -6.6494 | <0.00001 | *** |
| theta_2 | 0.273345 | 0.11415 | 2.3946 | 0.01664 | ** |

| | | | |
|---|---|---|---|
| Mean dependent var | 10.03143 | S.D. dependent var | 15.92138 |
| Mean of innovations | 0.127798 | S.D. of innovations | 15.25431 |
| Log-likelihood | -435.2311 | Akaike criterion | 880.4622 |
| Schwarz criterion | 893.7320 | Hannan-Quinn | 885.8394 |

| | | *Real* | *Imaginary* | *Modulus* | *Frequency* |
|---|---|---|---|---|---|
| AR | | | | | |
| | Root 1 | 1.2325 | 0.0000 | 1.2325 | 0.0000 |
| MA | | | | | |
| | Root 1 | 1.7254 | -0.8254 | 1.9127 | -0.0710 |
| | Root 2 | 1.7254 | 0.8254 | 1.9127 | 0.0710 |

**For Model 1.3:**

$$\Delta^{1d}Y_{2t} = c_2 + \varphi_1 \Delta^{1d}Y_{2,t-1} + \varepsilon_{2t} + \theta_1 \varepsilon_{2,t-1} + \theta_2 \varepsilon_{2,t-2} \quad (1.11)$$

Where $c_2$ *is a constant*, $\varepsilon_{2t}$ *is white noise*

$$\Delta^{1d}Y_{2t} = Y_{2t} - Y_{2,t-1} \quad (1.12)$$

Combine (1.11) and (1.12): $Y_{2t} - Y_{2,t-1} = c_2 + \varphi_1 (Y_{2,t-1} - Y_{2,t-2}) + \varepsilon_{2t} + \theta_1 \varepsilon_{2,t-1} + \theta_2 \varepsilon_{2,t-2}$

Replace the symbol with estimated parameters, we have:

$$Y_{2t} - Y_{2,t-1} = 9.7675 + 0.811342(Y_{2,t-1} - Y_{2,t-2}) + \varepsilon_{2t} - 0.943283\varepsilon_{2,t-1} + 0.273345\varepsilon_{2,t-2} \quad (1.13)$$

$$Y_{2t} = 9.7675 + 1.811342Y_{2,t-1} - 0.811342Y_{2,t-2} + \varepsilon_{2t} - 0.943283\varepsilon_{2,t-1} + 0.273345\varepsilon_{2,t-2} \quad (1.14)$$

More simply, let first difference of GCEC1Q be d_GCEC1Q; $Y_{d_GCEC1Q}$ be $\Delta^{1d}Y_2$, then:

$$\Delta^{1d}Y_{2t} = 9.7675 + 0.811342\Delta^{1d}Y_{2,t-1} + \varepsilon_{2t} - 0.943283\varepsilon_{2,t-1} + 0.273345\varepsilon_{2,t-2} \quad (1.15)$$

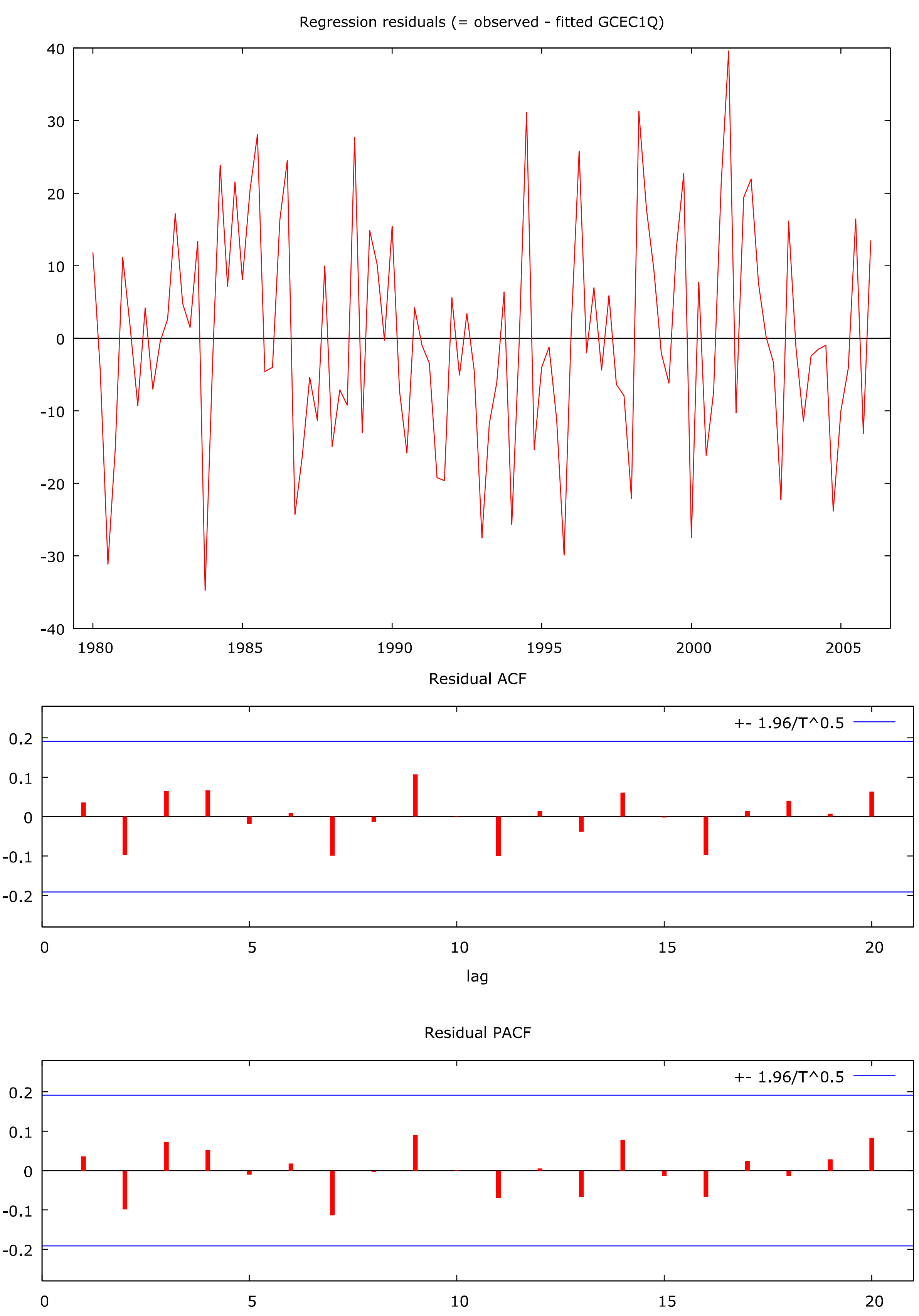

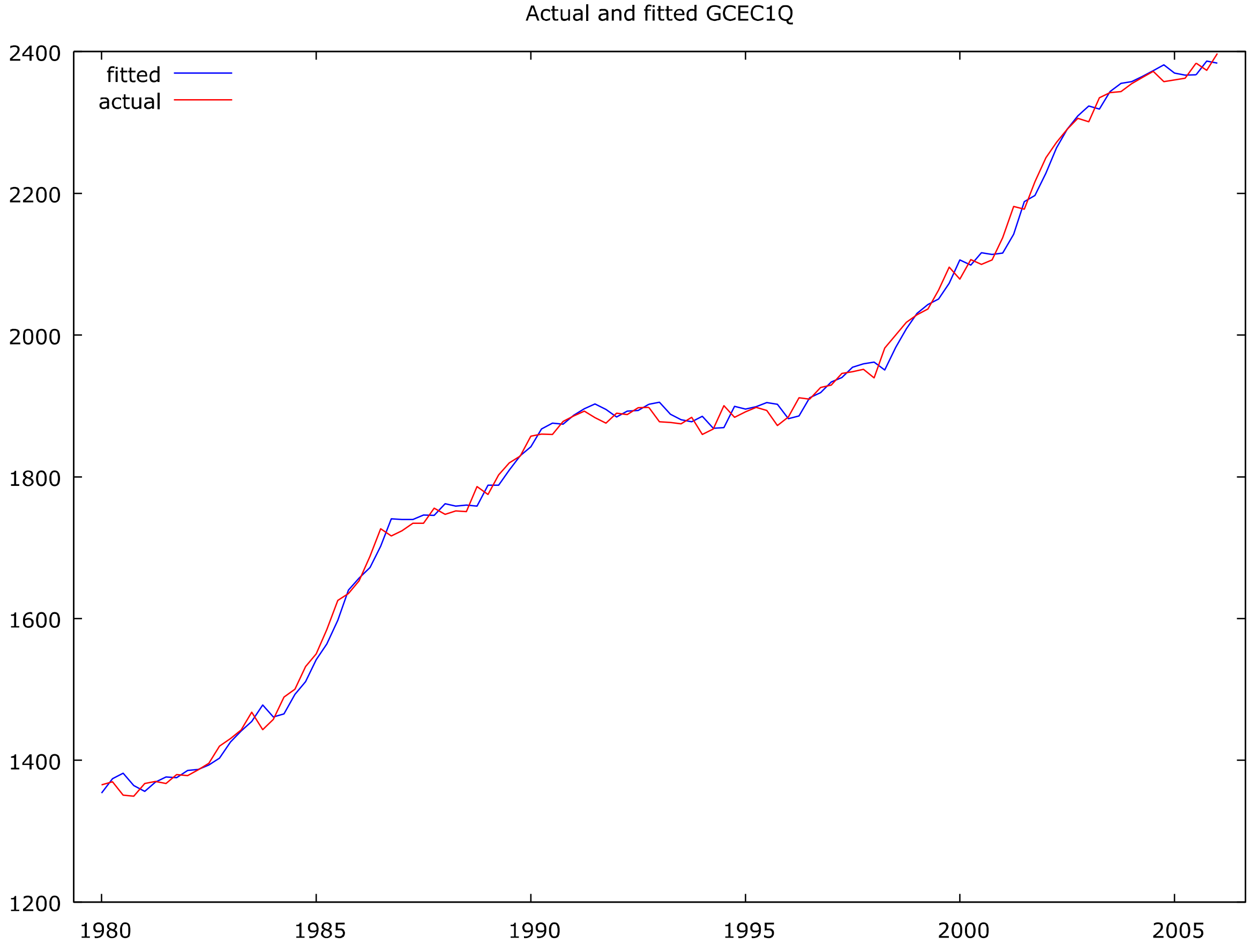

Model 1.3 estimation range: 1980:1 - 2006:1
Standard error of residuals = 15.2543

| | GCEC1Q | fitted | residual |
|---|---|---|---|
| 1980:1 | 1365.40 | 1353.57 | 11.8325 |
| 1980:2 | 1369.70 | 1374.22 | -4.51706 |
| 1980:3 | 1350.80 | 1381.94 | -31.1393 |
| 1980:4 | 1349.40 | 1364.51 | -15.1100 |
| 1981:1 | 1367.30 | 1356.14 | 11.1595 |
| 1981:2 | 1370.40 | 1369.12 | 1.27825 |
| 1981:3 | 1367.30 | 1376.59 | -9.28578 |
| 1981:4 | 1379.90 | 1375.74 | 4.16444 |
| 1982:1 | 1378.50 | 1385.50 | -6.99946 |
| 1982:2 | 1386.50 | 1386.95 | -0.447576 |
| 1982:3 | 1396.00 | 1393.34 | 2.65761 |
| 1982:4 | 1420.10 | 1402.92 | 17.1788 |
| 1983:1 | 1430.80 | 1426.02 | 4.78191 |
| 1983:2 | 1443.00 | 1441.51 | 1.49089 |
| 1983:3 | 1468.00 | 1454.64 | 13.3581 |
| 1983:4 | 1443.20 | 1477.93 | -34.7333 |
| 1984:1 | 1457.80 | 1461.34 | -3.53614 |
| 1984:2 | 1489.20 | 1465.33 | 23.8703 |
| 1984:3 | 1500.20 | 1493.04 | 7.16417 |
| 1984:4 | 1532.30 | 1510.73 | 21.5655 |
| 1985:1 | 1549.90 | 1541.80 | 8.09730 |
| 1985:2 | 1584.70 | 1564.28 | 20.4209 |
| 1985:3 | 1625.80 | 1597.73 | 28.0719 |
| 1985:4 | 1635.50 | 1640.09 | -4.59111 |
| 1986:1 | 1653.20 | 1657.22 | -4.01676 |
| 1986:2 | 1688.30 | 1671.94 | 16.3625 |
| 1986:3 | 1726.60 | 1702.09 | 24.5116 |
| 1986:4 | 1716.60 | 1740.87 | -24.2683 |
| 1987:1 | 1723.70 | 1739.92 | -16.2213 |
| 1987:2 | 1734.60 | 1739.97 | -5.37092 |
| 1987:3 | 1734.60 | 1745.92 | -11.3186 |
| 1987:4 | 1755.60 | 1745.65 | 9.94874 |
| 1988:1 | 1747.10 | 1762.00 | -14.9025 |
| 1988:2 | 1751.70 | 1758.82 | -7.12305 |
| 1988:3 | 1750.70 | 1759.92 | -9.22040 |
| 1988:4 | 1786.20 | 1758.48 | 27.7182 |
| 1989:1 | 1775.20 | 1788.18 | -12.9789 |
| 1989:2 | 1802.80 | 1787.94 | 14.8627 |
| 1989:3 | 1819.70 | 1809.47 | 10.2316 |
| 1989:4 | 1829.40 | 1829.67 | -0.265702 |
| 1990:1 | 1857.60 | 1842.16 | 15.4399 |
| 1990:2 | 1860.40 | 1867.69 | -7.28578 |
| 1990:3 | 1859.80 | 1875.61 | -15.8074 |
| 1990:4 | 1878.30 | 1874.08 | 4.22474 |
| 1991:1 | 1885.90 | 1886.85 | -0.946534 |
| 1991:2 | 1892.50 | 1895.96 | -3.45658 |
| 1991:3 | 1883.50 | 1902.70 | -19.1994 |
| 1991:4 | 1875.60 | 1895.21 | -19.6062 |
| 1992:1 | 1889.90 | 1884.28 | 5.62072 |
| 1992:2 | 1887.60 | 1892.68 | -5.08371 |
| 1992:3 | 1897.30 | 1893.91 | 3.39160 |
| 1992:4 | 1897.90 | 1902.42 | -4.52389 |
| 1993:1 | 1877.90 | 1905.42 | -27.5239 |
| 1993:2 | 1876.50 | 1888.24 | -11.7421 |
| 1993:3 | 1874.60 | 1880.76 | -6.15944 |
| 1993:4 | 1883.90 | 1877.50 | 6.39839 |
| 1994:1 | 1859.90 | 1885.57 | -25.6690 |
| 1994:2 | 1867.70 | 1868.23 | -0.532646 |
| 1994:3 | 1900.50 | 1869.36 | 31.1429 |
| 1994:4 | 1884.10 | 1899.43 | -15.3326 |

| | | | | |
|---|---|---|---|---|
| 1995:1 | 1891.60 | 1895.61 | -4.01243 | |
| 1995:2 | 1897.90 | 1899.12 | -1.22155 | |
| 1995:3 | 1893.70 | 1904.91 | -11.2097 | |
| 1995:4 | 1872.50 | 1902.38 | -29.8750 | |
| 1996:1 | 1884.50 | 1882.26 | 2.24122 | |
| 1996:2 | 1911.60 | 1885.80 | 25.8015 | |
| 1996:3 | 1909.70 | 1911.70 | -2.00461 | |
| 1996:4 | 1925.90 | 1918.94 | 6.95521 | |
| 1997:1 | 1929.40 | 1933.78 | -4.37778 | |
| 1997:2 | 1946.00 | 1940.11 | 5.88693 | |
| 1997:3 | 1948.20 | 1954.56 | -6.36131 | |
| 1997:4 | 1951.50 | 1959.44 | -7.93735 | |
| 1998:1 | 1939.70 | 1961.77 | -22.0685 | |
| 1998:2 | 1981.90 | 1950.62 | 31.2839 | |
| 1998:3 | 2000.20 | 1982.44 | 17.7606 | |
| 1998:4 | 2018.10 | 2008.69 | 9.41165 | |
| 1999:1 | 2028.40 | 2030.44 | -2.04266 | |
| 1999:2 | 2036.90 | 2043.10 | -6.19898 | |
| 1999:3 | 2063.30 | 2050.93 | 12.3718 | |
| 1999:4 | 2095.90 | 2073.20 | 22.7025 | |
| 2000:1 | 2078.70 | 2106.16 | -27.4594 | |
| 2000:2 | 2106.40 | 2098.70 | 7.70476 | |
| 2000:3 | 2099.80 | 2115.94 | -16.1432 | |
| 2000:4 | 2106.20 | 2113.62 | -7.42154 | |
| 2001:1 | 2137.30 | 2115.82 | 21.4768 | |
| 2001:2 | 2181.70 | 2142.09 | 39.6118 | * |
| 2001:3 | 2177.80 | 2188.07 | -10.2717 | |
| 2001:4 | 2216.40 | 2197.00 | 19.4047 | |
| 2002:1 | 2250.40 | 2228.45 | 21.9513 | |
| 2002:2 | 2272.00 | 2264.43 | 7.57376 | |
| 2002:3 | 2290.40 | 2290.22 | 0.176208 | |
| 2002:4 | 2305.70 | 2309.08 | -3.37545 | |
| 2003:1 | 2300.90 | 2323.19 | -22.2884 | |
| 2003:2 | 2335.10 | 2318.95 | 16.1501 | |
| 2003:3 | 2342.00 | 2343.36 | -1.36407 | |
| 2003:4 | 2343.70 | 2355.14 | -11.4422 | |
| 2004:1 | 2354.90 | 2357.34 | -2.44240 | |
| 2004:2 | 2363.50 | 2365.01 | -1.50594 | |
| 2004:3 | 2372.10 | 2373.07 | -0.973166 | |
| 2004:4 | 2357.60 | 2381.43 | -23.8266 | |
| 2005:1 | 2359.90 | 2369.89 | -9.98745 | |
| 2005:2 | 2362.40 | 2366.52 | -4.11691 | |
| 2005:3 | 2383.90 | 2367.42 | 16.4755 | |
| 2005:4 | 2373.40 | 2386.52 | -13.1201 | |
| 2006:1 | 2397.10 | 2383.60 | 13.4969 | |

Note: * denotes a residual in excess of 2.5 standard errors

## IV. Forecast

In this section, I use already defined models to forecast GDPC1Q and GCEC1Q from 2006 Q2 to 2013 Q1. I will also perform the forecasting accuracy assessment at the end of this session, comparing forecast with the actual data.

**For Model 1.1 GDPC1Q ARIMA (1, 1, 0):**

For 95% confidence intervals, z(0.025) = 1.96

| Obs | GDPC1Q | prediction | std. error | 95% interval |
|---|---|---|---|---|
| 2006:2 | 12948.7 | 12994.2 | 54.9179 | (12886.5, 13101.8) |
| 2006:3 | 12950.4 | 13071.2 | 91.2910 | (12892.3, 13250.2) |
| 2006:4 | 13038.4 | 13141.5 | 120.616 | (12905.1, 13377.9) |
| 2007:1 | 13056.1 | 13209.6 | 145.159 | (12925.1, 13494.1) |
| 2007:2 | 13173.6 | 13277.0 | 166.425 | (12950.8, 13603.2) |
| 2007:3 | 13269.8 | 13344.1 | 185.357 | (12980.8, 13707.4) |
| 2007:4 | 13326.0 | 13411.1 | 202.555 | (13014.1, 13808.1) |
| 2008:1 | 13266.8 | 13478.1 | 218.411 | (13050.0, 13906.2) |
| 2008:2 | 13310.5 | 13545.1 | 233.193 | (13088.1, 14002.2) |
| 2008:3 | 13186.9 | 13612.1 | 247.094 | (13127.8, 14096.4) |
| 2008:4 | 12883.5 | 13679.1 | 260.254 | (13169.0, 14189.2) |
| 2009:1 | 12711.0 | 13746.1 | 272.779 | (13211.5, 14280.7) |
| 2009:2 | 12701.0 | 13813.1 | 284.754 | (13255.0, 14371.2) |
| 2009:3 | 12746.7 | 13880.1 | 296.246 | (13299.5, 14460.7) |
| 2009:4 | 12873.1 | 13947.1 | 307.308 | (13344.8, 14549.4) |
| 2010:1 | 12947.6 | 14014.1 | 317.985 | (13390.8, 14637.3) |
| 2010:2 | 13019.6 | 14081.1 | 328.316 | (13437.6, 14724.6) |
| 2010:3 | 13103.5 | 14148.1 | 338.331 | (13484.9, 14811.2) |
| 2010:4 | 13181.2 | 14215.1 | 348.058 | (13532.9, 14897.2) |
| 2011:1 | 13183.8 | 14282.1 | 357.521 | (13581.3, 14982.8) |
| 2011:2 | 13264.7 | 14349.0 | 366.739 | (13630.3, 15067.8) |
| 2011:3 | 13306.9 | 14416.0 | 375.731 | (13679.6, 15152.5) |
| 2011:4 | 13441.0 | 14483.0 | 384.513 | (13729.4, 15236.7) |
| 2012:1 | 13506.4 | 14550.0 | 393.099 | (13779.6, 15320.5) |
| 2012:2 | 13548.5 | 14617.0 | 401.502 | (13830.1, 15404.0) |
| 2012:3 | 13652.5 | 14684.0 | 409.732 | (13881.0, 15487.1) |
| 2012:4 | 13665.4 | 14751.0 | 417.800 | (13932.1, 15569.9) |
| 2013:1 | 13750.1 | 14818.0 | 425.715 | (13983.6, 15652.4) |

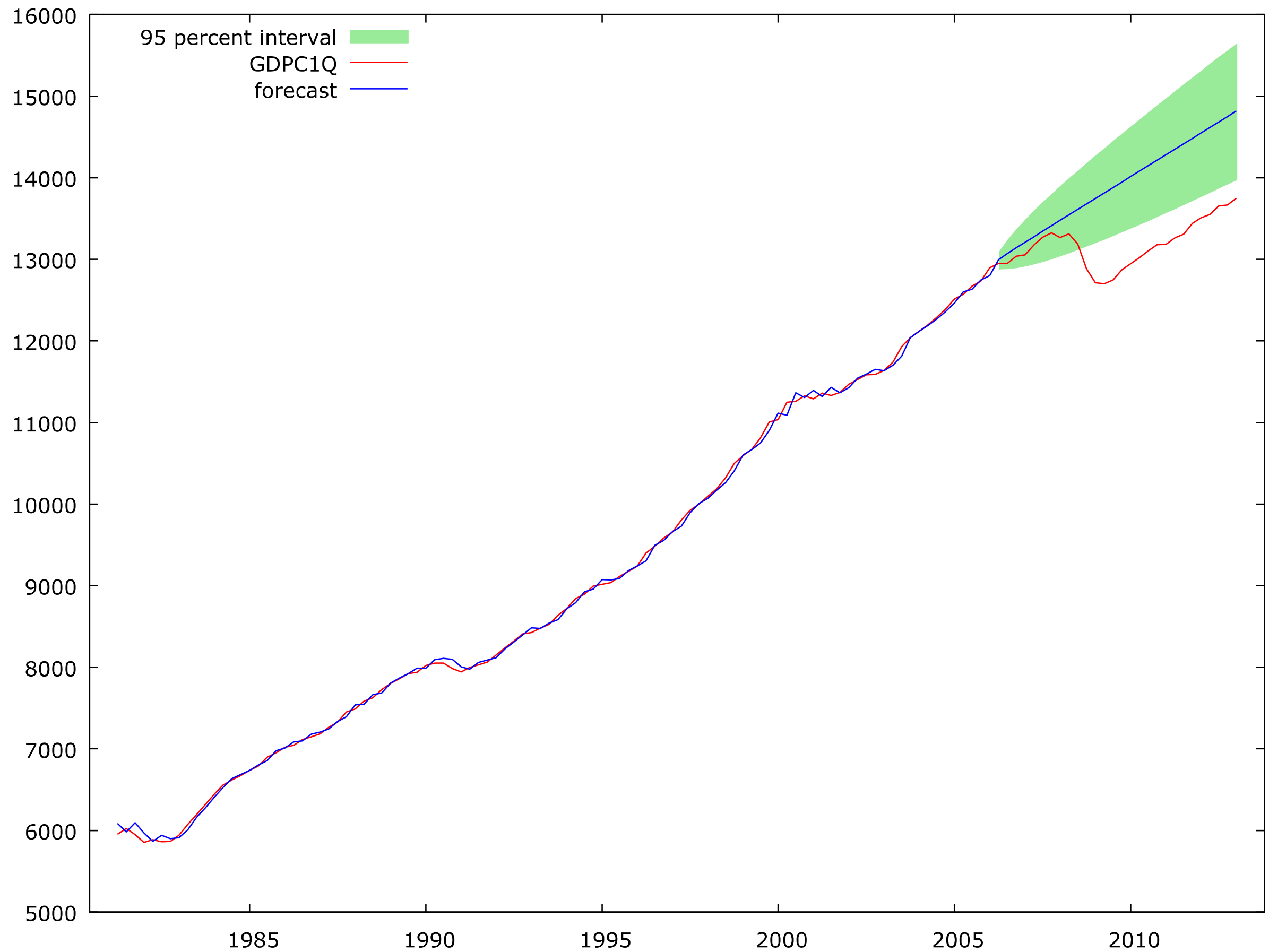

Forecast evaluation statistics
Mean Error -733.73
Mean Squared Error 7.3107e+005
Root Mean Squared Error 855.03
Mean Absolute Error 733.73
Mean Percentage Error -5.5643
**Mean Absolute Percentage Error 5.5643**
Theil's U 8.7131
Bias proportion, UM 0.73639
Regression proportion, UR 0.19322
Disturbance proportion, UD 0.070388

Overall, we can see that ARIMA (1, 1, 0) provides a good fit for GDPC1Q data. It gives a fairly accurate forecasting. However, although forecasts from 2006 Q2 to 2008 Q3 are within the 95% percent interval, the graph shows that the red line of actual data has gradually moving out of the confidence interval starting from 2007 Q4 and climbing back up toward the interval from 2009 Q1. Such trend exactly coincides with the way of how the economy has evolved since great recession of 2008. But, the weakness of ARIMA model is that it could not predict such trend but rather assume the same pattern from 1980 Q1-2006 Q1, that is, the ARIMA model is not good for volatility analysis, but GARCH is. Finally, GDP does strive back after a deep dip in 2009 Q1.

**For Model 1.2 GDPC1Q ARIMA (1, 1, 1):**

For 95% confidence intervals, z(0.025) = 1.96

| Obs | GDPC1Q | prediction | std. error | 95% interval |
|---|---|---|---|---|
| 2006:2 | 12948.7 | 12992.2 | 53.6190 | (12887.1, 13097.3) |
| 2006:3 | 12950.4 | 13080.9 | 87.1201 | (12910.1, 13251.6) |
| 2006:4 | 13038.4 | 13164.2 | 118.346 | (12932.2, 13396.1) |
| 2007:1 | 13056.1 | 13243.4 | 147.950 | (12953.4, 13533.4) |
| 2007:2 | 13173.6 | 13319.5 | 176.035 | (12974.5, 13664.5) |
| 2007:3 | 13269.8 | 13393.2 | 202.658 | (12996.0, 13790.4) |
| 2007:4 | 13326.0 | 13465.2 | 227.887 | (13018.5, 13911.8) |
| 2008:1 | 13266.8 | 13535.8 | 251.807 | (13042.2, 14029.3) |
| 2008:2 | 13310.5 | 13605.3 | 274.514 | (13067.3, 14143.3) |
| 2008:3 | 13186.9 | 13674.1 | 296.104 | (13093.7, 14254.4) |
| 2008:4 | 12883.5 | 13742.2 | 316.672 | (13121.6, 14362.9) |
| 2009:1 | 12711.0 | 13809.9 | 336.307 | (13150.8, 14469.1) |
| 2009:2 | 12701.0 | 13877.3 | 355.092 | (13181.3, 14573.3) |
| 2009:3 | 12746.7 | 13944.4 | 373.104 | (13213.1, 14675.6) |
| 2009:4 | 12873.1 | 14011.3 | 390.409 | (13246.1, 14776.5) |
| 2010:1 | 12947.6 | 14078.0 | 407.071 | (13280.2, 14875.9) |
| 2010:2 | 13019.6 | 14144.6 | 423.143 | (13315.3, 14974.0) |
| 2010:3 | 13103.5 | 14211.2 | 438.676 | (13351.4, 15071.0) |
| 2010:4 | 13181.2 | 14277.7 | 453.714 | (13388.4, 15166.9) |
| 2011:1 | 13183.8 | 14344.1 | 468.296 | (13426.2, 15261.9) |
| 2011:2 | 13264.7 | 14410.5 | 482.456 | (13464.9, 15356.1) |
| 2011:3 | 13306.9 | 14476.8 | 496.228 | (13504.2, 15449.4) |
| 2011:4 | 13441.0 | 14543.2 | 509.638 | (13544.3, 15542.0) |
| 2012:1 | 13506.4 | 14609.5 | 522.712 | (13585.0, 15634.0) |
| 2012:2 | 13548.5 | 14675.8 | 535.473 | (13626.3, 15725.3) |
| 2012:3 | 13652.5 | 14742.1 | 547.941 | (13668.1, 15816.0) |
| 2012:4 | 13665.4 | 14808.4 | 560.135 | (13710.5, 15906.2) |
| 2013:1 | 13750.1 | 14874.6 | 572.071 | (13753.4, 15995.9) |

Forecast evaluation statistics
Mean Error -787.19
Mean Squared Error 8.2292e+005
Root Mean Squared Error 907.15
Mean Absolute Error 787.19
Mean Percentage Error -5.9698
**Mean Absolute Percentage Error 5.9698**
Theil's U 9.2447
Bias proportion, UM 0.75302

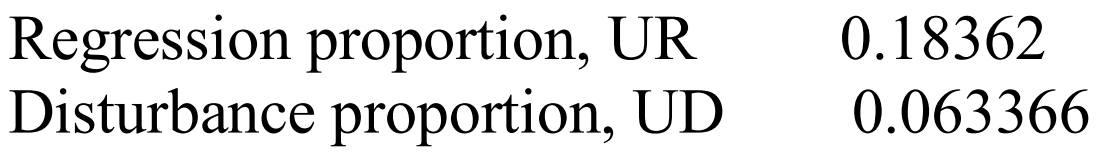
Regression proportion, UR 0.18362
Disturbance proportion, UD 0.063366

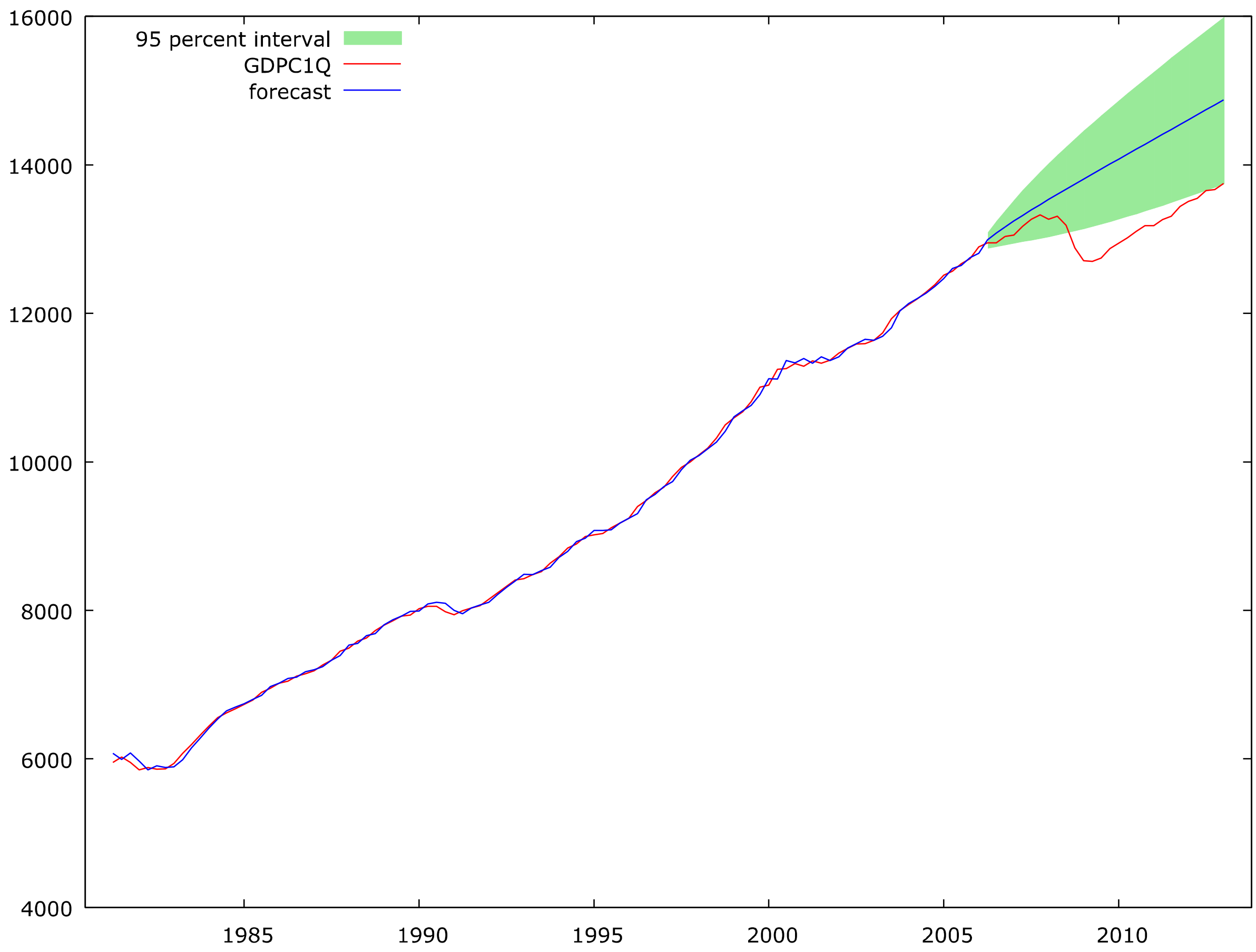

The above forecasting graph indicates that ARIMA (1, 1, 1) overall provides a good fit for model 1. Compared with ARIMA (1, 1, 0), it has wider confidence interval (green area). We can see that the actual data from 2013 Q1 is actually touching the interval, however there still is a gap between the red line and the interval in the ARIMA (1, 1, 0) forecast. The reason for the green interval not capturing the red line is because of the unexpected 2008 recession. Hence, I think that GARCH might have the advantage over ARIMA, since it has the ability and room for volatility forecasting. It means that GARCH would allow us to forecast data, which is expected to have large variations. I think that is the purpose for heteroskedasticity to exist within such data.

Based on the forecasting performance, testing and estimating results, it is very difficult to choose between ARIMA (1, 1, 0) and ARIMA (1, 1, 1). But the residual ACF and PACF of ARIMA (1, 1, 1) provides a slightly better result than that of ARIMA (1, 1, 0). We can see that ACF and PACF of ARIMA (1, 1, 0) has a spike crossing the $\pm 0.2$ boundary at lag 2, while nothing is crossing the boundary in ACF and PACF of ARIMA (1, 1, 1). It means that ARIMA (1, 1, 1) is slightly more stable than ARIMA (1, 1, 0), which probably provides better forecasting results. I also want to point out that although MAPE of ARIMA (1, 1, 0) is slightly better than ARIMA (1, 1, 1) (5.5643% compared with 5.9698%), it does not mean that ARIMA (1, 1, 0) is better than ARIMA (1, 1, 1), because of the stability issue mentioned above. Plus, there are only two residuals of ARIMA (1, 1, 1) in excess of 2.5 standard errors, compared with three of ARIMA (1, 1, 0). Therefore, I will fit GDPC1Q data with ARIMA (1, 1, 1).

**For Model 1.3 GCEC1Q ARIMA (1, 1, 2):**

For 95% confidence intervals, z(0.025) = 1.96

| Obs | GCEC1Q | prediction | std. error | 95% interval |
|---|---|---|---|---|
| 2006:2 | 2399.10 | 2401.85 | 15.2543 | (2371.96, 2431.75) |
| 2006:3 | 2402.70 | 2411.24 | 20.1999 | (2371.65, 2450.83) |
| 2006:4 | 2409.40 | 2420.70 | 25.6319 | (2370.47, 2470.94) |
| 2007:1 | 2406.70 | 2430.22 | 31.2272 | (2369.02, 2491.43) |
| 2007:2 | 2426.80 | 2439.79 | 36.8190 | (2367.62, 2511.95) |
| 2007:3 | 2447.90 | 2449.39 | 42.3182 | (2366.45, 2532.33) |
| 2007:4 | 2455.30 | 2459.02 | 47.6769 | (2365.58, 2552.47) |
| 2008:1 | 2473.90 | 2468.68 | 52.8708 | (2365.06, 2572.31) |
| 2008:2 | 2484.50 | 2478.36 | 57.8891 | (2364.90, 2591.82) |
| 2008:3 | 2510.70 | 2488.06 | 62.7296 | (2365.11, 2611.01) |
| 2008:4 | 2520.50 | 2497.77 | 67.3951 | (2365.68, 2629.86) |
| 2009:1 | 2531.60 | 2507.49 | 71.8915 | (2366.59, 2648.40) |
| 2009:2 | 2590.40 | 2517.22 | 76.2265 | (2367.82, 2666.62) |
| 2009:3 | 2614.30 | 2526.96 | 80.4084 | (2369.36, 2684.56) |
| 2009:4 | 2621.10 | 2536.70 | 84.4462 | (2371.19, 2702.21) |
| 2010:1 | 2600.40 | 2546.45 | 88.3487 | (2373.29, 2719.61) |
| 2010:2 | 2618.70 | 2556.20 | 92.1243 | (2375.64, 2736.76) |
| 2010:3 | 2616.70 | 2565.95 | 95.7812 | (2378.22, 2753.68) |
| 2010:4 | 2587.40 | 2575.71 | 99.3273 | (2381.03, 2770.39) |
| 2011:1 | 2540.70 | 2585.47 | 102.770 | (2384.04, 2786.89) |
| 2011:2 | 2535.40 | 2595.23 | 106.115 | (2387.25, 2803.21) |
| 2011:3 | 2516.60 | 2604.99 | 109.370 | (2390.63, 2819.35) |
| 2011:4 | 2502.70 | 2614.75 | 112.539 | (2394.18, 2835.33) |
| 2012:1 | 2483.70 | 2624.52 | 115.629 | (2397.89, 2851.15) |
| 2012:2 | 2479.40 | 2634.28 | 118.644 | (2401.74, 2866.82) |
| 2012:3 | 2503.10 | 2644.05 | 121.590 | (2405.73, 2882.36) |
| 2012:4 | 2458.10 | 2653.81 | 124.468 | (2409.86, 2897.77) |
| 2013:1 | 2427.10 | 2663.58 | 127.285 | (2414.10, 2913.05) |

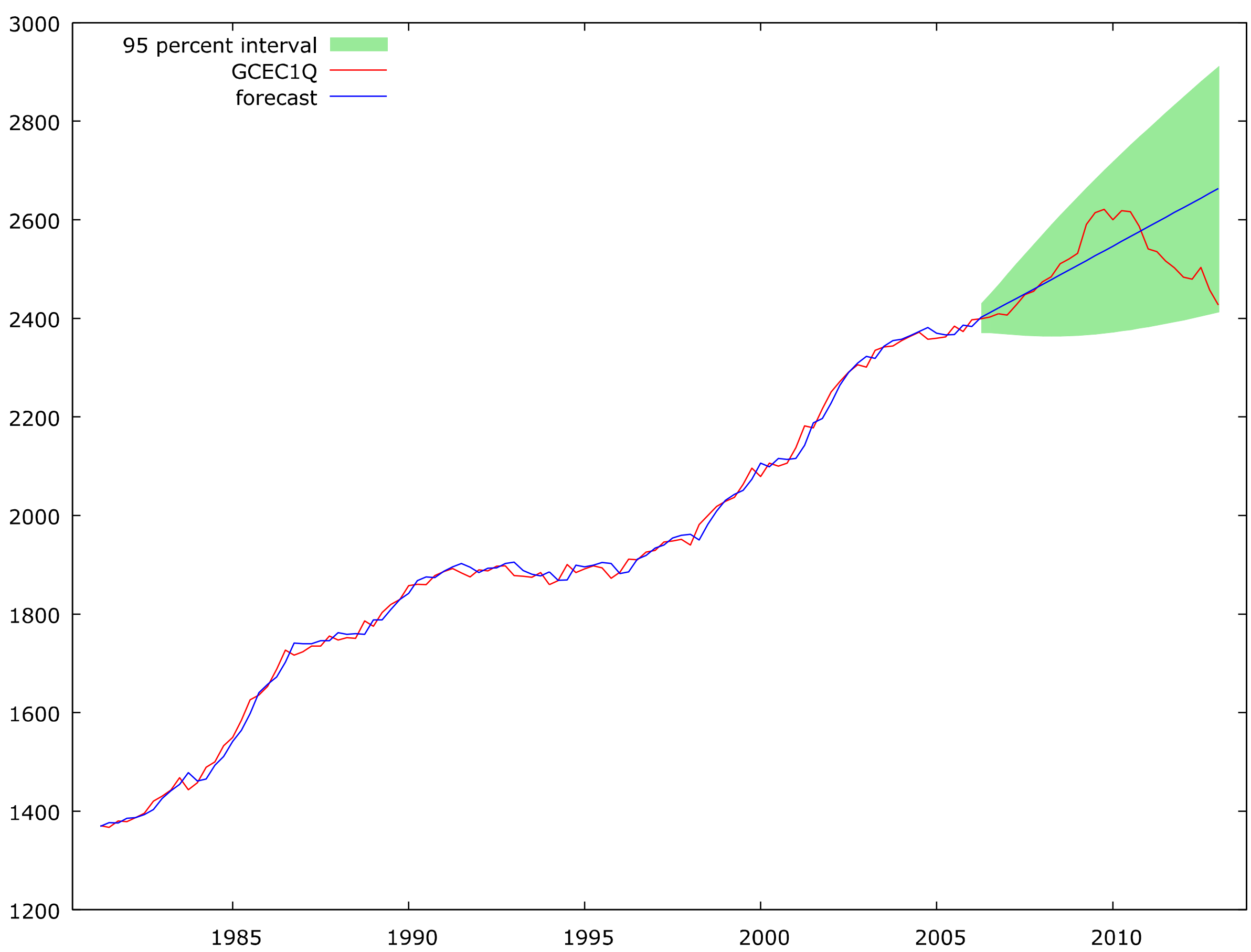

Forecast evaluation statistics
Mean Error -26.198
Mean Squared Error 7718.5
Root Mean Squared Error 87.855
Mean Absolute Error 62.244
Mean Percentage Error -1.092
**Mean Absolute Percentage Error 2.4814**
Theil's U 3.8702
Bias proportion, UM 0.088923
Regression proportion, UR 0.35153
Disturbance proportion, UD 0.55955

The forecast red line is within the 95% confidence interval, which is a good sign for accurate forecasting. And overall, we can find that the ARIMA (1, 1, 2) fits well for the GCEC1Q dataset. We do see that there is a sharp reduction in Government Consumption Expenditures & Gross Investment from 2010 Q2. Compared with the above forecast graph of GDPC1Q, we can see the confidence interval for GCEC1Q is much wider, meaning that the government spending each year has larger variations than GDP does and it becomes much more volatile.

***(2) The Tiao-Box (1981) Multiple Time Series (Multivariate VARMA) Approach:***

1980Q1-2006Q1 as the sample range; GDPC1Q and GCEC1Q as the observations.

Since we use ARIMA (1, 1, 1) for GDPC1Q and ARIMA (1, 1, 2) for GCEC1Q, we will first fit the sample data in either VARIMA (1, 1, 1) or VARIMA (1, 1, 2), then comparing the results and selecting the best one as our final model. VARIMA (1, 1, 0) is essentially VAR (1), which will be discussed in section 3.

Recall that we write univariate ARIMA (1, 1, 1) of GDPC1Q as:

$$\Delta^{1d} Y_t = \mu_0 + \varphi_1 \Delta^{1d} \mathrm{Y}_{t-1} + \varepsilon_t + \theta_1 \varepsilon_{t-1}$$

And univariate ARIMA (1, 1, 2) of GCEC1Q as:

$$\Delta^{1d} Y_t = \mu_0 + \varphi_1 \Delta^{1d} \mathrm{Y}_{t-1} + \varepsilon_t + \theta_1 \varepsilon_{t-1} + \theta_2 \varepsilon_{t-2}$$

Now, VARIMA (1, 1, 1) for both GDPC1Q ($Y_1$) and GCEC1Q ($Y_2$) can be written as:

$$\begin{bmatrix} \Delta^{1d} Y_{1t} \\ \Delta^{1d} Y_{2t} \end{bmatrix} = \begin{bmatrix} \mu_1 \\ \mu_2 \end{bmatrix} + \begin{bmatrix} \varphi_{1,1} & \varphi_{1,2} \\ \varphi_{2,1} & \varphi_{2,2} \end{bmatrix} \begin{bmatrix} \Delta^{1d} \mathrm{Y}_{1,t-1} \\ \Delta^{1d} \mathrm{Y}_{2,t-1} \end{bmatrix} + \begin{bmatrix} \varepsilon_{1,t} \\ \varepsilon_{2,t} \end{bmatrix} + \begin{bmatrix} \theta_{1,1} & \theta_{1,2} \\ \theta_{2,1} & \theta_{2,2} \end{bmatrix} \begin{bmatrix} \varepsilon_{1,t-1} \\ \varepsilon_{2,t-1} \end{bmatrix} \quad (2.1)$$

(2.1) can also be written as:

$$\Delta^{1d} Y_{1t} = \mu_1 + \varphi_{1,1} \Delta^{1d} \mathrm{Y}_{1,t-1} + \varphi_{1,2} \Delta^{1d} \mathrm{Y}_{2,t-1} + \varepsilon_{1,t} + \theta_{1,1} \varepsilon_{1,t-1} + \theta_{1,2} \varepsilon_{2,t-1} \quad (2.2)$$

$$\Delta^{1d} Y_{2t} = \mu_2 + \varphi_{2,1} \Delta^{1d} \mathrm{Y}_{1,t-1} + \varphi_{2,2} \Delta^{1d} \mathrm{Y}_{2,t-1} + \varepsilon_{2,t} + \theta_{2,1} \varepsilon_{1,t-1} + \theta_{2,2} \varepsilon_{2,t-1}$$

```
gretl version 1.9.14
Current session: 2014-01-17 21:03
? series x = diff(GDPC1Q)
Generated series x (ID 11)
? series y = diff(GCEC1Q)
Generated series y (ID 12)
? series a = uhat1
Generated series a (ID 13)
? series b = uhat2
Generated series b (ID 14)
? system
? equation x const x(-1) y(-1) a(-1) b(-1)
```

```
? equation y const x(-1) y(-1) a(-1) b(-1)
? end system
? estimate $system method=ols

Equation system, Ordinary Least Squares

Equation 1: OLS, using observations 1980:2-2006:1 (T = 104)
Dependent variable: x

             coefficient   std. error   t-ratio   p-value
  -------------------------------------------------------
  const       82.2102      129.947       0.6326   0.5284
  x_1          0.751179      0.236725    3.173    0.0020  ***
  y_1         -6.43447      12.6694     -0.5079   0.6127
  a_1         -0.460145      0.259285   -1.775    0.0790  *
  b_1          5.91783      12.6580      0.4675   0.6412

Mean dependent var   67.24038   S.D. dependent var   58.45773
Sum squared resid    294663.4   S.E. of regression   54.55638
R-squared            0.162846   Adjusted R-squared   0.129022

Equation 2: OLS, using observations 1980:2-2006:1 (T = 104)
Dependent variable: y

             coefficient   std. error   t-ratio    p-value
  --------------------------------------------------------
  const       8.59125      38.1638       0.2251    0.8224
  x_1         0.0601490     0.0695235    0.8652    0.3890
  y_1        -0.261946      3.72085     -0.07040   0.9440
  a_1        -0.105878      0.0761489   -1.390     0.1675
  b_1         0.279512      3.71751      0.07519   0.9402

Mean dependent var   9.920192   S.D. dependent var   15.95743
Sum squared resid    25415.58   S.E. of regression   16.02258
R-squared            0.030971   Adjusted R-squared  -0.008182

Cross-equation VCV for residuals
(correlations above the diagonal)

       2833.3      (0.246)
       204.94       244.38

log determinant = 13.3853
Breusch-Pagan test for diagonal covariance matrix:
  Chi-square(1) = 6.30863 [0.0120]
```

Now, we can write (2.1) and (2.2) as:

$$\begin{bmatrix}\Delta^{1d}Y_{1t}\\ \Delta^{1d}Y_{2t}\end{bmatrix} = \begin{bmatrix}82.2102\\ 8.59125\end{bmatrix} + \begin{bmatrix}0.751179 & -6.43447\\ 0.060149 & -0.261946\end{bmatrix}\begin{bmatrix}\Delta^{1d}Y_{1,t-1}\\ \Delta^{1d}Y_{2,t-1}\end{bmatrix} + \begin{bmatrix}\varepsilon_{1,t}\\ \varepsilon_{2,t}\end{bmatrix} + \begin{bmatrix}-0.460145 & 5.91783\\ -0.105878 & 0.279512\end{bmatrix}\begin{bmatrix}\varepsilon_{1,t-1}\\ \varepsilon_{2,t-1}\end{bmatrix} \qquad (2.1')$$

$$\Delta^{1d}Y_{1t} = 82.2102 + 0.751179\Delta^{1d}Y_{1,t-1} - 6.43447\Delta^{1d}Y_{2,t-1} + \varepsilon_{1,t} - 0.460145\varepsilon_{1,t-1} + 5.91783\varepsilon_{2,t-1} \qquad (2.2')$$

$$\Delta^{1d}Y_{2t} = 8.59125 + 0.060149\Delta^{1d}\mathrm{Y}_{1,t-1} - 0.261946\Delta^{1d}\mathrm{Y}_{2,t-1} + \varepsilon_{2,t} - 0.105878\varepsilon_{1,t-1} + 0.279512\varepsilon_{2,t-1}$$

Notes: error terms uhat1 and uhat2 are from ARIMA (1, 1, 1) of GDPC1Q and ARIMA (1, 1, 1) of GCEC1Q respectively.

In comparison, VARIMA (1, 1, 2) for both GDPC1Q ($Y_1$) and GCEC1Q ($Y_2$) can be written as:

$$\begin{bmatrix}\Delta^{1d}Y_{1t}\\ \Delta^{1d}Y_{2t}\end{bmatrix} = \begin{bmatrix}\mu_1\\ \mu_2\end{bmatrix} + \begin{bmatrix}\varphi_{1,1} & \varphi_{1,2}\\ \varphi_{2,1} & \varphi_{2,2}\end{bmatrix}\begin{bmatrix}\Delta^{1d}\mathrm{Y}_{1,t-1}\\ \Delta^{1d}\mathrm{Y}_{2,t-1}\end{bmatrix} + \begin{bmatrix}\varepsilon_{1,t}\\ \varepsilon_{2,t}\end{bmatrix} + \begin{bmatrix}\theta_{1,1} & \theta_{1,2}\\ \theta_{2,1} & \theta_{2,2}\end{bmatrix}\begin{bmatrix}\varepsilon_{1,t-1}\\ \varepsilon_{2,t-1}\end{bmatrix} + \begin{bmatrix}\vartheta_{1,1} & \vartheta_{1,2}\\ \vartheta_{2,1} & \vartheta_{2,2}\end{bmatrix}\begin{bmatrix}\varepsilon_{1,t-2}\\ \varepsilon_{2,t-2}\end{bmatrix} \quad (2.3)$$

(2.3) can also be written as:

$$\Delta^{1d}Y_{1t} = \mu_1 + \varphi_{1,1}\Delta^{1d}\mathrm{Y}_{1,t-1} + \varphi_{1,2}\Delta^{1d}\mathrm{Y}_{2,t-1} + \varepsilon_{1,t} + \theta_{1,1}\varepsilon_{1,t-1} + \theta_{1,2}\varepsilon_{2,t-1} + \vartheta_{1,1}\varepsilon_{1,t-2} + \vartheta_{1,2}\varepsilon_{2,t-2} \quad (2.4)$$

$$\Delta^{1d}Y_{2t} = \mu_2 + \varphi_{2,1}\Delta^{1d}\mathrm{Y}_{1,t-1} + \varphi_{2,2}\Delta^{1d}\mathrm{Y}_{2,t-1} + \varepsilon_{2,t} + \theta_{2,1}\varepsilon_{1,t-1} + \theta_{2,2}\varepsilon_{2,t-1} + \vartheta_{2,1}\varepsilon_{1,t-2} + \vartheta_{2,2}\varepsilon_{2,t-2}$$

```
gretl version 1.9.14
Current session: 2014-01-17 21:27
? series x = diff(GDPC1Q)
Generated series x (ID 11)
? series y = diff(GCEC1Q)
Generated series y (ID 12)
? series a = uhat1
Generated series a (ID 13)
? series b = uhat2
Generated series b (ID 14)
? system
? equation x const x(-1) y(-1) a(-1) b(-1) a(-2) b(-2)
? equation y const x(-1) y(-1) a(-1) b(-1) a(-2) b(-2)
? end system
? estimate $system method=ols

Equation system, Ordinary Least Squares

Equation 1: OLS, using observations 1980:3-2006:1 (T = 103)
Dependent variable: x

             coefficient   std. error   t-ratio   p-value
  -------------------------------------------------------
  const       46.7843      21.0741       2.220    0.0288  **
  x_1          0.369566     0.267391     1.382    0.1701
  y_1         -0.219727     1.18141     -0.1860   0.8528
  a_1         -0.113708     0.283210    -0.4015   0.6889
```

```
  b_1         -0.132601     1.22455     -0.1083   0.9140
  a_2          0.216516     0.124060     1.745    0.0841  *
  b_2         -0.104326     0.379611    -0.2748   0.7840

Mean dependent var   69.06796   S.D. dependent var   55.67800
Sum squared resid    263046.7   S.E. of regression   52.34568
R-squared            0.168111   Adjusted R-squared   0.116118

Equation 2: OLS, using observations 1980:3-2006:1 (T = 103)
Dependent variable: y

             coefficient   std. error   t-ratio    p-value
  --------------------------------------------------------
  const       0.156603     6.33361       0.02473   0.9803
  x_1         0.00582349   0.0803618     0.07247   0.9424
  y_1         0.950556     0.355062      2.677     0.0087  ***
  a_1        -0.0394123    0.0851159    -0.4630    0.6444
  b_1        -1.00433      0.368026     -2.729     0.0076  ***
  a_2         0.0111816    0.0372850     0.2999    0.7649
  b_2         0.165578     0.114089      1.451     0.1500

Mean dependent var   9.974757   S.D. dependent var   16.02571
Sum squared resid    23759.55   S.E. of regression   15.73198
R-squared            0.093008   Adjusted R-squared   0.036321

Cross-equation VCV for residuals
(correlations above the diagonal)

       2553.9      (0.264)
       202.99       230.68

log determinant = 13.2139
Breusch-Pagan test for diagonal covariance matrix:
  Chi-square(1) = 7.20435 [0.0073]
```

Now, we can write (2.3) and (2.4) as:

$$\begin{bmatrix}\Delta^{1d}Y_{1t}\\ \Delta^{1d}Y_{2t}\end{bmatrix} = \begin{bmatrix}46.7843\\ 0.156603\end{bmatrix} + \begin{bmatrix}0.369566 & -0.219727\\ 0.00582349 & 0.950556\end{bmatrix}\begin{bmatrix}\Delta^{1d}\mathrm{Y}_{1,t-1}\\ \Delta^{1d}\mathrm{Y}_{2,t-1}\end{bmatrix} + \begin{bmatrix}\varepsilon_{1,t}\\ \varepsilon_{2,t}\end{bmatrix} + \begin{bmatrix}-0.113708 & -0.132601\\ -0.0394123 & -1.00433\end{bmatrix}\begin{bmatrix}\varepsilon_{1,t-1}\\ \varepsilon_{2,t-1}\end{bmatrix} + \begin{bmatrix}0.216516 & -0.104326\\ 0.0111816 & 0.165578\end{bmatrix}\begin{bmatrix}\varepsilon_{1,t-2}\\ \varepsilon_{2,t-2}\end{bmatrix} \quad (2.3')$$

$$\Delta^{1d}Y_{1t} = 46.7843 + 0.369566\Delta^{1d}\mathrm{Y}_{1,t-1} - 0.219727\Delta^{1d}\mathrm{Y}_{2,t-1} + \varepsilon_{1,t} - 0.113708\varepsilon_{1,t-1} - 0.132601\varepsilon_{2,t-1} + 0.216516\varepsilon_{1,t-2} - 0.104326\varepsilon_{2,t-2} \quad (2.4')$$

$$\Delta^{1d}Y_{2t} = 0.156603 + 0.00582349\Delta^{1d}\text{Y}_{1,t-1} + 0.950556\Delta^{1d}\text{Y}_{2,t-1} + \varepsilon_{2,t} - 0.0394123\varepsilon_{1,t-1} - 1.00433\varepsilon_{2,t-1} + 0.0111816\varepsilon_{1,t-2} + 0.165578\varepsilon_{2,t-2}$$

Notes: error terms uhat1 and uhat2 are from ARIMA (1, 1, 2) of GDPC1Q and ARIMA (1, 1, 2) of GCEC1Q respectively.

Essentially, Multivariate VARMA (2.1) and (2.3) can be transformed back to Univariate ARMA, if:

$$\begin{bmatrix} \varphi_{1,1} & \varphi_{1,2} \\ \varphi_{2,1} & \varphi_{2,2} \end{bmatrix} = \begin{bmatrix} \varphi_{1,1} & 0 \\ 0 & \varphi_{2,2} \end{bmatrix} \textit{ and } \begin{bmatrix} \theta_{1,1} & \theta_{1,2} \\ \theta_{2,1} & \theta_{2,2} \end{bmatrix} = \begin{bmatrix} \theta_{1,1} & 0 \\ 0 & \theta_{2,2} \end{bmatrix} \textit{ and } \begin{bmatrix} \vartheta_{1,1} & \vartheta_{1,2} \\ \vartheta_{2,1} & \vartheta_{2,2} \end{bmatrix} = \begin{bmatrix} \vartheta_{1,1} & 0 \\ 0 & \vartheta_{2,2} \end{bmatrix}$$

We conclude that Multivariate VARMA is more generalized than Univariate ARMA.

The findings are not surprising at all that both VARIMA (1, 1, 1) and VARIMA (1, 1, 2) are not good at explaining the relationship between GDPC1Q and GCEC1Q and between GCEC1Q and GDPC1Q, either showing the statistical significance for such relationships. Thereby, we resort to VAR (1) analysis in the following section 3, because VAR (1) shares the same component AR with both ARIMA (1, 1, 1) of GDPC1Q and ARIMA (1, 1, 2) of GCEC1Q, in comparison VARIMA (1, 1, 1) does not share the same MA component with ARIMA (1, 1, 2) and neither does VARIMA (1, 1, 2) share the same MA component with ARIMA (1, 1, 1), which might be the reason that VAR (1) is more widely used in the field compared to VARMA that however is more generalized than VAR, because of which we can consider both Univariate ARMA, VAR and VMA as special cases of VARMA.

### *(3) The VAR (p) Model and Impulse-Response Function:*

1980Q1-2006Q1 as sample range; GDPC1Q and GCEC1Q as endogenous variables.

## I. Identification:

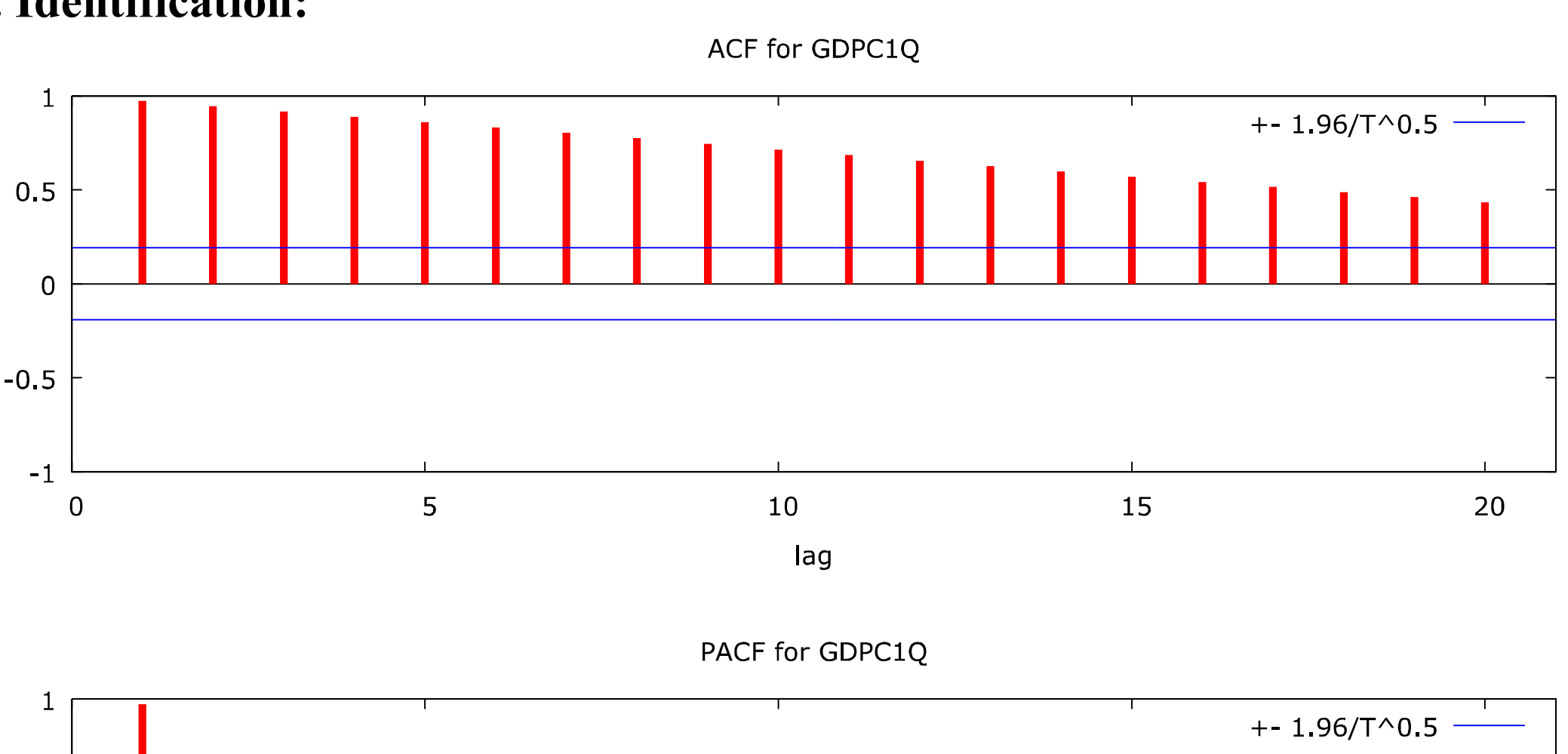

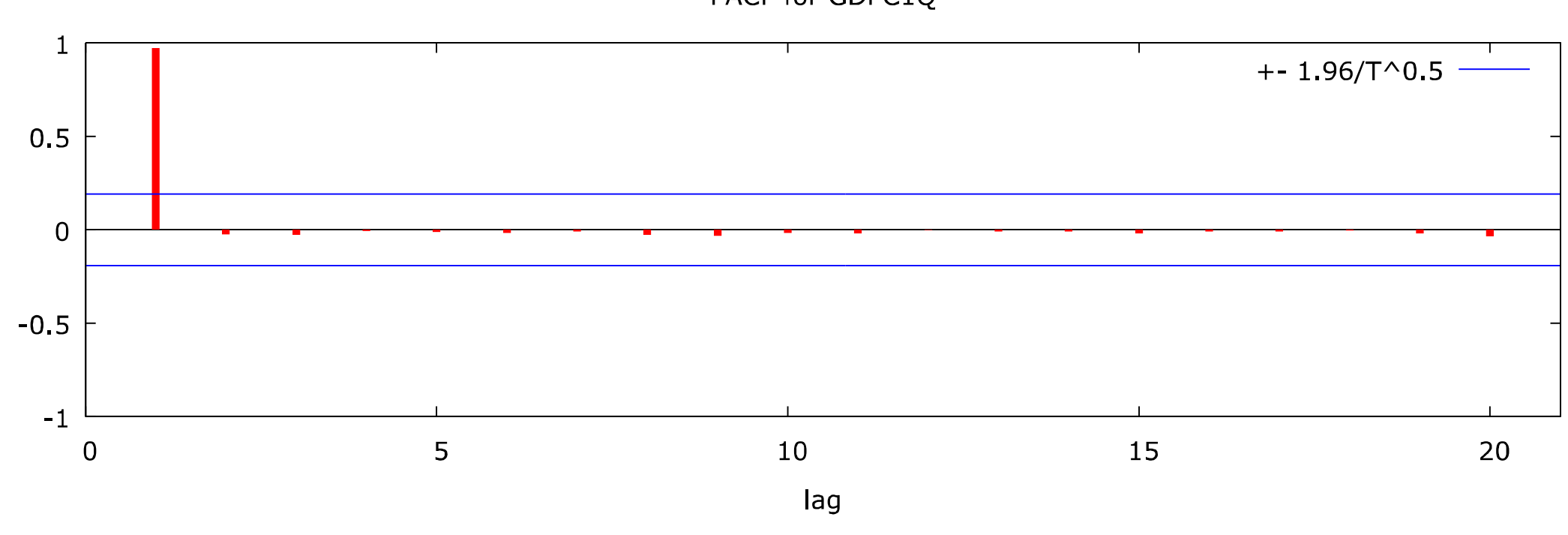

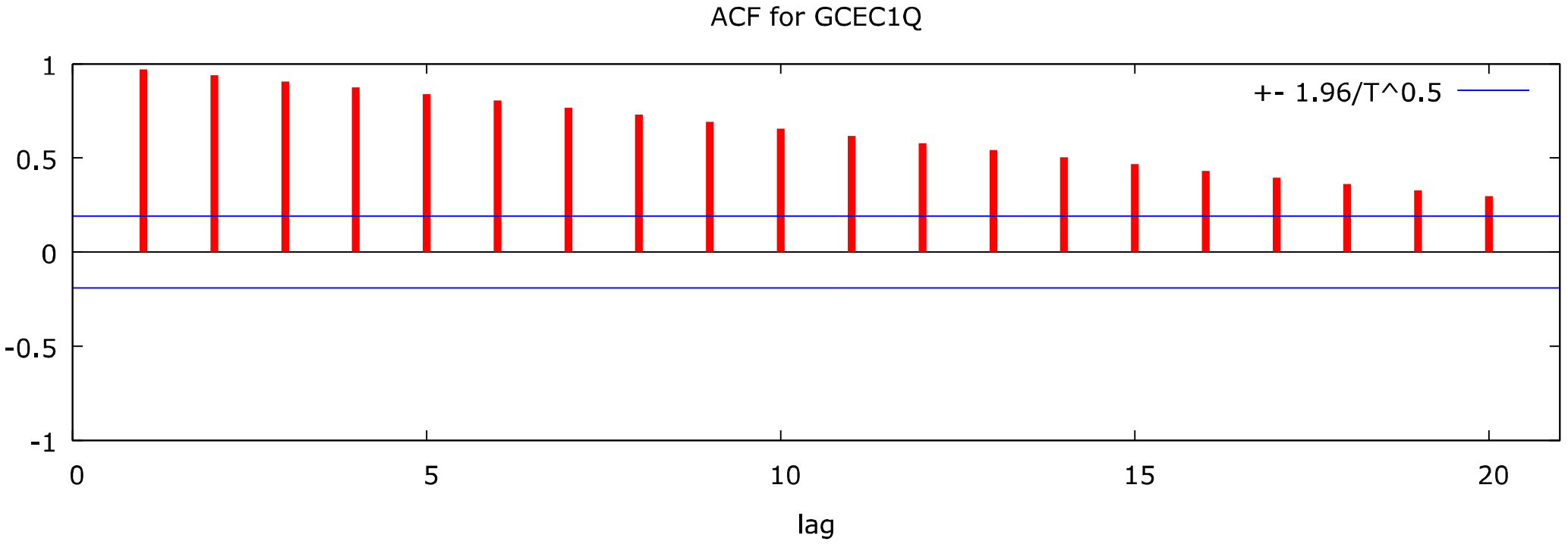

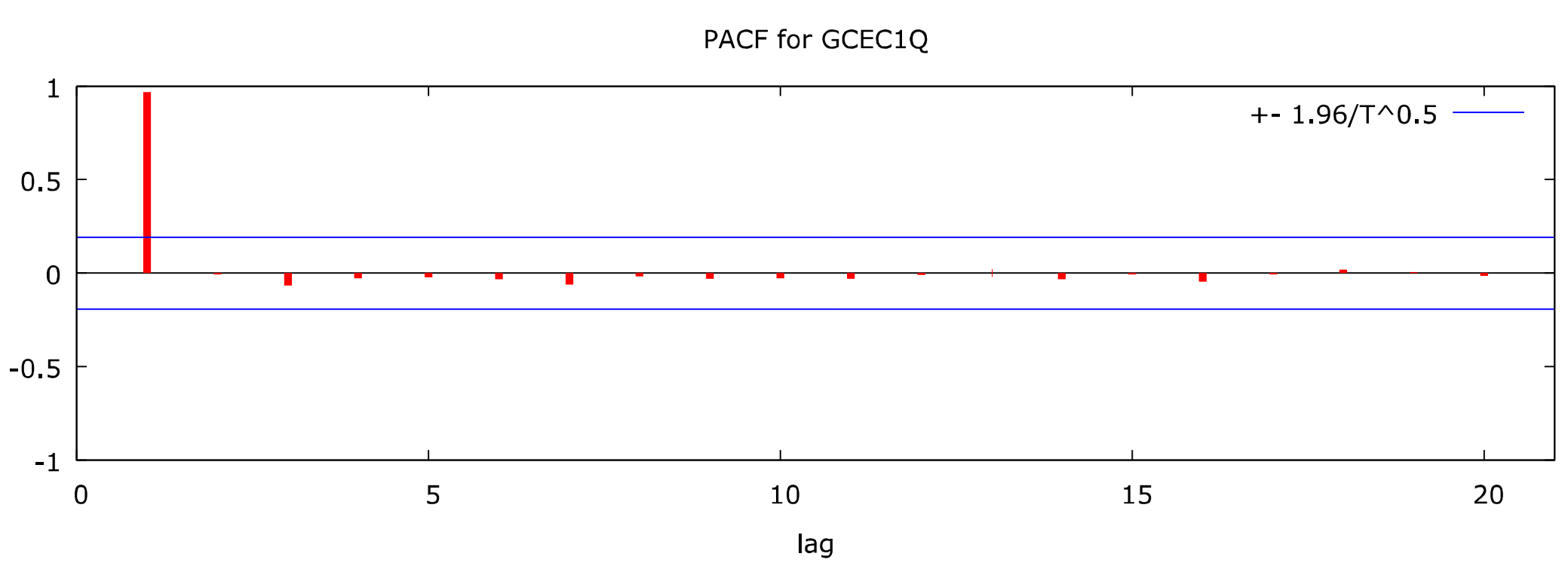

Since the above correlogram indicates nonstationarity, we have to difference GDPC1Q and GCEC1Q for at least once, denoting them as d_GDPC1Q and d_GCEC1Q.

```
VAR system, maximum lag order 8

The asterisks below indicate the best (that is, minimized) values
of the respective information criteria, AIC = Akaike criterion,
BIC = Schwarz Bayesian criterion and HQC = Hannan-Quinn criterion.

lags        loglik    p(LR)       AIC          BIC          HQC

   1    -914.45020            18.978355    19.137615*   19.042752*
   2    -910.05198  0.06639   18.970144    19.235578    19.077472
   3    -904.12837  0.01852   18.930482*   19.302090    19.080742
   4    -901.71218  0.30493   18.963138    19.440919    19.156329
   5    -900.28794  0.58349   19.016246    19.600201    19.252369
   6    -899.41488  0.78232   19.080719    19.770848    19.359773
   7    -897.98592  0.58188   19.133730    19.930033    19.455716
   8    -889.94431  0.00291   19.050398    19.952874    19.415315
```

The above analysis tells us to identify the model as VAR (1), because both BIC and HQC are significant at lag 1.

**II. Estimation:**

VAR system, lag order 1
OLS estimates, observations 1980:2-2006:1 (T = 104)
Log-likelihood = -993.60761
Determinant of covariance matrix = 681522.47
AIC = 19.2232
BIC = 19.3758
HQC = 19.2850
Portmanteau test: LB(26) = 109.69, df = 100 [0.2385]

Equation 1: d_GDPC1Q

| | *Coefficient* | *Std. Error* | *t-ratio* | *p-value* | |
|---|---|---|---|---|---|
| const | 49.3131 | 8.4876 | 5.8100 | <0.00001 | *** |
| d_GDPC1Q_1 | 0.360705 | 0.0954613 | 3.7785 | 0.00027 | *** |
| d_GCEC1Q_1 | -0.589345 | 0.346763 | -1.6996 | 0.09229 | * |

| | | | |
|---|---|---|---|
| Mean dependent var | 67.24038 | S.D. dependent var | 58.45773 |
| Sum squared resid | 305827.1 | S.E. of regression | 55.02718 |
| R-squared | 0.131130 | Adjusted R-squared | 0.113925 |
| F(2, 101) | 7.621464 | P-value(F) | 0.000826 |
| rho | -0.096450 | Durbin-Watson | 2.078413 |

F-tests of zero restrictions:
All lags of d_GDPC1Q F(1, 101) = 14.277 [0.0003]
All lags of d_GCEC1Q F(1, 101) = 2.8885 [0.0923]

Equation 2: d_GCEC1Q

| | *Coefficient* | *Std. Error* | *t-ratio* | *p-value* | |
|---|---|---|---|---|---|
| const | 11.8267 | 2.47176 | 4.7847 | <0.00001 | *** |
| d_GDPC1Q_1 | -0.0290698 | 0.0278002 | -1.0457 | 0.29821 | |
| d_GCEC1Q_1 | 0.000855439 | 0.100984 | 0.0085 | 0.99326 | |

| | | | |
|---|---|---|---|
| Mean dependent var | 9.920192 | S.D. dependent var | 15.95743 |
| Sum squared resid | 25936.91 | S.E. of regression | 16.02501 |
| R-squared | 0.011093 | Adjusted R-squared | -0.008489 |
| F(2, 101) | 0.566501 | P-value(F) | 0.569301 |
| rho | 0.014121 | Durbin-Watson | 1.962761 |

F-tests of zero restrictions:
All lags of d_GDPC1Q F(1, 101) = 1.0934 [0.2982]
All lags of d_GCEC1Q F(1, 101) = 7.1758e-005 [0.9933]

Let $y_{d_GDPC1Q,t}$ be $\Delta^{1d} y_{1,t} = y_{1t} - y_{1,t-1}$ and $y_{d_GCEC1Q,t}$ be $\Delta^{1d} y_{2,t} = y_{2t} - y_{2,t-1}$, then we have VAR (1) model in matrix form:

$$\begin{bmatrix} \Delta^{1d} y_{1,t} \\ \Delta^{1d} y_{2,t} \end{bmatrix} = \begin{bmatrix} c_1 \\ c_2 \end{bmatrix} + \begin{bmatrix} A_{1,1} & A_{1,2} \\ A_{2,1} & A_{2,2} \end{bmatrix} \begin{bmatrix} \Delta^{1d} y_{1,t-1} \\ \Delta^{1d} y_{2,t-1} \end{bmatrix} + \begin{bmatrix} e_{1,t} \\ e_{2,t} \end{bmatrix} \tag{3.1}$$

Equation (3.1) can also be written as:

$$\begin{aligned} \Delta^{1d} y_{1,t} &= c_1 + A_{1,1} \Delta^{1d} y_{1,t-1} + A_{1,2} \Delta^{1d} y_{2,t-1} + e_{1,t} \\ \Delta^{1d} y_{2,t} &= c_2 + A_{2,1} \Delta^{1d} y_{1,t-1} + A_{2,2} \Delta^{1d} y_{2,t-1} + e_{2,t} \end{aligned} \tag{3.2}$$

Inputting the estimated parameters into equation (3.1) and (3.2), we have:

$$\begin{bmatrix} \Delta^{1d} y_{1,t} \\ \Delta^{1d} y_{2,t} \end{bmatrix} = \begin{bmatrix} 49.3131 \\ 11.8267 \end{bmatrix} + \begin{bmatrix} 0.360705 & -0.589345 \\ -0.0290698 & 0.000855439 \end{bmatrix} \begin{bmatrix} \Delta^{1d} y_{1,t-1} \\ \Delta^{1d} y_{2,t-1} \end{bmatrix} + \begin{bmatrix} e_{1,t} \\ e_{2,t} \end{bmatrix} \tag{3.3}$$

$$\begin{aligned} \Delta^{1d} y_{1,t} &= 49.3131 + 0.360705 \Delta^{1d} y_{1,t-1} - 0.589345 \Delta^{1d} y_{2,t-1} + e_{1,t} \\ \Delta^{1d} y_{2,t} &= 11.8267 - 0.0290698 \Delta^{1d} y_{1,t-1} + 0.000855439 \Delta^{1d} y_{2,t-1} + e_{2,t} \end{aligned} \tag{3.4}$$

Impulse Responses:

Responses to a one-standard error shock in d_GDPC1Q

| period | d_GDPC1Q | d_GCEC1Q |
|---|---|---|
| 1 | 54.228 | 4.1993 |

| 2 | 17.085 | -1.5728 |
|---|---|---|
| 3 | 7.0897 | -0.49801 |
| 4 | 2.8508 | -0.20652 |
| 5 | 1.15 | -0.083049 |
| 6 | 0.46376 | -0.033502 |
| 7 | 0.18702 | -0.01351 |
| 8 | 0.075423 | -0.0054483 |
| 9 | 0.030416 | -0.0021972 |
| 10 | 0.012266 | -0.00088608 |
| 11 | 0.0049467 | -0.00035734 |
| 12 | 0.0019949 | -0.00014411 |
| 13 | 0.0008045 | -5.8115e-005 |
| 14 | 0.00032444 | -2.3436e-005 |
| 15 | 0.00013084 | -9.4513e-006 |
| 16 | 5.2764e-005 | -3.8115e-006 |
| 17 | 2.1279e-005 | -1.5371e-006 |
| 18 | 8.5812e-006 | -6.1988e-007 |
| 19 | 3.4606e-006 | -2.4998e-007 |
| 20 | 1.3956e-006 | -1.0081e-007 |

Responses to a one-standard error shock in d_GCEC1Q

| period | d_GDPC1Q | d_GCEC1Q |
|---|---|---|
| 1 | 0 | 15.224 |
| 2 | -8.972 | 0.013023 |
| 3 | -3.2439 | 0.26082 |
| 4 | -1.3238 | 0.094523 |
| 5 | -0.53321 | 0.038564 |
| 6 | -0.21506 | 0.015533 |
| 7 | -0.086728 | 0.006265 |
| 8 | -0.034976 | 0.0025265 |
| 9 | -0.014105 | 0.0010189 |
| 10 | -0.0056882 | 0.0004109 |
| 11 | -0.0022939 | 0.00016571 |
| 12 | -0.00092509 | 6.6825e-005 |
| 13 | -0.00037307 | 2.6949e-005 |
| 14 | -0.00015045 | 1.0868e-005 |
| 15 | -6.0673e-005 | 4.3828e-006 |
| 16 | -2.4468e-005 | 1.7675e-006 |
| 17 | -9.8674e-006 | 7.1279e-007 |
| 18 | -3.9793e-006 | 2.8745e-007 |
| 19 | -1.6048e-006 | 1.1592e-007 |
| 20 | -6.4717e-007 | 4.6749e-008 |

Decomposition of variance for d_GDPC1Q

| period | std. error | d_GDPC1Q | d_GCEC1Q |
|---|---|---|---|
| 1 | 54.2277 | 100.0000 | 0.0000 |
| 2 | 57.5591 | 97.5703 | 2.4297 |
| 3 | 58.0848 | 97.3022 | 2.6978 |
| 4 | 58.1697 | 97.2583 | 2.7417 |
| 5 | 58.1836 | 97.2512 | 2.7488 |
| 6 | 58.1858 | 97.2500 | 2.7500 |
| 7 | 58.1862 | 97.2498 | 2.7502 |
| 8 | 58.1862 | 97.2498 | 2.7502 |
| 9 | 58.1862 | 97.2498 | 2.7502 |
| 10 | 58.1862 | 97.2498 | 2.7502 |
| 11 | 58.1862 | 97.2498 | 2.7502 |
| 12 | 58.1862 | 97.2498 | 2.7502 |
| 13 | 58.1862 | 97.2498 | 2.7502 |
| 14 | 58.1862 | 97.2498 | 2.7502 |
| 15 | 58.1862 | 97.2498 | 2.7502 |
| 16 | 58.1862 | 97.2498 | 2.7502 |
| 17 | 58.1862 | 97.2498 | 2.7502 |
| 18 | 58.1862 | 97.2498 | 2.7502 |
| 19 | 58.1862 | 97.2498 | 2.7502 |
| 20 | 58.1862 | 97.2498 | 2.7502 |

Decomposition of variance for d_GCEC1Q

| period | std. error | d_GDPC1Q | d_GCEC1Q |
|---|---|---|---|
| 1 | 15.7922 | 7.0707 | 92.9293 |
| 2 | 15.8703 | 7.9834 | 92.0166 |
| 3 | 15.8803 | 8.0717 | 91.9283 |
| 4 | 15.8819 | 8.0870 | 91.9130 |
| 5 | 15.8822 | 8.0895 | 91.9105 |
| 6 | 15.8822 | 8.0899 | 91.9101 |
| 7 | 15.8822 | 8.0899 | 91.9101 |
| 8 | 15.8822 | 8.0899 | 91.9101 |
| 9 | 15.8822 | 8.0899 | 91.9101 |
| 10 | 15.8822 | 8.0899 | 91.9101 |
| 11 | 15.8822 | 8.0899 | 91.9101 |
| 12 | 15.8822 | 8.0899 | 91.9101 |
| 13 | 15.8822 | 8.0899 | 91.9101 |
| 14 | 15.8822 | 8.0899 | 91.9101 |
| 15 | 15.8822 | 8.0899 | 91.9101 |
| 16 | 15.8822 | 8.0899 | 91.9101 |
| 17 | 15.8822 | 8.0899 | 91.9101 |
| 18 | 15.8822 | 8.0899 | 91.9101 |
| 19 | 15.8822 | 8.0899 | 91.9101 |
| 20 | 15.8822 | 8.0899 | 91.9101 |

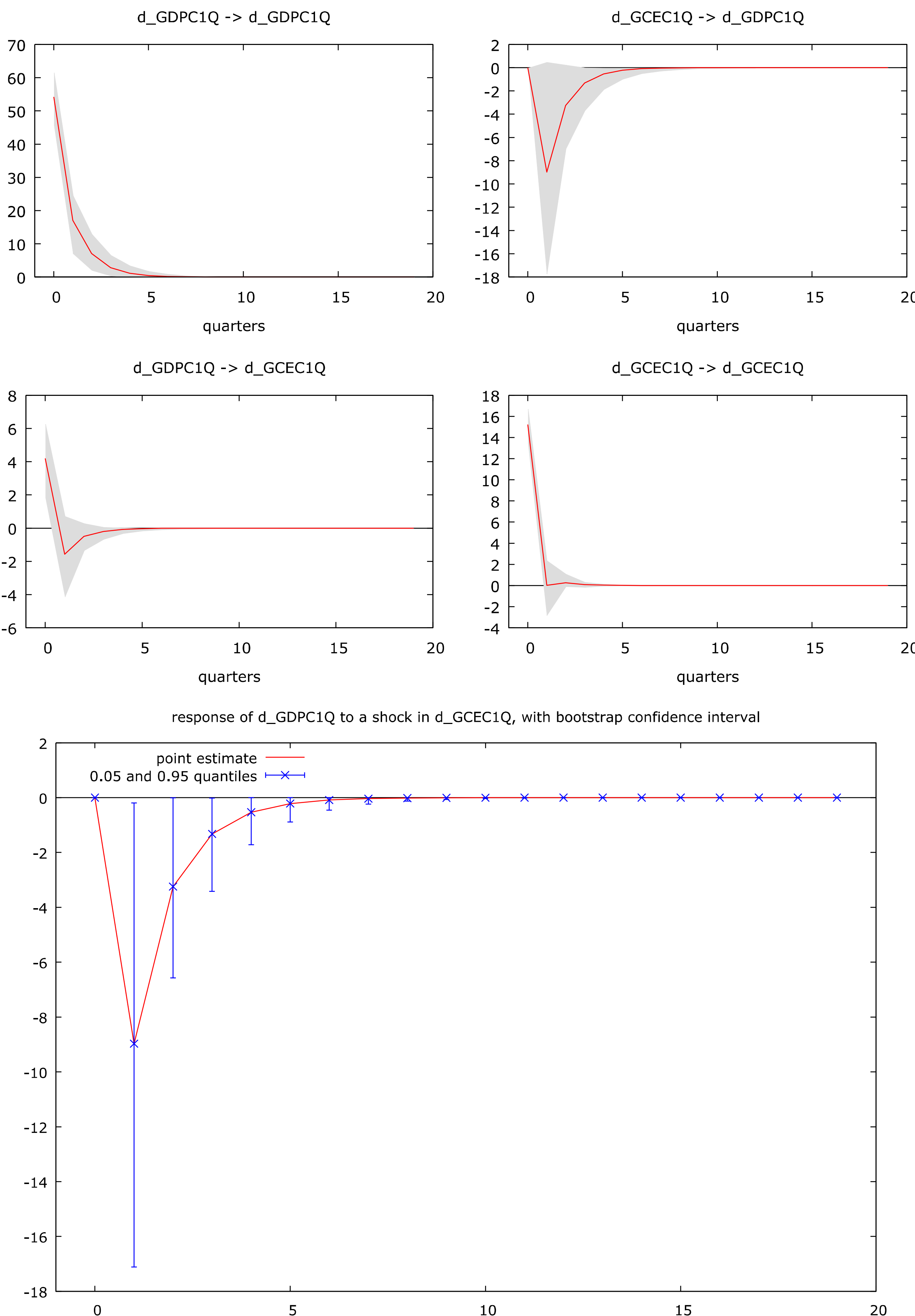

The above analysis tells us that d_GDPC1Q reacts positively to a shock in d_GCEC1Q (both decrease) between period 1 and 2, positively (both increase) from 2 to 3, and negatively (d_GCEC1Q decrease, d_GDPC1Q increase) from 3 to 15. But, starting from 15, the shock of d_GCEC1Q fades away. The overall sign of impulse response is negative (except for period 1), which conforms to the sign of coefficient $(-0.589345)$ of $y_{2,t-1}$ in equation 1 of (3.4). It implies that a shock in government consumption expenditure and gross investment can either decrease or increase GDP temporarily, however GDP will react negatively to such shock in the long run. Hence, the results suggest that government should keep a steady rather than volatile fiscal policy. Since the shock of GCEC on GDP fades out after certain period of time, influencing GDP merely through government expenditure is not optimal in the long run. Further researches are needed for a more comprehensive and effective approach. Final equilibrium is reached when government expenditure shock has little to no impact on GDP.

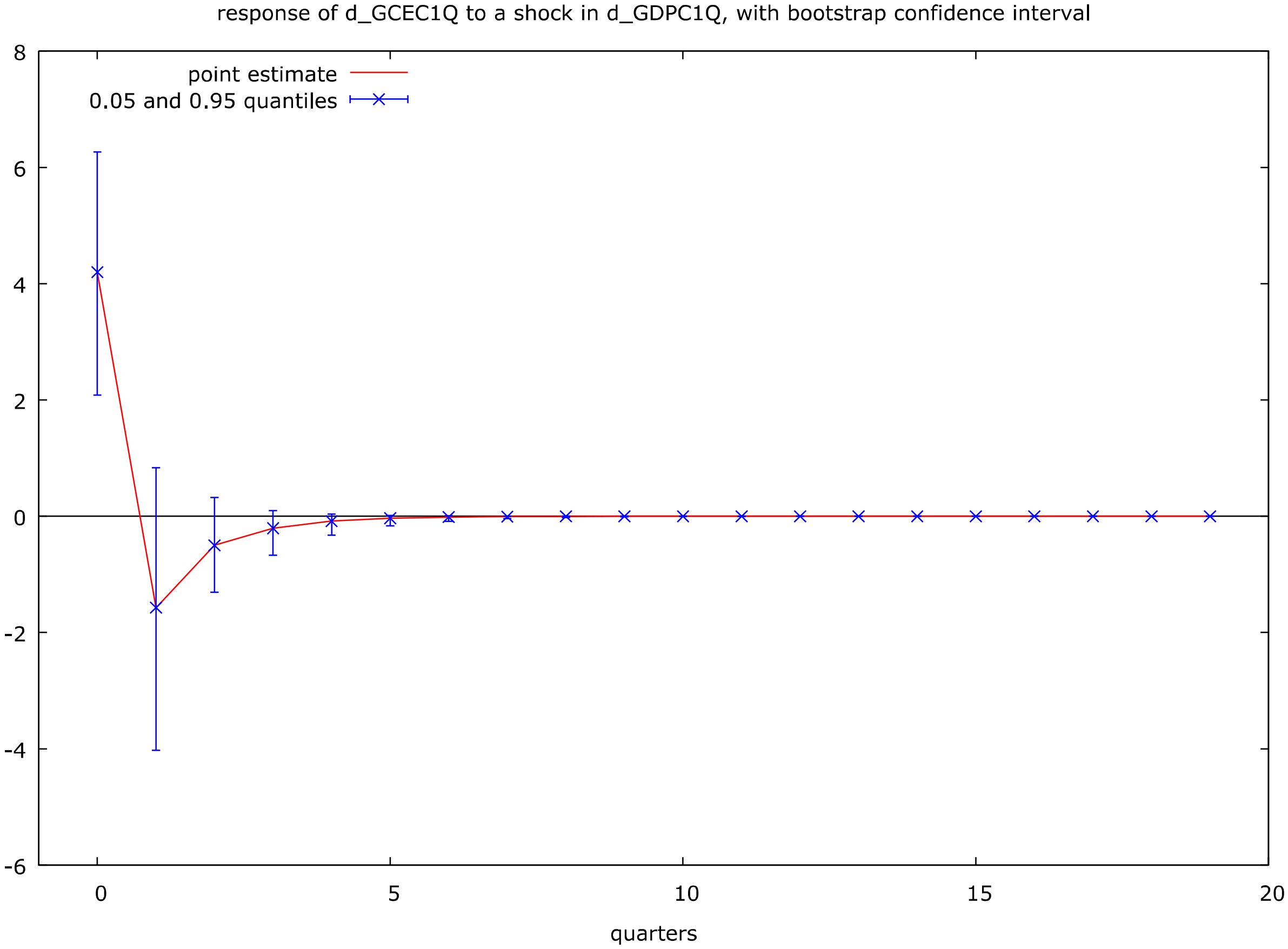

In contrast, d_GCEC1Q reacts positively to a shock in d_GCEC1Q (both decrease) between period 1 and 2, negatively from 2 to 3 (d_GDPC1Q decrease, d_GCEC1Q increase), and negatively from 3 to 15 (d_GDPC1Q decrease, d_GCEC1Q increase). However, a shock in d_GDPC1Q after 15 is fading out. The overall sign of impulse response is negative (except for period 1), which corresponds to the sign of coefficient $(-0.0290698)$ of $y_{1,t-1}$in equation 2 of (3.4). The results suggest that GDP does influence government spending negatively in the long run, but its negative impact force gradually reduces over time. Interestingly, government spending response negatively to GDP shock from period 2 to 3. One theorizes that government try to stimulate weak economy through spending, but unexpectedly continuous expenditure will hurt GDP in the long term. But equilibrium is reached when GDP shock has little to no effect on fiscal policy.

In comparison, one finds that a shock of government spending has more profound influence on GDP than a shock of GDP on expenditure, because the absolute value of coefficient of $y_{2,t-1}$in equation 1 is larger than that of $y_{1,t-1}$ in equation 2 $(0.589345 > 0.0290698)$. The response impulse graphs also hints that response of d_GDPC1Q to a shock in d_GCEC1Q has a deeper valley (curve) than response of d_GCEC1Q to d_GDPC1Q.

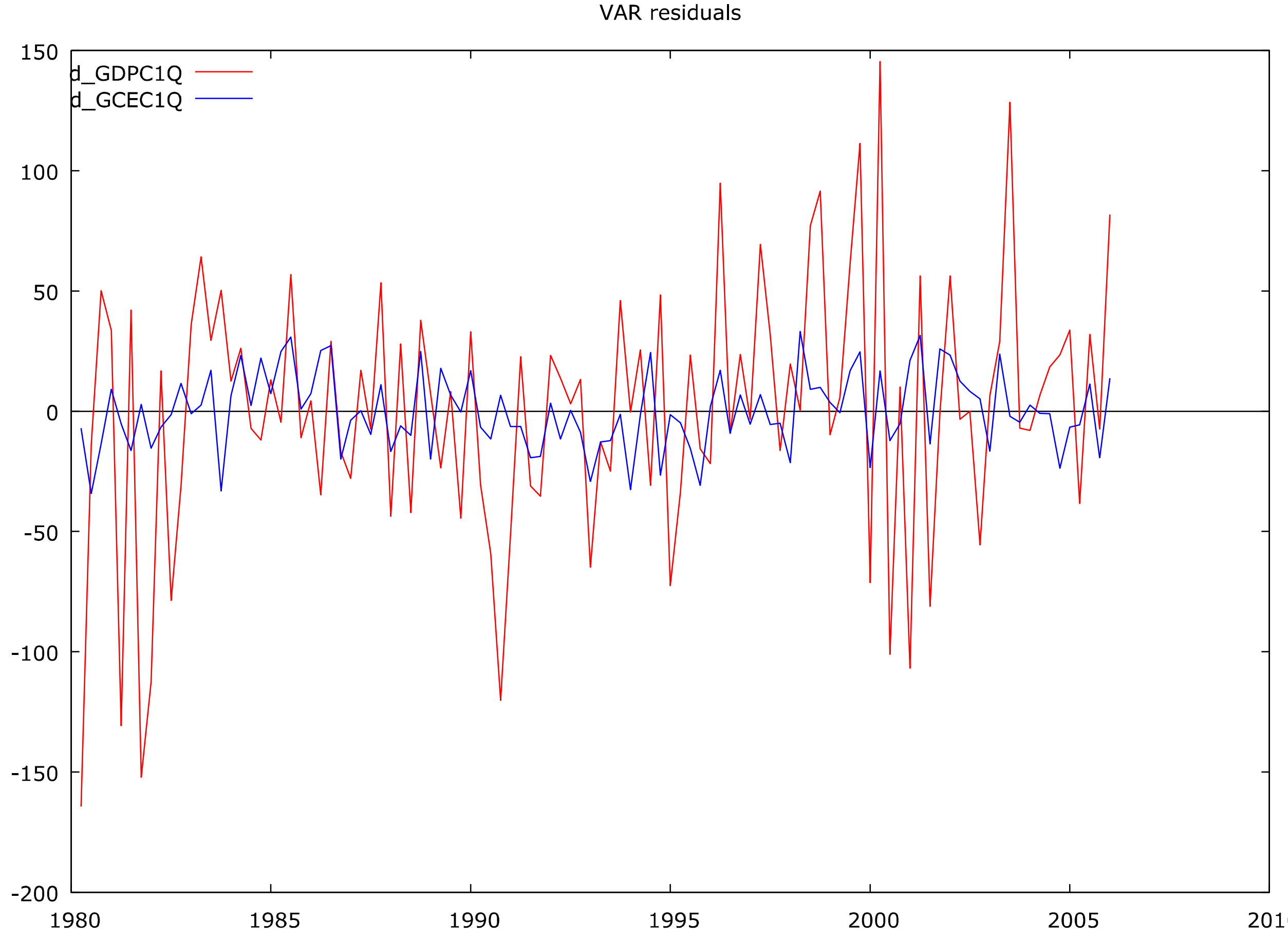

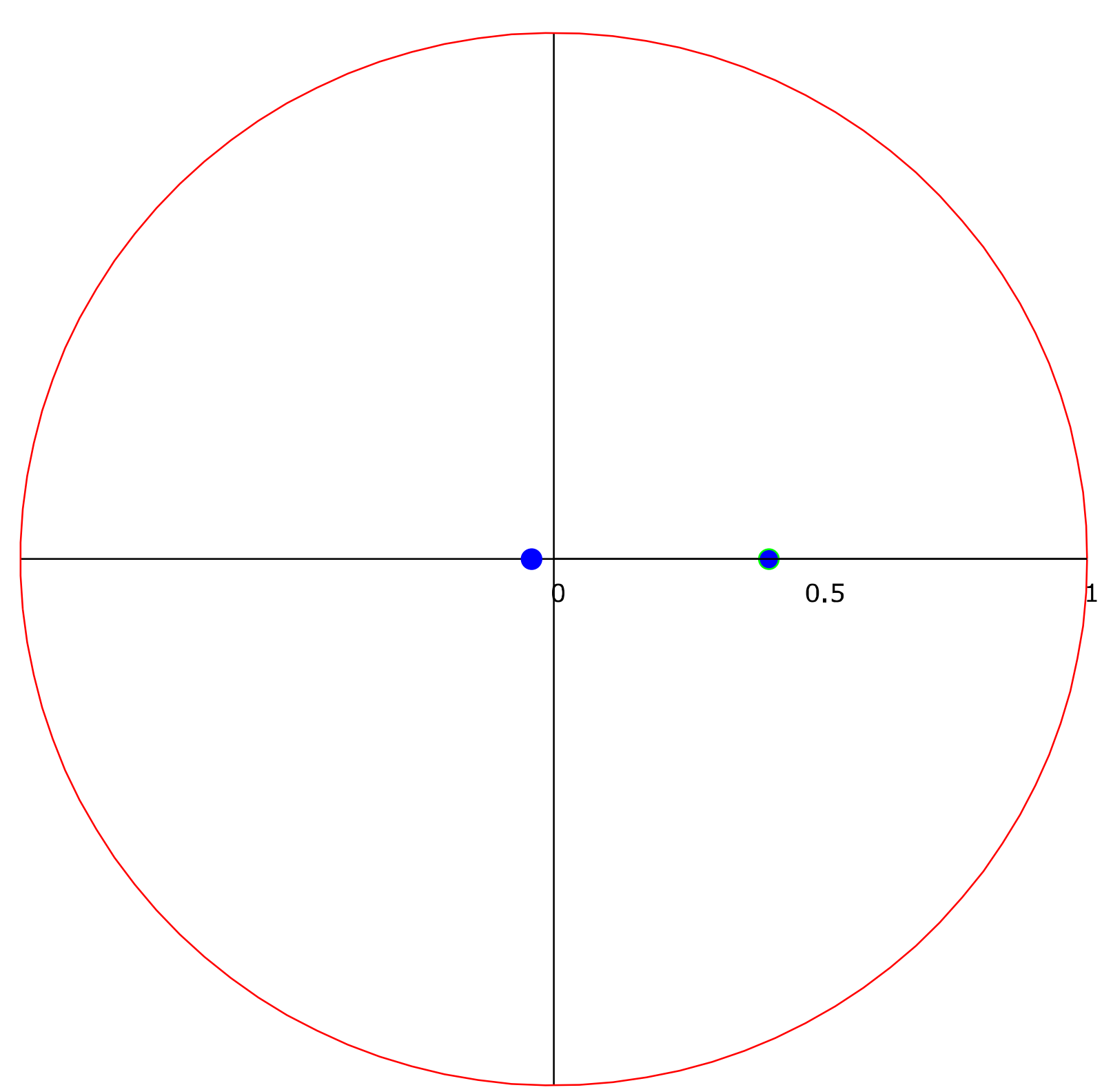

From the VAR residuals time plot, we can see that residuals of equation $y_{2,t}$(d_GCEC1Q) has narrower bandwidth than $y_{1,t}$(d_GDPC1Q). The VAR inverse roots chart shows that VAR (1) is a stable system with both roots within the unit circle.

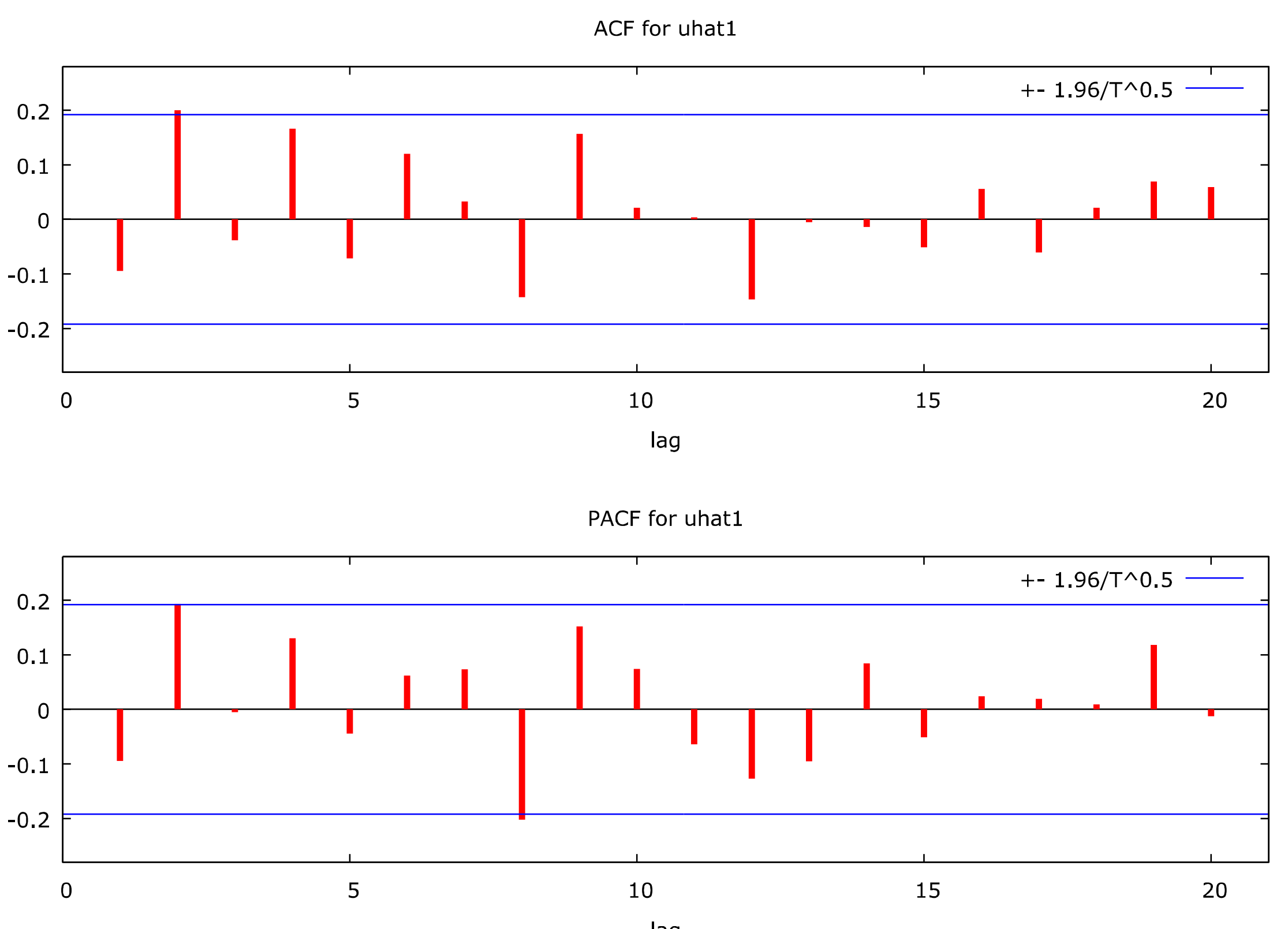

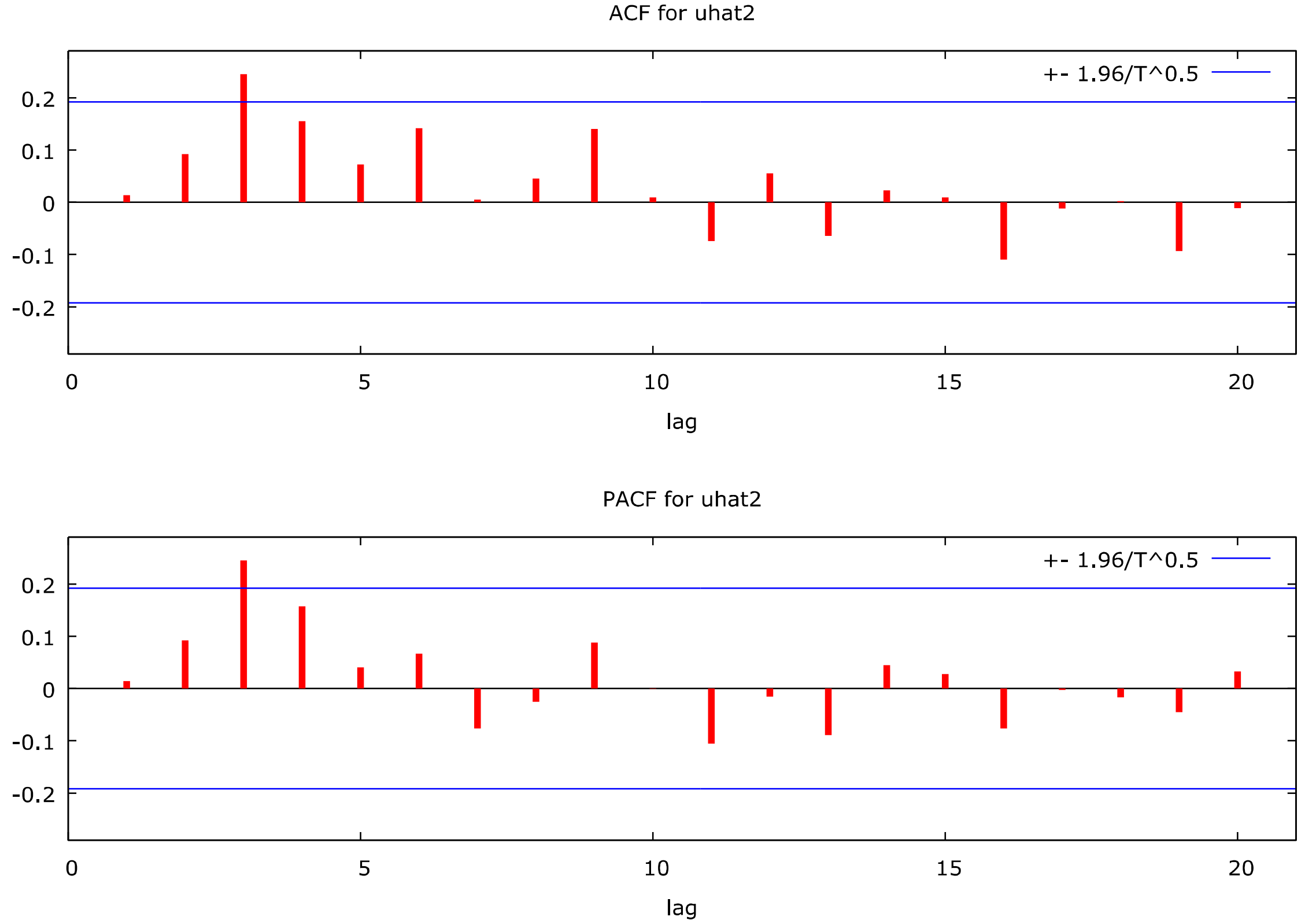

From the above correlograms, we can find that uhat1 (residual from VAR system, equation 1) appears slightly more stable than uhat2 (residual from VAR system, equation 2), because uhat2 is over the upper bound 0.2 at lag 3.

## III. Diagnostic Checking:

1) Autocorrelation test:

```
Equation 1:
Ljung-Box Q' = 8.5046 with p-value = P(Chi-square(4) > 8.5046) = 0.0747

Equation 2:
Ljung-Box Q' = 10.2071 with p-value = P(Chi-square(4) > 10.2071) = 0.0371
```

Autocorrelation test passes for both equation 1 and 2 at p-value of 0.0747 and 0.0371 respectively. The null hypothesis of no autocorrelation cannot be rejected.

2) ARCH test:

```
Test for ARCH of order 4

Equation 1:
             coefficient    std. error    t-ratio    p-value
  ---------------------------------------------------------
  alpha(0)    1472.71        618.846       2.380      0.0193  **
  alpha(1)       0.110710      0.102541    1.080      0.2830
```

```
  alpha(2)      0.168133      0.102610    1.639     0.1046
  alpha(3)      0.117233      0.102445    1.144     0.2554
  alpha(4)      0.0687508     0.0896131   0.7672    0.4449

  Null hypothesis: no ARCH effect is present
  Test statistic: LM = 9.00216
  with p-value = P(Chi-square(4) > 9.00216) = 0.0610456

Equation 2:
             coefficient   std. error   t-ratio   p-value
  -------------------------------------------------------
  alpha(0)   261.313       59.3986       4.399    2.84e-05 ***
  alpha(1)     0.0471413    0.102408     0.4603   0.6463
  alpha(2)    -0.0705335    0.101950    -0.6918   0.4907
  alpha(3)    -0.100824     0.0972352   -1.037    0.3024
  alpha(4)     0.0560263    0.0976747    0.5736   0.5676

  Null hypothesis: no ARCH effect is present
  Test statistic: LM = 2.29443
  with p-value = P(Chi-square(4) > 2.29443) = 0.681783
```

ARCH test passes for both equation 1 and 2 with p-value of 0.0610456 and 0.681783 respectively. The null hypothesis of no ARCH effect cannot be rejected.

3) Multivariate normality test:

```
Residual correlation matrix, C (2 x 2)

      1.0000      0.26591
     0.26591       1.0000

Eigenvalues of C

   0.734092
    1.26591

Doornik-Hansen test
 Chi-square(4) = 7.23579 [0.1239]
```

Doornik-Hansen test shows p value of 0.1239. The null hypothesis of normality cannot be rejected at any significant level.

4) Cointegration test (lag order 4):

```
Step 1: testing for a unit root in d_GDPC1Q

Augmented Dickey-Fuller test for d_GDPC1Q
including 3 lags of (1-L)d_GDPC1Q
(max was 4, criterion modified AIC)
sample size 100
unit-root null hypothesis: a = 1

   test with constant
   model: (1-L)y = b0 + (a-1)*y(-1) + ... + e
   1st-order autocorrelation coeff. for e: 0.041
   lagged differences: F(3, 95) = 4.212 [0.0076]
   estimated value of (a - 1): -0.444352
```

```
   test statistic: tau_c(1) = -3.33419
   asymptotic p-value 0.01345

Step 2: testing for a unit root in d_GCEC1Q

Augmented Dickey-Fuller test for d_GCEC1Q
including 4 lags of (1-L)d_GCEC1Q
(max was 4, criterion modified AIC)
sample size 100
unit-root null hypothesis: a = 1

   test with constant
   model: (1-L)y = b0 + (a-1)*y(-1) + ... + e
   1st-order autocorrelation coeff. for e: -0.003
   lagged differences: F(4, 94) = 2.567 [0.0431]
   estimated value of (a - 1): -0.566423
   test statistic: tau_c(1) = -2.98579
   asymptotic p-value 0.03625

Step 3: cointegrating regression

Cointegrating regression -
OLS, using observations 1980:1-2006:1 (T = 105)
Dependent variable: d_GDPC1Q

             coefficient   std. error   t-ratio    p-value
  ---------------------------------------------------------
  const       59.2415       6.63033      8.935    1.72e-014 ***
  d_GCEC1Q     0.751492     0.353546     2.126    0.0359    **

Mean dependent var   66.78000   S.D. dependent var   58.36696
Sum squared resid    339408.8   S.E. of regression   57.40410
R-squared            0.042022   Adjusted R-squared   0.032721
Log-likelihood      -573.2411   Akaike criterion     1150.482
Schwarz criterion    1155.790   Hannan-Quinn         1152.633
rho                  0.383388   Durbin-Watson        1.218974

Step 4: testing for a unit root in uhat

Augmented Dickey-Fuller test for uhat
including one lag of (1-L)uhat
(max was 4, criterion modified AIC)
sample size 100
unit-root null hypothesis: a = 1

   model: (1-L)y = (a-1)*y(-1) + ... + e
   1st-order autocorrelation coeff. for e: 0.038
   estimated value of (a - 1): -0.442103
   test statistic: tau_c(2) = -4.05898
   asymptotic p-value 0.005862

There is evidence for a cointegrating relationship if:
(a) The unit-root hypothesis is not rejected for the individual variables.
(b) The unit-root hypothesis is rejected for the residuals (uhat) from the
    cointegrating regression.
```

The cointegration test results indicate that (a) The unit-root hypothesis is rejected for d_GDPC1Q and d_GCEC1Q; (b) The unit-root hypothesis is rejected for the residuals (uhat) from the cointegrating regression. Thereby, cointegration is not present for variables d_GDPC1Q and d_GCEC1Q. VAR(1) is adequate and VECM is not needed. Additionally, residuals plot for both equations has relatively stable trend. Based on the results of diagnostic checking, the fitted model is appropriate.

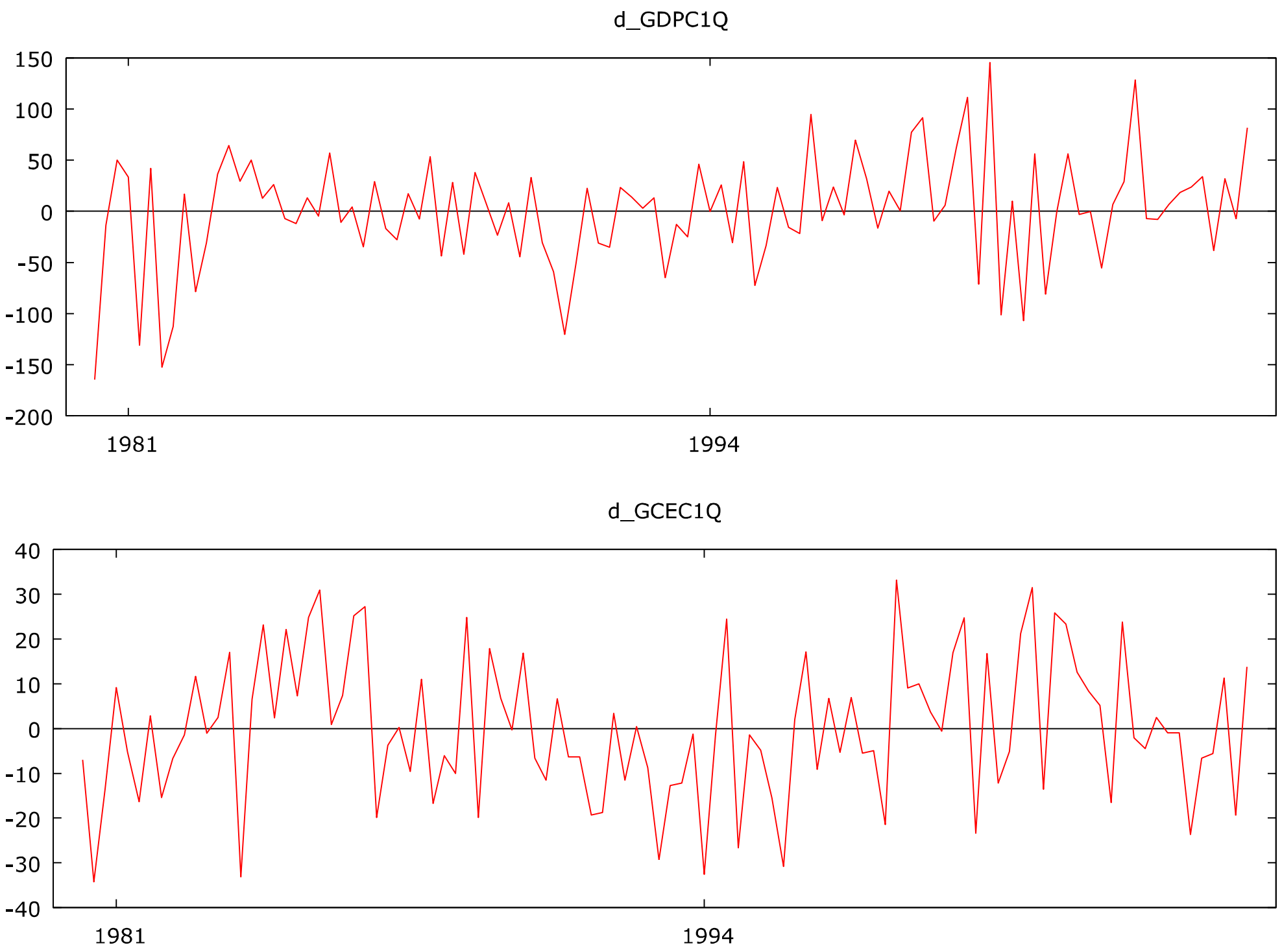

## IV. Forecast:

For 95% confidence intervals, t(101, 0.025) = 1.984

| Obs | d_GDPC1Q | prediction | std. error | 95% interval |
|---|---|---|---|---|
| 2006:2 | undefined | 93.3470 | 54.2277 | (-14.2262, 200.920) |
| 2006:3 | undefined | 78.7567 | 57.5591 | (-35.4251, 192.939) |
| 2006:4 | undefined | 72.3467 | 58.0848 | (-42.8779, 187.571) |
| 2007:1 | undefined | 69.7836 | 58.1697 | (-45.6096, 185.177) |
| 2007:2 | undefined | 68.7490 | 58.1836 | (-46.6715, 184.170) |
| 2007:3 | undefined | 68.3318 | 58.1858 | (-47.0931, 183.757) |
| 2007:4 | undefined | 68.1636 | 58.1862 | (-47.2621, 183.589) |
| 2008:1 | undefined | 68.0958 | 58.1862 | (-47.3301, 183.522) |
| 2008:2 | undefined | 68.0684 | 58.1862 | (-47.3574, 183.494) |
| 2008:3 | undefined | 68.0574 | 58.1862 | (-47.3685, 183.483) |
| 2008:4 | undefined | 68.0529 | 58.1862 | (-47.3729, 183.479) |
| 2009:1 | undefined | 68.0511 | 58.1862 | (-47.3747, 183.477) |
| 2009:2 | undefined | 68.0504 | 58.1862 | (-47.3754, 183.476) |
| 2009:3 | undefined | 68.0501 | 58.1862 | (-47.3757, 183.476) |
| 2009:4 | undefined | 68.0500 | 58.1862 | (-47.3759, 183.476) |
| 2010:1 | undefined | 68.0499 | 58.1862 | (-47.3759, 183.476) |

| 2010:2 | undefined | 68.0499 | 58.1862 | (-47.3759, 183.476) |
|---|---|---|---|---|
| 2010:3 | undefined | 68.0499 | 58.1862 | (-47.3759, 183.476) |
| 2010:4 | undefined | 68.0499 | 58.1862 | (-47.3759, 183.476) |
| 2011:1 | undefined | 68.0499 | 58.1862 | (-47.3759, 183.476) |
| 2011:2 | undefined | 68.0499 | 58.1862 | (-47.3759, 183.476) |
| 2011:3 | undefined | 68.0499 | 58.1862 | (-47.3759, 183.476) |
| 2011:4 | undefined | 68.0499 | 58.1862 | (-47.3759, 183.476) |
| 2012:1 | undefined | 68.0499 | 58.1862 | (-47.3759, 183.476) |
| 2012:2 | undefined | 68.0499 | 58.1862 | (-47.3759, 183.476) |
| 2012:3 | undefined | 68.0499 | 58.1862 | (-47.3759, 183.476) |
| 2012:4 | undefined | 68.0499 | 58.1862 | (-47.3759, 183.476) |
| 2013:1 | undefined | 68.0499 | 58.1862 | (-47.3759, 183.476) |

For 95% confidence intervals, t(101, 0.025) = 1.984

| Obs | d_GCEC1Q | prediction | std. error | 95% interval |
|---|---|---|---|---|
| 2006:2 | undefined | 7.17258 | 15.7922 | (-24.1549, 38.5000) |
| 2006:3 | undefined | 9.11928 | 15.8703 | (-22.3632, 40.6017) |
| 2006:4 | undefined | 9.54508 | 15.8803 | (-21.9571, 41.0473) |
| 2007:1 | undefined | 9.73179 | 15.8819 | (-21.7736, 41.2372) |
| 2007:2 | undefined | 9.80646 | 15.8822 | (-21.6995, 41.3124) |
| 2007:3 | undefined | 9.83659 | 15.8822 | (-21.6694, 41.3426) |
| 2007:4 | undefined | 9.84875 | 15.8822 | (-21.6573, 41.3548) |
| 2008:1 | undefined | 9.85365 | 15.8822 | (-21.6524, 41.3597) |
| 2008:2 | undefined | 9.85562 | 15.8822 | (-21.6504, 41.3617) |
| 2008:3 | undefined | 9.85642 | 15.8822 | (-21.6496, 41.3625) |
| 2008:4 | undefined | 9.85674 | 15.8822 | (-21.6493, 41.3628) |
| 2009:1 | undefined | 9.85687 | 15.8822 | (-21.6492, 41.3629) |
| 2009:2 | undefined | 9.85692 | 15.8822 | (-21.6491, 41.3630) |
| 2009:3 | undefined | 9.85695 | 15.8822 | (-21.6491, 41.3630) |
| 2009:4 | undefined | 9.85695 | 15.8822 | (-21.6491, 41.3630) |
| 2010:1 | undefined | 9.85696 | 15.8822 | (-21.6491, 41.3630) |
| 2010:2 | undefined | 9.85696 | 15.8822 | (-21.6491, 41.3630) |
| 2010:3 | undefined | 9.85696 | 15.8822 | (-21.6491, 41.3630) |
| 2010:4 | undefined | 9.85696 | 15.8822 | (-21.6491, 41.3630) |
| 2011:1 | undefined | 9.85696 | 15.8822 | (-21.6491, 41.3630) |
| 2011:2 | undefined | 9.85696 | 15.8822 | (-21.6491, 41.3630) |
| 2011:3 | undefined | 9.85696 | 15.8822 | (-21.6491, 41.3630) |
| 2011:4 | undefined | 9.85696 | 15.8822 | (-21.6491, 41.3630) |
| 2012:1 | undefined | 9.85696 | 15.8822 | (-21.6491, 41.3630) |
| 2012:2 | undefined | 9.85696 | 15.8822 | (-21.6491, 41.3630) |
| 2012:3 | undefined | 9.85696 | 15.8822 | (-21.6491, 41.3630) |
| 2012:4 | undefined | 9.85696 | 15.8822 | (-21.6491, 41.3630) |
| 2013:1 | undefined | 9.85696 | 15.8822 | (-21.6491, 41.3630) |

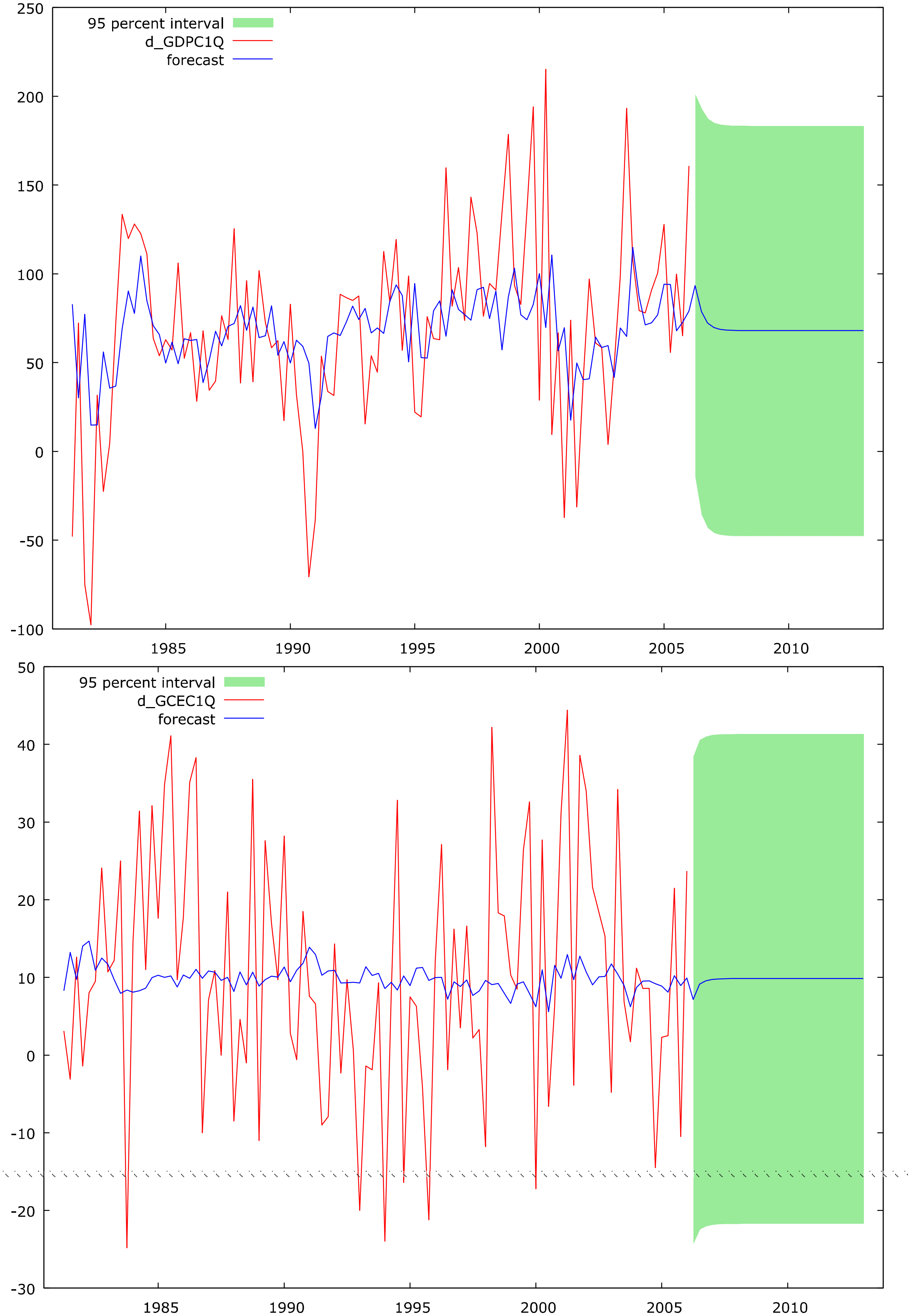

VAR (1) forecast of d_GDPC1Q shows downward trend, compared with upward of d_GCEC1Q. Interestingly, both forecasts are moving exactly the opposite direction at the initial stage, but only the decreasing rate of d_GDPC1Q is much larger than the increasing of d_GCEC1Q. In one's opinion, empirical cause and effect is very evident for those two relations, GDP and Government. It is fruitful for future researchers to explore such phenomenon, critically important to the fiscal policy maker.

---

***(4) A State-Space Model with A Kalman Filter (ARMA estimation via Kalman example)***

For Model 1.2 GDPC1Q ARIMA (1, 1, 1):

Recall that:

First difference of GDPC1Q $(I, 1)$=d_GDPC1Q and $Y_{d_GDPC1Q} = \Delta^{1d} Y_1 = Y_1 - Y_{1,t-1}$

$$\Delta^{1d} Y_{1t} = c_1 + \varphi_1 \Delta^{1d} Y_{1,t-1} + \varepsilon_{1t} + \theta_1 \varepsilon_{1,t-1}$$

$$\Delta^{1d} Y_{1t} = 66.2629 + 0.759918 \Delta^{1d} Y_{1,t-1} + \varepsilon_{1t} - 0.479305 \varepsilon_{1,t-1}$$

Based on James Hamilton's with slight variations, a state-space model can be written as:

$$\xi_{t+1} = \mathrm{F}_t \xi_t + \mathrm{v}_t \qquad (4.1)$$

$$y_t = A_t' \mathrm{x}_t + \mathrm{H}_t' \xi_t + \mathrm{w}_t \qquad (4.2)$$

Where (4.1) is the state transition equation and (4.2) is the observation or measurement equation. The state vector, $\xi_t$, is $(r \times 1)$ and the vector of observables, $y_t$, is $(n \times 1)$; $\mathrm{x}_t$ is a $(k \times 1)$vector of exogenous variables. The $(r \times 1)$ vector $\mathrm{v}_t$ and the $(n \times 1)$ vector $\mathrm{w}_t$ are assumed to be vector white noise:

$$E(\mathrm{v}_t \mathrm{v}_s') = \mathrm{Q}_t \; for \; t = s, otherwise \; 0$$

$$E(\mathrm{w}_t \mathrm{w}_s') = \mathrm{R}_t \; for \; t = s, otherwise \; 0$$

The Kalman filter provides a very efficient yet accurate recursive way to compute the likelihood of ARMA models. ARMA (1, 1) can be written as:

$$y_t = \phi y_{t-1} + \varepsilon_t + \theta\varepsilon_{t-1} \tag{4.3}$$

In order for equation (4.3) to be transformed into state-space form, we can define a latent process $\xi_t = (1 - \phi L)^{-1}\varepsilon_t$. Then, the observation equation (4.3) can be rewritten as:

$$y_t = \xi_t + \theta\xi_{t-1} \tag{4.4}$$

Construct the following two-equation system:

$$\begin{aligned} \xi_t &= \phi\xi_{t-1} + \varepsilon_t \\ \xi_{t-1} &= \xi_{t-1} + 0\xi_{t-2} + 0 \end{aligned} \tag{4.5}$$

And the state transition equation—(4.5) in matrix form will be:

$$\begin{bmatrix} \xi_t \\ \xi_{t-1} \end{bmatrix} = \begin{bmatrix} \phi & 0 \\ 1 & 0 \end{bmatrix} \begin{bmatrix} \xi_{t-1} \\ \xi_{t-2} \end{bmatrix} + \begin{bmatrix} \varepsilon_t \\ 0 \end{bmatrix} \tag{4.6}$$

In comparison, for ARIMA (1, 1, 1), we just need to modify (4.3) as:

$$\Delta^{1d} y_t = \phi\Delta^{1d} y_{t-1} + \varepsilon_t + \theta\varepsilon_{t-1} \tag{4.7}$$

Modify (4.4) as:

$$\Delta^{1d} y_t = \xi_t + \theta\xi_{t-1} \tag{4.8}$$

Equation (4.5) and (4.6) are the same. We first build the filter, then use Maximum Likelihood (ML) to estimate the parameters of ARIMA (1, 1, 1) with the sample range of 1980 Q1-2006 Q1. The numerical results are demonstrated below:

(gretl scripts)

```
series y = diff(GDPC1Q)

/* parameter initalization */
phi = 0
theta = 0
sigma = 1

/* Kalman filter setup */
matrix H = {1; theta}
matrix F = {phi, 0; 1, 0}
matrix Q = {sigma^2, 0; 0, 0}
```

```
kalman
   obsy y
   obsymat H
   statemat F
   statevar Q
end kalman

/* maximum likelihood estimation */
mle logl = ERR ? NA : $kalman_llt
   H[2] = theta
   F[1,1] = phi
   Q[1,1] = sigma^2
   ERR = kfilter()
   params phi theta sigma
end mle -h
```

(gretl results)
Using numerical derivatives

Tolerance = 1.81899e-012

Function evaluations: 120

Evaluations of gradient: 50

Model 1: ML, using observations 1980:1-2006:1 (T = 105)

logl = ERR ? NA : $kalman_llt

Standard errors based on Hessian

| | estimate | std. error | z | p-value | |
|---|---|---|---|---|---|
| phi | 0.975190 | 0.0252938 | 38.55 | 0.0000 | *** |
| theta | -1.43617 | 0.231869 | -6.194 | 5.87e-010 | *** |
| sigma | 38.3644 | 6.70031 | 5.726 | 1.03e-08 | *** |

Log-likelihood -570.6483 Akaike criterion 1147.297

Schwarz criterion 1155.258 Hannan-Quinn 1150.523

Also, we are able to get the results on the following scalar H, F, Q:

$$H = \begin{bmatrix} 1 \\ -1.43617 \end{bmatrix}; F = \begin{bmatrix} 0.97519 & 0 \\ 1 & 0 \end{bmatrix}; Q = \begin{bmatrix} 1471.79253 & 0 \\ 0 & 0 \end{bmatrix}$$

In conclusion, we can write:

$$\xi_{t+1} = \begin{bmatrix} 0.97519 & 0 \\ 1 & 0 \end{bmatrix} \xi_t + \mathrm{v}_t \tag{4.1'}$$

$$y_t = \begin{bmatrix} 1 \\ -1.43617 \end{bmatrix} \xi_t \tag{4.2'}$$

$$where\ E(\mathrm{v}_t \mathrm{v}_s') = \begin{bmatrix} 1471.79253 & 0 \\ 0 & 0 \end{bmatrix} for\ t = s, otherwise\ 0$$

And,

$$\Delta^{1d} y_t = 0.975190\, \Delta^{1d} y_{t-1} + \varepsilon_t - 1.43617\, \varepsilon_{t-1} \tag{4.7'}$$

$$\Delta^{1d} y_t = \xi_t - 1.43617 \xi_{t-1} \tag{4.8'}$$

$$\begin{bmatrix} \xi_t \\ \xi_{t-1} \end{bmatrix} = \begin{bmatrix} 0.975190 & 0 \\ 1 & 0 \end{bmatrix} \begin{bmatrix} \xi_{t-1} \\ \xi_{t-2} \end{bmatrix} + \begin{bmatrix} \varepsilon_t \\ 0 \end{bmatrix} \tag{4.6'}$$

We find that ARIMA (1, 1, 1) of GDPC1Q without a constant has the same positive sign on $\varphi$ and negative sign on $\theta$, plus significant parameters, compared to ARIMA (1, 1, 1) with a constant.

---

***(5) Transfer-Function and Intervention Models (VAR and ARMAX)***

Sample range: 1980Q1-2006Q1

*Times series models with exogenous variables are also known as *transfer function models.*

VAR (1) modeled in section 3 is one kind of the intervention model. Recall that:

Let $y_{d_GDPC1Q,t}$ be $\Delta^{1d} y_{1,t} = y_{1t} - y_{1,t-1}$ and $y_{d_GCEC1Q,t}$ be $\Delta^{1d} y_{2,t} = y_{2t} - y_{2,t-1}$, then we have VAR (1) model in matrix form:

$$\begin{aligned} \Delta^{1d} y_{1,t} &= 49.3131 + 0.360705 \Delta^{1d} y_{1,t-1} - 0.589345 \Delta^{1d} y_{2,t-1} + e_{1,t} \\ \Delta^{1d} y_{2,t} &= 11.8267 - 0.0290698 \Delta^{1d} y_{1,t-1} + 0.000855439 \Delta^{1d} y_{2,t-1} + e_{2,t} \end{aligned} \tag{3.4}$$

In the above model, the intervention government expenditure ($\Delta^{1d} y_{2,t-1}$) negatively affects GDP ($\Delta^{1d} y_{1,t}$) permanently and in return, $\Delta^{1d} y_{2,t}$ reacts negatively to a shock in $\Delta^{1d} y_{1,t-1}$ temporarily. Detailed impulse response analysis can be found in section 3.

Another kind of the intervention model is to add a regressor in Model 1 d_GDPC1Q_ARIMA (1, 1, 1) or Model 2 d_GCEC1Q_ARIMA (1, 1, 2), namely ARMAX. Recall that:

$$\Delta^{1d}Y_{1t} = c_1 + \varphi_1 \Delta^{1d}Y_{1,t-1} + \varepsilon_{1t} + \theta_1 \varepsilon_{1,t-1} \tag{1.16}$$

$$\Delta^{1d}Y_{2t} = c_2 + \varphi_1 \Delta^{1d}Y_{2,t-1} + \varepsilon_{2t} + \theta_1 \varepsilon_{2,t-1} + \theta_2 \varepsilon_{2,t-2} \tag{1.11}$$

Now modify equation (1.6) and (1.11) as follows:

$$\Delta^{1d}y_{1t} = c_1 + \varphi_1 \Delta^{1d}y_{1,t-1} + \xi_1 \Delta^{1d}y_{2t} + \varepsilon_{1t} + \theta_1 \varepsilon_{1,t-1} \tag{5.1}$$

$$\Delta^{1d}y_{2t} = c_2 + \varphi_1 \Delta^{1d}y_{2,t-1} + \xi_1 \Delta^{1d}y_{1t} + \varepsilon_{2t} + \theta_1 \varepsilon_{2,t-1} + \theta_2 \varepsilon_{2,t-2} \tag{5.2}$$

Equation (5.1) can be then estimated:

Model 5.1: ARMAX, using observations 1980:1-2006:1 (T = 105)
Dependent variable: d_GDPC1Q
Standard errors based on Hessian

| | *Coefficient* | *Std. Error* | *z* | *p-value* | |
|---|---|---|---|---|---|
| const | 57.0849 | 11.2986 | 5.0524 | <0.00001 | *** |
| phi_1 | 0.720079 | 0.137865 | 5.2231 | <0.00001 | *** |
| theta_1 | -0.383438 | 0.174517 | -2.1971 | 0.02801 | ** |
| d_GCEC1Q | 0.922759 | 0.304911 | 3.0263 | 0.00248 | *** |

| | | | |
|---|---|---|---|
| Mean dependent var | 66.78000 | S.D. dependent var | 58.36696 |
| Mean of innovations | 0.777440 | S.D. of innovations | 51.48378 |
| Log-likelihood | -562.9434 | Akaike criterion | 1135.887 |
| Schwarz criterion | 1149.157 | Hannan-Quinn | 1141.264 |

| | *Real* | *Imaginary* | *Modulus* | *Frequency* |
|---|---|---|---|---|
| AR | | | | |
| Root 1 | 1.3887 | 0.0000 | 1.3887 | 0.0000 |
| MA | | | | |
| Root 1 | 2.6080 | 0.0000 | 2.6080 | 0.0000 |

$$\Delta^{1d}y_{1t} = 57.0849 + 0.720079\Delta^{1d}y_{1,t-1} + 0.922759\Delta^{1d}y_{2t} + \varepsilon_{1t} - 0.383438\varepsilon_{1,t-1} \tag{5.3}$$

```
Frequency distribution for uhat11, obs 133-237
number of bins = 11, mean = 0.77744, sd = 52.613

       interval          midpt   frequency    rel.     cum.

          < -143.79   -158.24        1      0.95%    0.95%
   -143.79 - -114.89   -129.34        4      3.81%    4.76% *
   -114.89 - -85.988   -100.44        2      1.90%    6.67%
   -85.988 - -57.086   -71.537        6      5.71%   12.38% **
   -57.086 - -28.185   -42.636       13     12.38%   24.76% ****
```

```
   -28.185 -   0.71708  -13.734       26      24.76%   49.52% ********
   0.71708 -   29.619    15.168       22      20.95%   70.48% *******
    29.619 -   58.521    44.070       21      20.00%   90.48% *******
    58.521 -   87.422    72.972        7       6.67%   97.14% **
    87.422 -   116.32    101.87        1       0.95%   98.10%
          >=   116.32    130.78        2       1.90%  100.00%

Test for null hypothesis of normal distribution:
Chi-square(2) = 5.861 with p-value 0.05338
```

Equation (5.3) passes the normality test for the null hypothesis cannot be rejected with p-value of 0.05338.

```
Test for autocorrelation up to order 4

Ljung-Box Q' = 1.58435,
with p-value = P(Chi-square(2) > 1.58435) = 0.4529
```

Model 5.1 passes the Ljung-Box autocorrelation test as the null hypothesis cannot be rejected with p-value of 0.4529.

```
Test for ARCH of order 4

             coefficient    std. error    t-ratio   p-value
  ---------------------------------------------------------
  alpha(0)   1295.21        538.403        2.406     0.0181   **
  alpha(1)      0.0370509     0.101975     0.3633    0.7172
  alpha(2)      0.229126      0.101382     2.260     0.0261   **
  alpha(3)      0.0790132     0.0885744    0.8921    0.3746
  alpha(4)      0.0875882     0.0889496    0.9847    0.3272

  Null hypothesis: no ARCH effect is present
  Test statistic: LM = 9.78853
  with p-value = P(Chi-square(4) > 9.78853) = 0.0441445
```

Model 5.1 has no ARCH effect because of p-value 0.0441445.

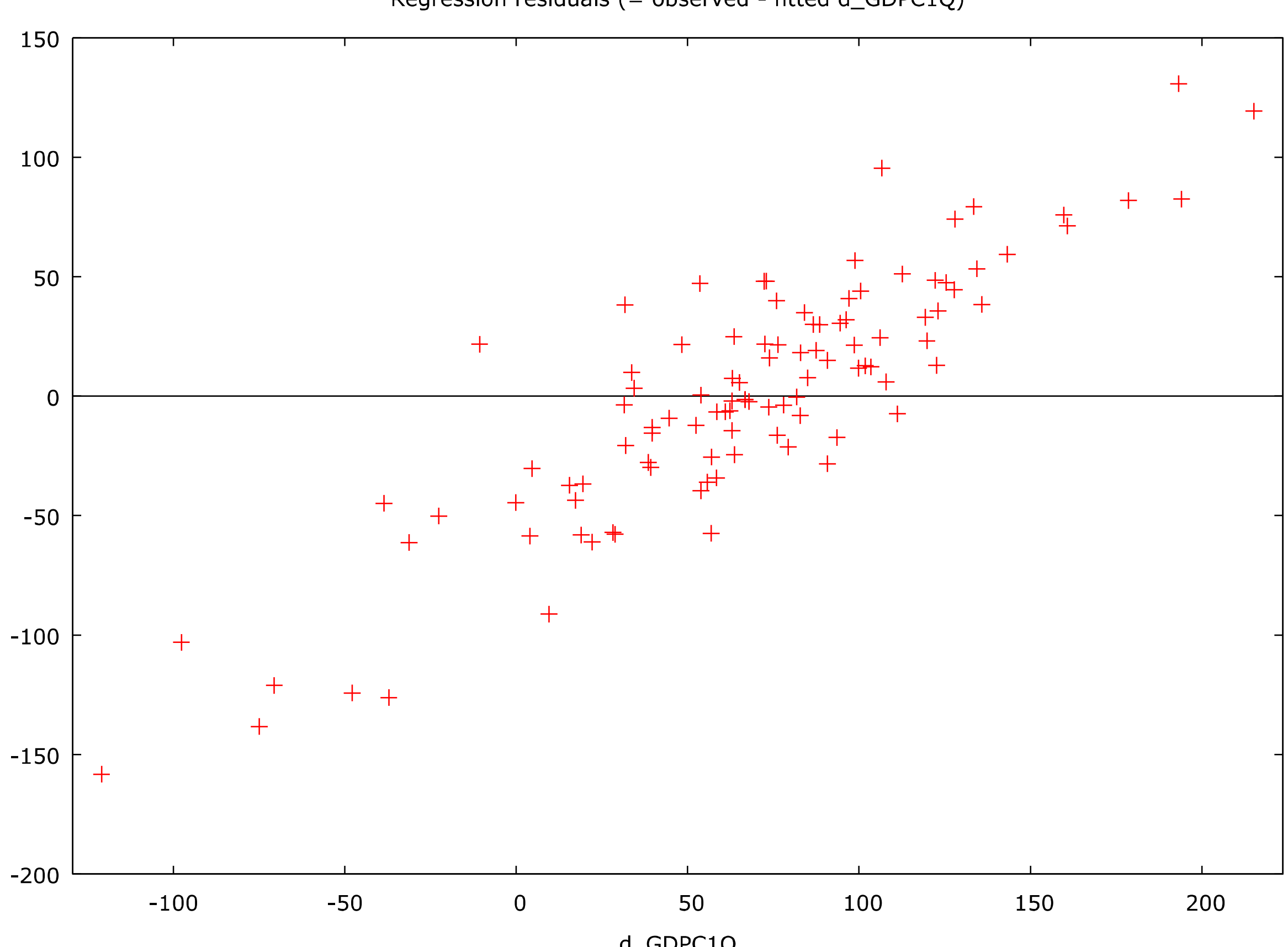

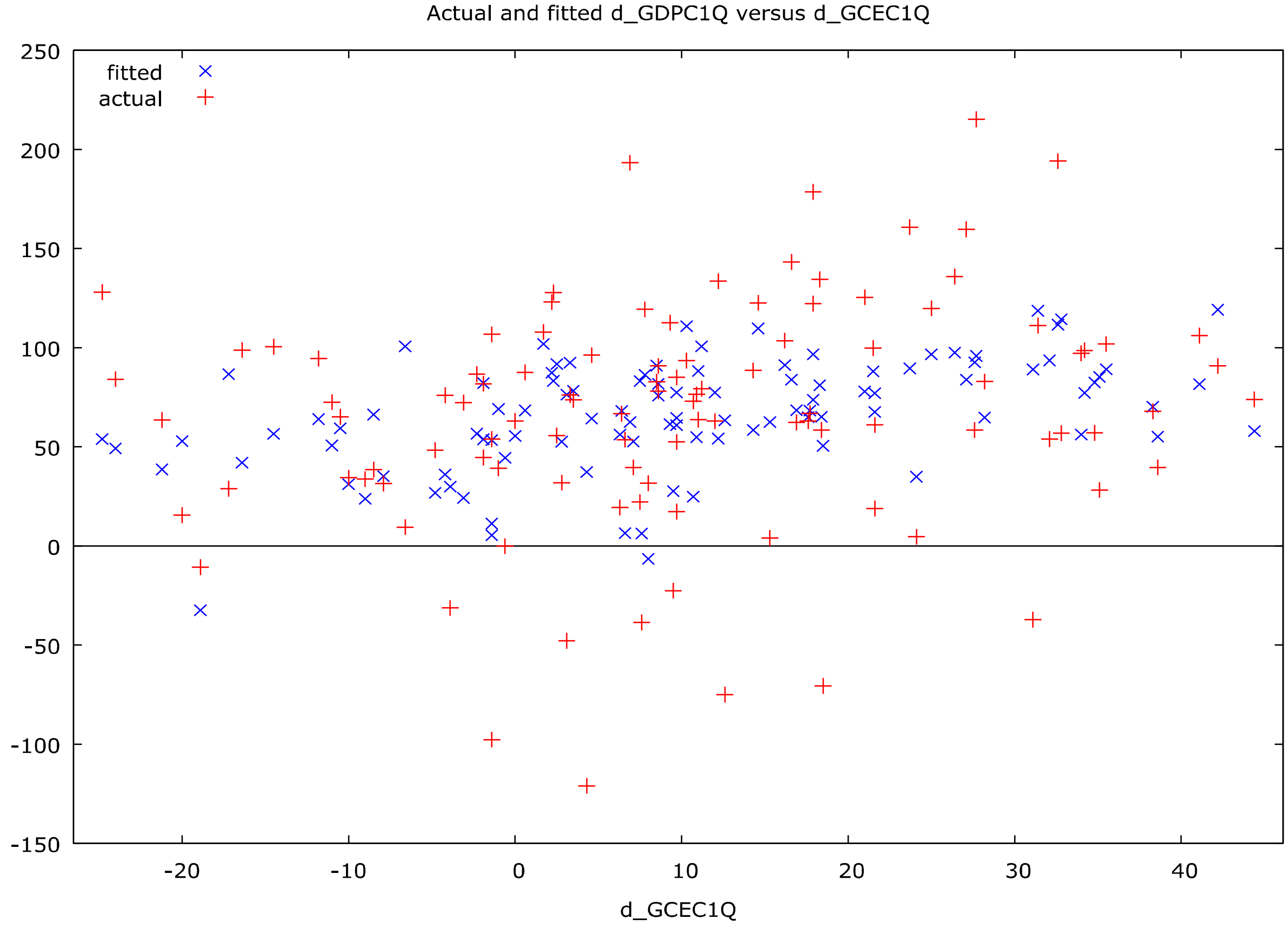

The above graphs show a positive relationship between d_GDPC1Q and residual. Also, there seems to be a positive relationship between d_GDPC1Q and d_GCEC1Q concurrently, corresponding to d_GCEC1Q coefficient of 0.922759 with p-value 0.00248, which implies that GDP reacts very positively to changes in government spending at the same spatial time. This is exactly the way of how economic stimulus package works. But its long term effect on GDP is different as discussed in the VAR analysis.

Equation (5.2) can also be estimated:

Model 5.2: ARMAX, using observations 1980:1-2006:1 (T = 105)
Dependent variable: d_GCEC1Q
Standard errors based on Hessian

| | *Coefficient* | *Std. Error* | *z* | *p-value* | |
|---|---|---|---|---|---|
| const | 6.6791 | 3.01259 | 2.2171 | 0.02662 | ** |
| phi_1 | 0.777622 | 0.132586 | 5.8650 | <0.00001 | *** |
| theta_1 | -0.862196 | 0.159308 | -5.4121 | <0.00001 | *** |
| theta_2 | 0.239026 | 0.127766 | 1.8708 | 0.06137 | * |
| d_GDPC1Q | 0.0483137 | 0.0276438 | 1.7477 | 0.08051 | * |

| | | | |
|---|---|---|---|
| Mean dependent var | 10.03143 | S.D. dependent var | 15.92138 |
| Mean of innovations | 0.021340 | S.D. of innovations | 15.03032 |
| Log-likelihood | -433.6534 | Akaike criterion | 879.3068 |
| Schwarz criterion | 895.2305 | Hannan-Quinn | 885.7594 |

| | *Real* | *Imaginary* | *Modulus* | *Frequency* |
|---|---|---|---|---|
| AR | | | | |
| Root 1 | 1.2860 | 0.0000 | 1.2860 | 0.0000 |
| MA | | | | |
| Root 1 | 1.8036 | -0.9648 | 2.0454 | -0.0782 |
| Root 2 | 1.8036 | 0.9648 | 2.0454 | 0.0782 |

$$\Delta^{1d} y_{2t} = 6.6791 + 0.777622\Delta^{1d} y_{2,t-1} + 0.0483137\Delta^{1d} y_{1t} + \varepsilon_{2t} - 0.862196\varepsilon_{2,t-1} + 0.239026\varepsilon_{2,t-2} \quad (5.4)$$

```
Test for null hypothesis of normal distribution:
Chi-square(2) = 0.191 with p-value 0.90911

Test for autocorrelation up to order 4
Ljung-Box Q' = 3.01165,
with p-value = P(Chi-square(1) > 3.01165) = 0.08267

Test for ARCH of order 4

             coefficient    std. error   t-ratio    p-value
  ---------------------------------------------------------
  alpha(0)   262.456        57.2692       4.583     1.38e-05 ***
  alpha(1)     0.00688222    0.102188     0.06735   0.9464
  alpha(2)    -0.0306509     0.0993920   -0.3084    0.7585
  alpha(3)    -0.153295      0.0991503   -1.546     0.1254
  alpha(4)     0.00261747    0.100694     0.02599   0.9793

  Null hypothesis: no ARCH effect is present
  Test statistic: LM = 2.56683
  with p-value = P(Chi-square(4) > 2.56683) = 0.63271
```

Diagnostic checks out O.K. for Model 5.2. The model passes all the tests, with normally distribution residuals, without autocorrelation and ARCH effect.

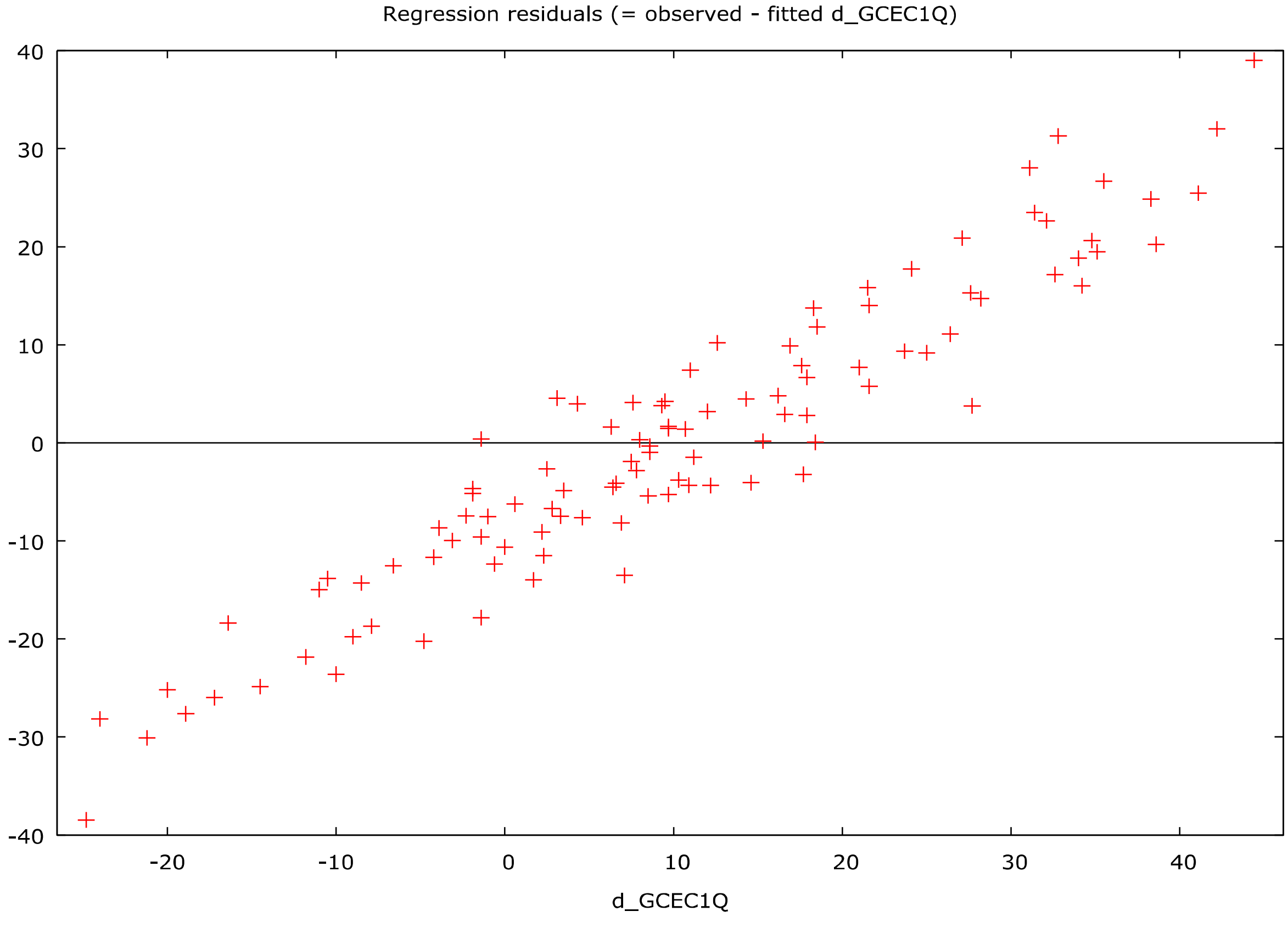

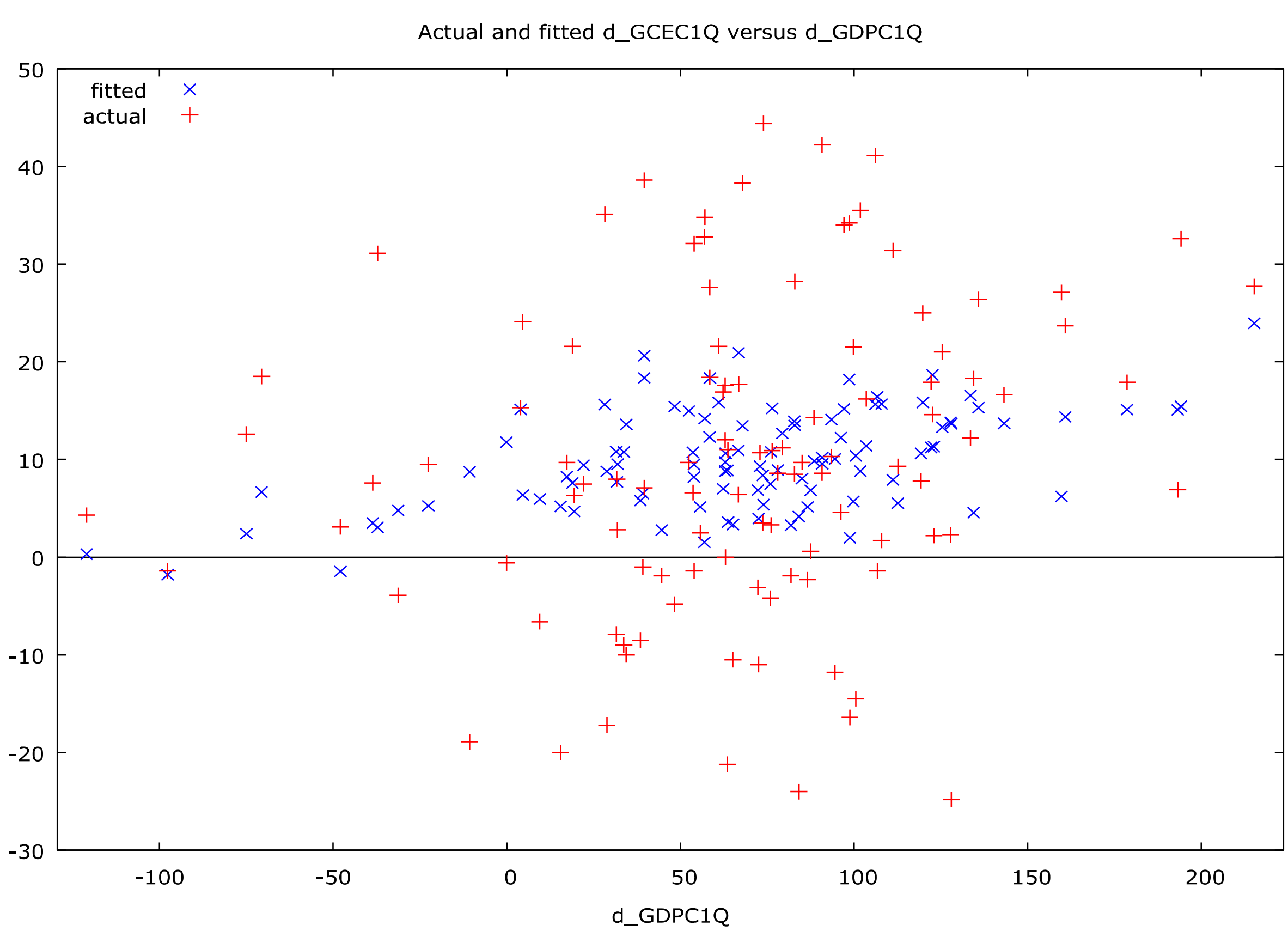

d_GCEC1Q shows positive relationship with residual. There's no obvious pattern how d_GCEC1Q reacts to d_GDPC1Q concurrently. But it appears that the fitted values is clustering around coordinates (70, 9). This conforms to the numerical results that d_GDPC1Q has little positive impact or significance (coefficient 0.0483137 with p-value 0.08051) on d_GCEC1Q, which also implies that the response of government expenditure to changes in GDP is minimal concurrently. But its long term impact on government spending is different as shown in the VAR analysis.

Equation (5.1) and (5.2) can be also modified as (5.5) and (5.6) respectively and estimated jointly as equation (5.7):

$$\Delta^{1d} y_{1t} = c_1 + \varphi_{1,1}\Delta^{1d} y_{1,t-1} + \xi_{1,1}\Delta^{1d} y_{2t} + \varepsilon_{1t} + \theta_{1,1}\varepsilon_{1,t-1} \tag{5.5}$$

$$\Delta^{1d} y_{2t} = c_2 + \varphi_{2,1}\Delta^{1d} y_{2,t-1} + \xi_{2,1}\Delta^{1d} y_{1t} + \varepsilon_{2t} + \theta_{2,1}\varepsilon_{2,t-1} \tag{5.6}$$

$$\begin{bmatrix}\Delta^{1d} y_{1t}\\ \Delta^{1d} y_{2t}\end{bmatrix} = \begin{bmatrix}c_1\\ c_2\end{bmatrix} + \begin{bmatrix}\varphi_{1,1} & \varphi_{1,2}\\ \varphi_{2,1} & \varphi_{2,2}\end{bmatrix}\begin{bmatrix}\Delta^{1d} y_{1,t-1}\\ \Delta^{1d} y_{2,t-1}\end{bmatrix} + \begin{bmatrix}\xi_{1,1} & \xi_{1,2}\\ \xi_{2,1} & \xi_{2,2}\end{bmatrix}\begin{bmatrix}\Delta^{1d} y_{2t}\\ \Delta^{1d} y_{1t}\end{bmatrix} + \begin{bmatrix}\varepsilon_{1t}\\ \varepsilon_{2t}\end{bmatrix} + \begin{bmatrix}\theta_{1,1} & \theta_{1,2}\\ \theta_{2,1} & \theta_{2,2}\end{bmatrix}\begin{bmatrix}\varepsilon_{1,t-1}\\ \varepsilon_{2,t-1}\end{bmatrix} \tag{5.7}$$

Parameters to be estimated: $\begin{bmatrix}c_1\\ c_2\end{bmatrix}$, $\begin{bmatrix}\varphi_{1,1} & \varphi_{1,2}\\ \varphi_{2,1} & \varphi_{2,2}\end{bmatrix}$, $\begin{bmatrix}\xi_{1,1} & \xi_{1,2}\\ \xi_{2,1} & \xi_{2,2}\end{bmatrix}$, $\sigma^2\left(\begin{bmatrix}\varepsilon_{1t}\\ \varepsilon_{2t}\end{bmatrix}\right)$, $\begin{bmatrix}\theta_{1,1} & \theta_{1,2}\\ \theta_{2,1} & \theta_{2,2}\end{bmatrix}$

Equation (5.7) can be considered Multivariate VARMAX process, which is a special case of B-J Multivariate VARMA.

---

### *(6) Unit Root Tests (Augmented Dickey-Fuller test)*

Sample range 1980Q1-2006Q1; Lag order 12 for ADF test.

### For GDPC1Q Data:

```
Augmented Dickey-Fuller test for GDPC1Q
including 2 lags of (1-L)GDPC1Q
(max was 12, criterion modified AIC)
sample size 105
unit-root null hypothesis: a = 1

   test with constant
   model: (1-L)y = b0 + (a-1)*y(-1) + ... + e
   1st-order autocorrelation coeff. for e: 0.020
   lagged differences: F(2, 101) = 6.669 [0.0019]
   estimated value of (a - 1): 0.0045083
   test statistic: tau_c(1) = 1.68139
   asymptotic p-value 0.9997
```

The test results indicate that GDPC1Q has unit root with significantly high asymptotic p-value 0.9997. This tells us that Model 1 data is non-stationary and we need to difference it at least once.

**For GCEC1Q Data:**

```
Augmented Dickey-Fuller test for GCEC1Q
including one lag of (1-L)GCEC1Q
(max was 12, criterion modified AIC)
sample size 105
unit-root null hypothesis: a = 1

   test with constant
   model: (1-L)y = b0 + (a-1)*y(-1) + ... + e
   1st-order autocorrelation coeff. for e: 0.002
   estimated value of (a - 1): -0.000652051
   test statistic: tau_c(1) = -0.124135
   asymptotic p-value 0.9451
```

The test results show that Model 2 GCEC1Q has unit root with relatively high asymptotic p-value 0.9451, which means that Model 2 data is non-stationary and we have to difference it at least once.

***(7) Cointegration Test (Engle-Granger test) and Error Correction Model (if Contegration is Present)***

Sample range 1980Q1-2006Q1; Variables GDPC1Q and GCEC1Q; Lag order 4.

```
Step 1: testing for a unit root in GDPC1Q

Augmented Dickey-Fuller test for GDPC1Q
including 2 lags of (1-L)GDPC1Q
(max was 4, criterion modified AIC)
sample size 100
unit-root null hypothesis: a = 1

   test with constant
   model: (1-L)y = b0 + (a-1)*y(-1) + ... + e
   1st-order autocorrelation coeff. for e: 0.062
   lagged differences: F(2, 96) = 8.226 [0.0005]
   estimated value of (a - 1): 0.00430732
   test statistic: tau_c(1) = 1.66308
   asymptotic p-value 0.9996

Step 2: testing for a unit root in GCEC1Q

Augmented Dickey-Fuller test for GCEC1Q
including one lag of (1-L)GCEC1Q
(max was 4, criterion modified AIC)
sample size 100
unit-root null hypothesis: a = 1
```

```
   test with constant
   model: (1-L)y = b0 + (a-1)*y(-1) + ... + e
   1st-order autocorrelation coeff. for e: 0.006
   estimated value of (a - 1): -0.00249536
   test statistic: tau_c(1) = -0.439981
   asymptotic p-value 0.9

Step 3: cointegrating regression

Cointegrating regression -
OLS, using observations 1980:1-2006:1 (T = 105)
Dependent variable: GDPC1Q

             coefficient   std. error   t-ratio    p-value
  ---------------------------------------------------------
  const      -4023.80      312.349      -12.88    3.38e-023 ***
  GCEC1Q         6.88632     0.165023    41.73    2.38e-066 ***

Mean dependent var   8846.749   S.D. dependent var   2128.680
Sum squared resid    26317802   S.E. of regression   505.4826
R-squared            0.944154   Adjusted R-squared   0.943611
Log-likelihood      -801.6578   Akaike criterion     1607.316
Schwarz criterion    1612.624   Hannan-Quinn         1609.467
rho                  0.972847   Durbin-Watson        0.049978

Step 4: testing for a unit root in uhat

Augmented Dickey-Fuller test for uhat
including 3 lags of (1-L)uhat
(max was 4, criterion modified AIC)
sample size 100
unit-root null hypothesis: a = 1

   model: (1-L)y = (a-1)*y(-1) + ... + e
   1st-order autocorrelation coeff. for e: -0.026
   lagged differences: F(3, 96) = 4.967 [0.0030]
   estimated value of (a - 1): -0.0468461
   test statistic: tau_c(2) = -2.16499
   asymptotic p-value 0.4417

There is evidence for a cointegrating relationship if:
(a) The unit-root hypothesis is not rejected for the individual variables.
(b) The unit-root hypothesis is rejected for the residuals (uhat) from the
    cointegrating regression.
```

Test results with constant indicate that variables GDPC1Q and GCEC1Q are not cointegrated, because a) The unit-root hypothesis is not rejected for the individual variables; b) However, the cointegrating regression has unit root with 0.4417 p-value, therefore the unit-root hypothesis is NOT rejected for the residuals (uhat) from the cointegrating regression. The estimated cointegrating system is shown in step 3.

---

***(8) Volatility Tests by the ARCH, GARCH, ARCH-M, GARCH-M, EGARCH and other GARCH variants***

Sample (full) range: 1947Q1-2013Q1

In order to obtain comparable results (AIC, BIC, HQC), we use 100*ldiff(GDPC1Q) [100 times log first difference of GDPC1Q] for all ARCH, GARCH, ARCH-M, GARCH-M, TS-GARCH, GJR, TARCH, NARCH, APARCH, EGARCH. Let treated endogenous variable 100*ldiff(GDPC1Q) be $y_t$.

**Model 8.1: ARCH (1)**

I. Identification:

LM tests show that ARIMA (1, 1, 1) of GDPC1Q has ARCH effect with p-value of 0.0147128, however ARIMA (1, 1, 2) of GCEC1Q does not has such effect with p-value of 0.611714. Therefore, dependent variable GDPC1Q will be tested for volatility.

II. Estimation and Diagnostic Checking:

Model 8.1: ARCH (1), using observations 1947:2-2013:1 (T = 264)
Dependent variable: 100*ldiff(GDPC1Q)
Standard errors based on Hessian

| | *Coefficient* | *Std. Error* | *z* | *p-value* | |
|---|---|---|---|---|---|
| const | 0.798974 | 0.0587239 | 13.6056 | <0.00001 | *** |
| alpha(0) | 0.632159 | 0.0764556 | 8.2683 | <0.00001 | *** |
| alpha(1) | 0.36451 | 0.109357 | 3.3332 | 0.00086 | *** |

| | | | |
|---|---|---|---|
| Mean dependent var | 0.776391 | S.D. dependent var | 0.983733 |
| Log-likelihood | -357.1958 | Akaike criterion | 722.3915 |
| Schwarz criterion | 736.6953 | Hannan-Quinn | 728.1392 |

Unconditional error variance = 0.994759

$$y_t = 0.798974 + e_t \tag{8.1}$$

$$e_t | I_{t-1} \sim N(0, h_t) \tag{8.2}$$

$$h_t = 0.632159 + 0.36451 e_{t-1}^2 \tag{8.3}$$

```
Frequency distribution for uhat11, obs 2-265
number of bins = 17, mean = -0.0225837, sd = 0.987495

       interval          midpt   frequency    rel.     cum.

          < -3.3342   -3.5438        1      0.38%    0.38%
   -3.3342 - -2.9151   -3.1247        1      0.38%    0.76%
   -2.9151 - -2.4960   -2.7055        1      0.38%    1.14%
   -2.4960 - -2.0768   -2.2864        5      1.89%    3.03%
```

```
  -2.0768 - -1.6577   -1.8673        9      3.41%    6.44% *
  -1.6577 - -1.2385   -1.4481        9      3.41%    9.85% *
  -1.2385 - -0.81940  -1.0290       14      5.30%   15.15% *
 -0.81940 - -0.40027  -0.60984      39     14.77%   29.92% *****
 -0.40027 -  0.018873 -0.19070      67     25.38%   55.30% *********
 0.018873 -  0.43801   0.22844      44     16.67%   71.97% ******
  0.43801 -  0.85715   0.64758      27     10.23%   82.20% ***
  0.85715 -  1.2763    1.0667       28     10.61%   92.80% ***
   1.2763 -  1.6954    1.4859       10      3.79%   96.59% *
   1.6954 -  2.1146    1.9050        4      1.52%   98.11%
   2.1146 -  2.5337    2.3241        2      0.76%   98.86%
   2.5337 -  2.9528    2.7433        0      0.00%   98.86%
         >=  2.9528    3.1624        3      1.14%  100.00%

Test for null hypothesis of normal distribution:
Chi-square(2) = 19.654 with p-value 0.00005
```

Normality (Doornik-Hansen) test of $H_0$ is rejected with extremely small p-value 0.00005.

Residual definitely is not normally distributed.

III. Analysis:

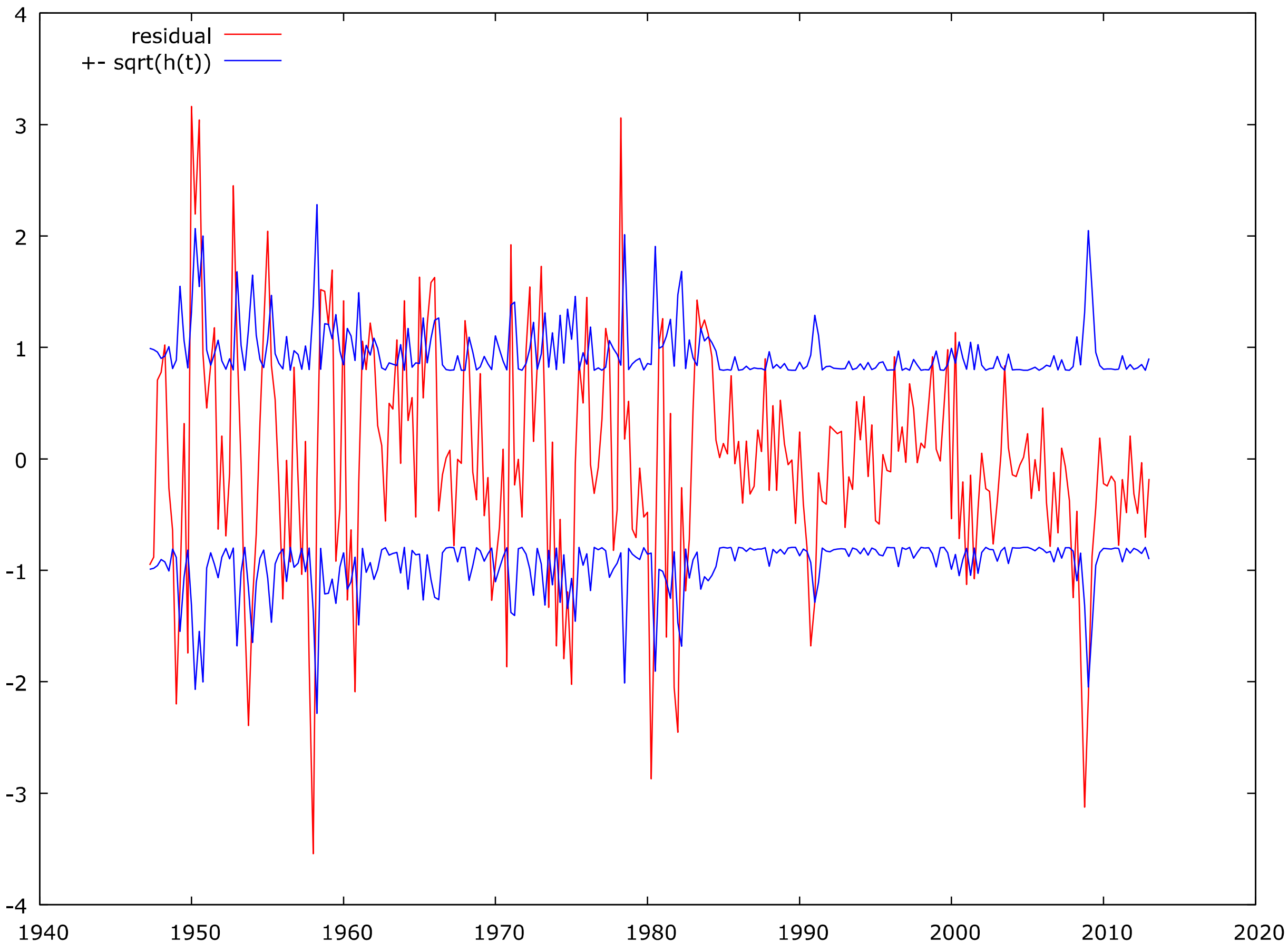

The residual plot shows approximately 33 GDP spikes outside of the blue upper and lower boundaries. There are large spikes occurring around the year of 1949, 1950, 1951, 1953, 1954,

1955, 1958, 1961, 1978, 1980, 1982, 1991, and 2009. From a historian point of view, the analysis accurately catches behaviors of the economy during major recessions, for example 1981-1982, 1990-1991, 2001, especially the great recession 2008. More interestingly, the analysis also sends out ex-ante and ex-post warning signals to policy makers. For example, 1980, just before 1981-1982 recession shows volatility (huge downward spike) in GDP followed by 1983, 1984. Additionally, 1988, 2000, and 2007 exhibit instable economic situation just before several major recessions. In conclusion, proper volatility analysis can help the government not only monitor economy but also prevent major recessions through proper economic policy adjustment.

**Model 8.2: GARCH (1, 1)**

Model 8.2: GARCH (1, 1), using observations 1947:2-2013:1 (T = 264)
Dependent variable: 100*ldiff(GDPC1Q)
Standard errors based on Hessian

| | *Coefficient* | *Std. Error* | *z* | *p-value* | |
|---|---|---|---|---|---|
| const | 0.820552 | 0.0556729 | 14.7388 | <0.00001 | *** |
| alpha(0) | 0.0385705 | 0.02223 | 1.7351 | 0.08273 | * |
| alpha(1) | 0.255331 | 0.0749903 | 3.4048 | 0.00066 | *** |
| beta(1) | 0.734709 | 0.0607463 | 12.0947 | <0.00001 | *** |

| | | | |
|---|---|---|---|
| Mean dependent var | 0.776391 | S.D. dependent var | 0.983733 |
| Log-likelihood | -346.8643 | Akaike criterion | 703.7286 |
| Schwarz criterion | 721.6084 | Hannan-Quinn | 710.9132 |

Unconditional error variance = 3.87253

The GARCH (1, 1) model can be expressed as:

$$y_t = 0.820552 + e_t \tag{8.4}$$

$$e_t | I_{t-1} \sim N(0, h_t) \tag{8.5}$$

$$h_t = 0.0385705 + 0.255331 e_{t-1}^2 + 0.734709 h_{t-1} \tag{8.6}$$

GARCH (1, 1) shows definite nonnormality with p value of 0.00005, same as ARCH (1).

III. Analysis:

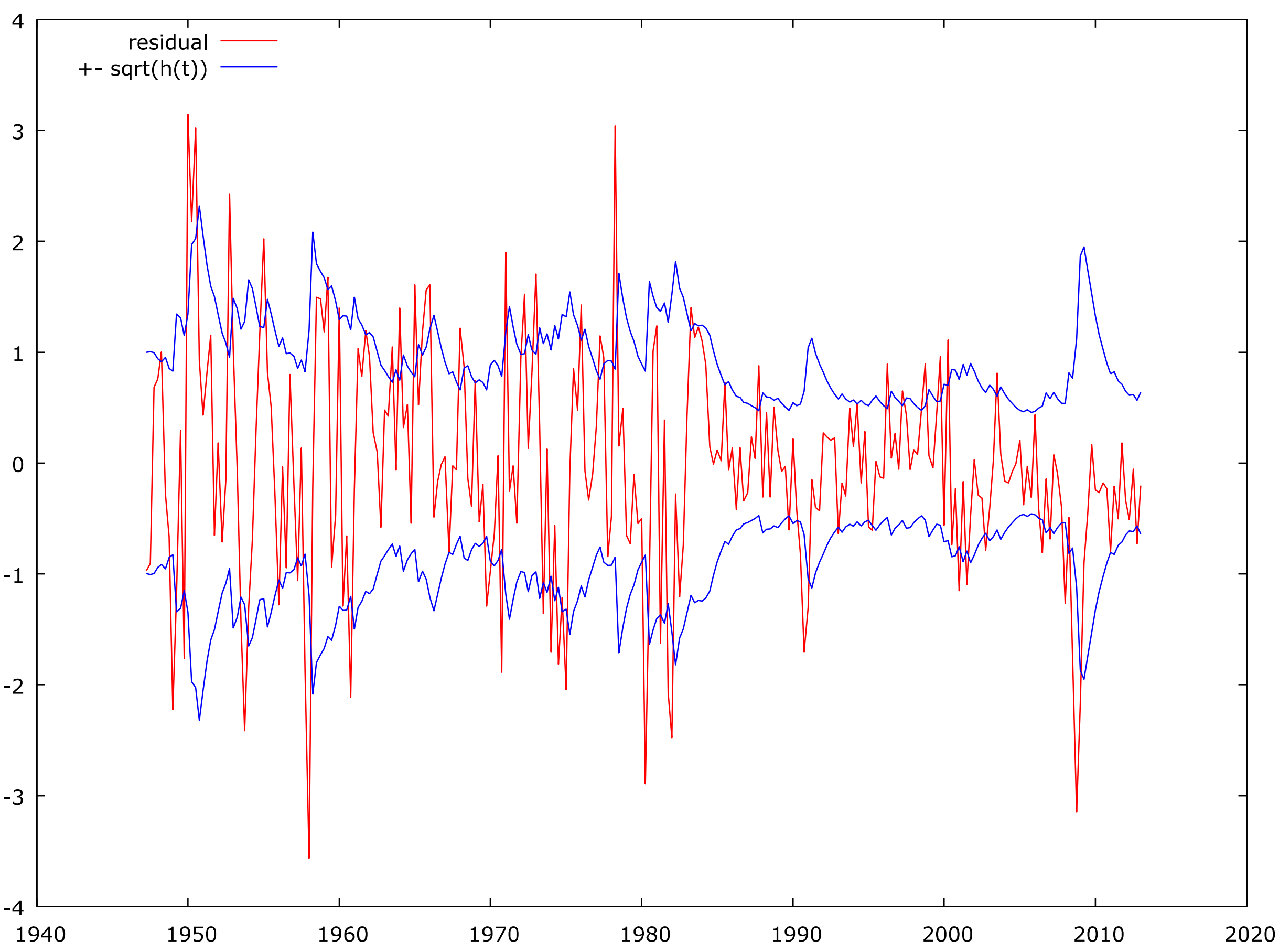

The residual plot of GARCH (1, 1) is almost identical to ARCH (1), but slightly different from it. For instance, GARCH (1, 1)'s upper and lower boundaries are more adaptive and volatile than ARCH (1)'s, therefore GARCH includes all the traits of ARCH, but also catches what ARCH misses. In other words, GARCH is more generalized than ARCH and thus it reflects volatility more accurately.

**Model 8.3: ARCH-M (1)**

(gretl scripts)

```
series y = 100*ldiff(GDPC1Q)
scalar mu = 0.0
scalar delta = 0.04
scalar alpha = 0.4
scalar beta = 0.0
scalar theta =0.0001
```

```
mle ll = -0.5*(log(h) + (e^2)/h)
  series h = var(y)
  series e = y - theta*h
  series h = delta + alpha*(e(-1))^2
  params theta delta alpha
end mle
```

(gretl script results)
Using numerical derivatives

Tolerance = 1.81899e-012

Function evaluations: 214

Evaluations of gradient: 57

Model 19: ML, using observations 1947:3-2013:1 (T = 263)

ll = -0.5*(log(h) + (e^2)/h)

Standard errors based on Outer Products matrix

| | estimate | std. error | z | p-value | |
|---|---|---|---|---|---|
| theta | 0.829080 | 0.0566959 | 14.62 | 2.00e-048 | *** |
| delta | 0.632092 | 0.0602920 | 10.48 | 1.02e-025 | *** |
| alpha | 0.365035 | 0.102237 | 3.570 | 0.0004 | *** |

| | | | |
|---|---|---|---|
| Log-likelihood | -114.1423 | Akaike criterion | 234.2847 |
| Schwarz criterion | 245.0011 | Hannan-Quinn | 238.5914 |

$$y_t = \theta h_t + e_t \tag{8.7}$$

$$h_t = \delta + \alpha_1 e_{t-1}^2 \tag{8.8}$$

Input the parameters into equation 8.7 and 8.8, we get:

$$y_t = 0.82908 h_t + e_t \tag{8.9}$$

$$h_t = 0.632092 + 0.365035 e_{t-1}^2 \tag{8.10}$$

**Model 8.4: GARCH-M (1, 1)**

(gretl scripts)

```
series y = 100*ldiff(GDPC1Q)
scalar mu = 0.0
scalar delta = 0.04
```

```
scalar alpha = 0.4
scalar beta = 0.5
scalar theta =0.0001

mle ll = -0.5*(log(h) + (e^2)/h)
  series h = var(y)
  series e = y - theta*h
  series h = delta + alpha*(e(-1))^2 + beta*h(-1)
  params theta delta alpha beta
end mle
```

(gretl script results)

Model 13: ML, using observations 1947:2-2013:1 (T = 264)

ll = -0.5*(log(h) + (e^2)/h)

Standard errors based on Outer Products matrix

| | estimate | std. error | z | p-value | |
|---|---|---|---|---|---|
| theta | 0.847852 | 0.0563744 | 15.04 | 4.04e-051 | *** |
| delta | 0.0386463 | 0.0211140 | 1.830 | 0.0672 | * |
| alpha | 0.255461 | 0.0558469 | 4.574 | 4.78e-06 | *** |
| beta | 0.734511 | 0.0412182 | 17.82 | 4.94e-071 | *** |

Log-likelihood -104.2686 Akaike criterion 216.5371

Schwarz criterion 230.8409 Hannan-Quinn 222.2848

$$y_t = \theta h_t + e_t \tag{8.11}$$

$$h_t = \delta + \alpha_1 e_{t-1}^2 + \beta_1 h_{t-1} \tag{8.12}$$

Input the parameters into equation 8.11 and 8.12, we get:

$$y_t = 0.847852 h_t + e_t \tag{8.13}$$

$$h_t = 0.0386463 + 0.255461 e_{t-1}^2 + 0.734511 h_{t-1} \tag{8.14}$$

## Model 8.5: Taylor-Schwert GARCH

```
Model: Taylor/Schwert's GARCH(1,1) (Normal)
Dependent variable: y
Sample: 1947:2-2013:1 (T = 264), VCV method: Hessian

    Conditional mean equation

              coefficient   std. error     z     p-value
  ---------------------------------------------------------
  const       0.836373     0.00279419   299.3   0.0000  ***
```

```
    Conditional variance equation

             coefficient   std. error     z       p-value
  ---------------------------------------------------------
  omega       0.0662126    0.0345328     1.917   0.0552    *
  alpha       0.252868     0.0588503     4.297   1.73e-05  ***
  beta        0.738908     0.0643768    11.48    1.70e-030 ***

       Llik:   -345.48651   AIC:    698.97303
       BIC:     713.27682   HQC:    704.72073
```

## Model 8.6: GJR (The Glosten Jagannathan Runkle GARCH)

```
Model: GJR(1,1) [Glosten et al.] (Normal)*
Dependent variable: y
Sample: 1947:2-2013:1 (T = 264), VCV method: OPG

    Conditional mean equation

             coefficient   std. error     z      p-value
  --------------------------------------------------------
  const       0.850098     0.0518322    16.40   1.88e-060 ***

    Conditional variance equation

             coefficient   std. error     z      p-value
  --------------------------------------------------------
  omega       0.0858183    0.0455942    1.882   0.0598    *
  alpha       0.341302     0.100939     3.381   0.0007    ***
  gamma       0.313410     0.114741     2.731   0.0063    ***
  beta        0.592662     0.0887584    6.677   2.43e-011 ***

   (alt. parametrization)

             coefficient   std. error     z      p-value
  --------------------------------------------------------
  delta       0.0858183    0.0455942    1.882   0.0598    *
  alpha       0.160892     0.0863810    1.863   0.0625    *
  gamma       0.427870     0.150627     2.841   0.0045    ***
  beta        0.592662     0.0887584    6.677   2.43e-011 ***

       Llik:   -344.38603   AIC:    698.77205
       BIC:     716.65180   HQC:    705.95668
```

## Model 8.7: TARCH

```
Model: TARCH(1,1) [Zakoian] (Normal)
Dependent variable: y
Sample: 1947:2-2013:1 (T = 264), VCV method: Hessian

    Conditional mean equation

             coefficient   std. error     z      p-value
  --------------------------------------------------------
  const       0.830636     0.0490712    16.93   2.84e-064 ***

    Conditional variance equation

             coefficient   std. error     z      p-value
```

```
  -------------------------------------------------------
  omega       0.107087     0.0500759    2.138   0.0325    **
  alpha       0.284205     0.0661465    4.297   1.73e-05  ***
  gamma       0.363679     0.130245     2.792   0.0052    ***
  beta        0.664701     0.0864832    7.686   1.52e-014 ***

      Llik:   -341.62854   AIC:    693.25708
      BIC:     711.13683   HQC:    700.44171
```

## Model 8.8: NARCH

```
Model: NARCH(1,1) [Higgins and Bera] (Normal)
Dependent variable: y
Sample: 1947:2-2013:1 (T = 264), VCV method: Hessian

    Conditional mean equation

             coefficient   std. error        z       p-value
  ---------------------------------------------------------
  const       0.843591     3.95938e-05   2.131e+04   0.0000  ***

    Conditional variance equation

             coefficient   std. error      z       p-value
  --------------------------------------------------------
  omega       0.0811524    0.0349671     2.321    0.0203    **
  alpha       0.156068     0.0728736     2.142    0.0322    **
  beta        0.779823     0.0701164    11.12     9.82e-029 ***
  delta       0.246768     0.282203      0.8744   0.3819

      Llik:   -345.06132   AIC:    700.12265
      BIC:     718.00239   HQC:    707.30727
```

## Model 8.9: APARCH

```
Model: APARCH(1,1) [Ding] (Normal)
Dependent variable: y
Sample: 1947:2-2013:1 (T = 264), VCV method: Hessian

    Conditional mean equation

             coefficient   std. error     z      p-value
  -------------------------------------------------------
  const       0.829406     0.0553688    14.98   9.97e-051 ***

    Conditional variance equation

             coefficient   std. error     z      p-value
  -------------------------------------------------------
  omega       0.112231     0.0510380    2.199   0.0279    **
  alpha       0.247851     0.0790212    3.137   0.0017    ***
  gamma       0.387464     0.142873     2.712   0.0067    ***
  beta        0.687873     0.0880695    7.811   5.69e-015 ***
  delta       0.690535     0.419295     1.647   0.0996    *

      Llik:   -341.37515   AIC:    694.75029
      BIC:     716.20599   HQC:    703.37184
```

### Model 8.10: EGARCH (1, 1)

```
Model: EGARCH(1,1) [Nelson] (Normal)
Dependent variable: y
Sample: 1947:2-2013:1 (T = 264), VCV method: Hessian

    Conditional mean equation

             coefficient   std. error     z      p-value
  --------------------------------------------------------
  const        0.836103     0.0500604    16.70   1.27e-062 ***

    Conditional variance equation

             coefficient   std. error     z       p-value
  ---------------------------------------------------------
  omega       -0.464418     0.0937466    -4.954   7.27e-07  ***
  alpha        0.541685     0.110848      4.887   1.03e-06  ***
  gamma       -0.196858     0.0756444    -2.602   0.0093    ***
  beta         0.843356     0.0583697    14.45    2.56e-047 ***

       Llik:   -340.43681   AIC:     690.87362
       BIC:     708.75337   HQC:     698.05825
```

### Goodness-of-Fit Comparison

| Model / Criteria | AIC | BIC | HQC |
|---|---|---|---|
| **ARCH (1)*** | 722.3915 | 736.6953 | 728.1392 |
| **GARCH (1, 1)*** | 703.7286 | 721.6084 | 710.9132 |
| **ARCH-M (1)*** | 234.2847 | 245.0011 | 238.5914 |
| **GARCH-M (1, 1)*** | 216.5371 | 230.8409 | 222.2848 |
| **TS-GARCH (1, 1)*** | 698.97303 | 713.27682 | 704.72073 |
| **GJR (1, 1)*** | 698.77205 | 716.65180 | 705.95668 |
| **TARCH (1, 1)*** | 693.25708 | 711.13683 | 700.44171 |
| **NARCH (1, 1)** | 700.12265 | 718.00239 | 707.30727 |
| **APARCH (1, 1)*** | 694.75029 | 716.20599 | 703.37184 |
| **EGARCH (1, 1)*** | 690.87362 | 708.75337 | 698.05825 |

Residuals are normally distributed.
* denotes that parameters (all the coefficients) are significant.

### Discussion of the Best Volatility Model

From the above goodness of fit comparison chart, we learn that GARCH-M (1, 1) definitely has the lowest AIC, BIC, HQC. For a more comprehensive comparison, we now assume that the normality assumption of residual distribution is violated and those distributions (t, GED—

Generalized Error Distribution, skewed t, skewed GED) will be used to construct comparable models. The general technique for obtaining the best volatility model is to experiment with models of nonnormal distribution that results in the lowest possible overall fitness criterion, which we compare with those of normal distribution with the lowest criterion. More specifically, the algorithm of obtaining the best volatility model is to first perform self-comparison within the models of nonnormal distribution; second perform cross-comparison between the models of nonnormal distribution and those of normal distribution. (The reason to first perform self-comparison within the models of nonnormal distribution is that those models have overall lower AIC, BIC, HQC than those of normal distribution, as results shown in the following table). For example, first compare fitness criterion among t, GED, Skewed t, Skewed GED, get the lowest result; second compare the model of nonnormal distribution with the lowest fitness criterion with its counterpart of normal distribution, get the lowest result as the best fitted model. Please see the following experiment results:

| | **AIC** | **BIC** | **HQC** |
|---|---|---|---|
| *ARCH (1)* | | | |
| **t*** | 712.58939 | 726.89319 | 718.33709 |
| **GED*** | 706.94010 | 721.24390 | 712.68780 |
| **Skewed t** | 714.58926 | 732.46901 | 721.77389 |
| **Skewed GED** | N/A | N/A | N/A |
| *GARCH (1, 1)* | | | |
| **t** | 694.76479 | 712.64453 | 701.94941 |
| **GED** | 691.90814 | 709.78789 | 699.09277 |
| **Skewed t** | 696.76243 | 718.21812 | 705.38398 |
| **Skewed GED** | 693.86115 | 715.31684 | 702.48270 |
| *TS-GARCH (1, 1)* | | | |
| **t*** | 693.75727 | 711.63702 | 700.94190 |
| **GED** | 691.12038 | 709.00012 | 698.30500 |
| **Skewed t** | 695.72381 | 717.17951 | 704.34536 |
| **Skewed GED** | 692.97543 | 714.43113 | 701.59698 |
| *GJR (1, 1)* | | | |
| **t*** | 694.17832 | 715.63401 | 702.79987 |
| **GED** | 691.54820 | 713.00390 | 700.16975 |
| **Skewed t** | 695.53312 | 720.56476 | 705.59159 |
| **Skewed GED** | 692.19620 | 717.22785 | 702.25468 |
| *TARCH (1, 1)* | | | |
| **t** | 692.16754 | 713.62323 | 700.78909 |
| **GED*** | 689.33261 | 710.78830 | 697.95416 |
| **Skewed t** | 693.37396 | 718.40561 | 703.43244 |
| **Skewed GED** | 689.77211 | 714.80376 | 699.83059 |
| *NARCH (1, 1)* | | | |
| **t** | 695.74212 | 717.19781 | 704.36367 |
| **GED** | 693.08054 | 714.53624 | 701.70209 |
| **Skewed t** | 697.72246 | 722.75410 | 707.78094 |
| **Skewed GED** | 694.97241 | 720.00406 | 705.03089 |
| *APARCH (1, 1)* | | | |
| **t** | 694.15953 | 719.19118 | 704.21801 |
| **GED*** | 691.33184 | 716.36349 | 701.39032 |
| **Skewed t** | 695.35481 | 723.96241 | 706.85022 |
| **Skewed GED** | 691.70526 | 720.31285 | 703.20066 |
| *EGARCH (1, 1)* | | | |
| **t** | 690.32769 | 711.78338 | 698.94924 |
| **GED** | N/A | N/A | N/A |
| **Skewed t** | 690.99794 | 716.02958 | 701.05642 |
| **Skewed GED** | N/A | N/A | N/A |

For the purpose of comparison, models with nonnormal distribution use the same covariance estimator as those with normal distribution. Results seem not to vary significantly from using different covariance estimation (Sandwich, Hessian, OPG).

For the purpose of comparable results, models with nonnormal distribution (ARCH, GARCH, TS-GARCH, GJR, TARCH, NARCH, APARCH, EGARCH) use 100*ldiff(GDPC1Q).

N/A means that calculation error occurs for associated distribution to estimate the corresponding model.

* means that parameters (all the coefficients) are significant.

From the above experiment results, we can find that GED (General Error Distribution) works the best for all the models of nonnormal distribution with the lowest overall fitness criterion. TARCH (1, 1) GED is the best model within the nonnormal group. APARCH (1, 1) GED is the second. TS-GARCH (1, 1) t and GJR (1, 1) t stands the third for the former has slightly lower overall fitness criterion than the latter. On the other hand, within the group of normal distribution, we find that GARCH-M (1, 1) and ARCH-M (1, 1) are very different from the rest with the lowest AIC, BIC, HQC in the 200 range. GARCH-M is slightly better than ARCH-M. In the 700 range, EGARCH is the first; TARCH and APARCH are the second (TARCH is slightly better than APARCH); TS-GARCH and GJR are the third (TS-GARCH is slightly better than GJR). Therefore, we conclude that GARCH-M and ARCH-M models are the best for volatility test of GDPC1Q. TARCH (1, 1) GED is the second best. EGARCH (1, 1) takes the third.

## SECTION III: DISCUSSION

### *1) Summary of Major Research Findings and Their Managerial Implications*

GDP—GDPC1Q ($Y_{1t}$), Government Expenditure—GCEC1Q ($Y_{2t}$)

Sample range:1980Q1—2006Q1 for (i)—(vii); Sample range:1947Q1-2013Q1 for (viii)

**(i) Univariate ARIMA**

$$\Delta^{1d}Y_{1t} = 66.2629 + 0.759918\Delta^{1d}Y_{1,t-1} + \varepsilon_{1t} - 0.479305\varepsilon_{1,t-1} \tag{1.10}$$

$$\Delta^{1d}Y_{2t} = 9.7675 + 0.811342\Delta^{1d}Y_{2,t-1} + \varepsilon_{2t} - 0.943283\varepsilon_{2,t-1} + 0.273345\varepsilon_{2,t-2} \tag{1.15}$$

Through Univariate ARMA analysis, we find that each observable value for GDP and Government Expenditure of current year are significantly positively associated with the value of past year, meaning that the future trend of both GDP and Government Spending can be accurately predicted via their past pattern, which is critical for policy makers to monitor and control the direction of overall economy. The forecasting results show a promising bouncing back trend of both GDP and Government Expenditure.

**(ii) Multivariate ARIMA**

The VARIMA (1, 1, 1) and VARIMA (1, 1, 2) analysis does not give us any useful insights regarding the interrelationship between GDP and Government Expenditure and between Government Expenditure and GDP because of statistically insignificant corresponding coefficients. However, we do find that both GDPC1Q and GCEC1Q from current year reacts positively to those from previous year, which is consistent with the findings from the univariate ARIMA analysis.

**(iii) VAR and Impulse Response Function**

$$\begin{bmatrix}\Delta^{1d}y_{1,t}\\ \Delta^{1d}y_{2,t}\end{bmatrix} = \begin{bmatrix}49.3131\\ 11.8267\end{bmatrix} + \begin{bmatrix}0.360705 & -0.589345\\ -0.0290698 & 0.000855439\end{bmatrix}\begin{bmatrix}\Delta^{1d}y_{1,t-1}\\ \Delta^{1d}y_{2,t-1}\end{bmatrix} + \begin{bmatrix}e_{1,t}\\ e_{2,t}\end{bmatrix} \quad (3.3)$$

$$\Delta^{1d}y_{1,t} = 49.3131 + 0.360705\Delta^{1d}y_{1,t-1} - 0.589345\Delta^{1d}y_{2,t-1} + e_{1,t} \quad (3.4)$$
$$\Delta^{1d}y_{2,t} = 11.8267 - 0.0290698\Delta^{1d}y_{1,t-1} + 0.000855439\Delta^{1d}y_{2,t-1} + e_{2,t}$$

Through VAR and impulse response, we find that the negative response of GDP from current year to a shock of Government Expenditure from last year is statistically significant, while the negative response of Government Spending from current year to a shock of GDP from last year is statistically insignificant. For example, a column of insignificant coefficients means that the empirical shocks of the corresponding variable have only exhibited temporary effects on the variables of the system, whereas a column of significant coefficients indicates permanent effects on the system (Katarina Juselius "The Cointegrated VAR Model: Methodology and Applications"). Based on Juselius's finding, we conclude the VAR analysis that a shock of last-year Government Expenditure on current-year GDP is not only negative but also permanent, while a shock of last-year GDP on current-year Government Expenditure is not necessarily negative but also temporary implies that substantially large government expenditure from the past will negatively impact GDP in the long-run, while temporary increase of government spending alone in the future might not necessarily be the only answer to a weak economy from the past. Thereby, we suggest that policy makers should be cautious about boosting economy via increasing government expenditure alone, rather government spending should be adjusted accordingly with consideration of GDP not only in the short-run but in the long-run as well.

**(iv) State-Space Model of ARMA via the Kalman Filter and Maximum Likelihood Estimation**

$$\xi_{t+1} = \begin{bmatrix}0.97519 & 0\\ 1 & 0\end{bmatrix}\xi_t + \mathrm{v}_t \quad (4.1')$$

$$y_t = \begin{bmatrix}1\\ -1.43617\end{bmatrix}\xi_t \quad (4.2')$$

$$where\ E(\mathrm{v}_t\mathrm{v}_s') = \begin{bmatrix} 1471.79253 & 0 \\ 0 & 0 \end{bmatrix} for\ t = s, otherwise\ 0$$

And,

$$\Delta^{1d} y_t = 0.975190\ \Delta^{1d} y_{t-1} + \varepsilon_t - 1.43617\ \varepsilon_{t-1} \quad (4.7')$$

$$\Delta^{1d} y_t = \xi_t - 1.43617\xi_{t-1} \quad (4.8')$$

$$\begin{bmatrix} \xi_t \\ \xi_{t-1} \end{bmatrix} = \begin{bmatrix} 0.975190 & 0 \\ 1 & 0 \end{bmatrix} \begin{bmatrix} \xi_{t-1} \\ \xi_{t-2} \end{bmatrix} + \begin{bmatrix} \varepsilon_t \\ 0 \end{bmatrix} \quad (4.6')$$

We find that ARIMA (1, 1, 1) of GDPC1Q without a constant has the same positive sign on $\varphi$ and negative sign on $\theta$, plus significant parameters, compared to ARIMA (1, 1, 1) with a constant, meaning that there is a statistically significant positive relationship between last year GDP and current year GDP. This provides policy makers with a very accurate yet powerful tool to monitor, control and predict the direction of overall economy.

**(v) Transfer Function and Intervention Model—ARMAX**

$$\Delta^{1d} y_{1t} = 57.0849 + 0.720079\Delta^{1d} y_{1,t-1} + 0.922759\Delta^{1d} y_{2t} + \varepsilon_{1t} - 0.383438\varepsilon_{1,t-1} \quad (5.3)$$

$$\Delta^{1d} y_{2t} = 6.6791 + 0.777622\Delta^{1d} y_{2,t-1} + 0.0483137\Delta^{1d} y_{1t} + \varepsilon_{2t} - 0.862196\varepsilon_{2,t-1} + 0.239026\varepsilon_{2,t-2} \quad (5.4)$$

Through ARMAX results, we find that both concurrent relationship between Government Expenditure and GDP and between GDP and Government Expenditure are statistically significantly positive, meaning that by increasing current-year Government Spending, policy makers can boost current-year GDP, while strong current-year GDP stimulates increasing current-year Government Expenditure in return, which is consistent with the classic Keynesian theory that increasing government investment in infrastructure can boost the economy in the short run.

**(vi) Unit Root Test (ADF Test)**

The ADF tests show that sample data of both GDPC1Q and GCEC1Q are not stationary, meaning that we have to difference the sample data at least for once in order for its suitability for further ARMA analysis.

**(vii) Cointegration Test (Engle-Granger test)**

The Engle-Granger tests show no sign of conintegration between GDPC1Q and GCEC1Q, meaning that it is not necessary for us to use error-corrected VAR (VEC).

**(viii) Volatility Tests via ARCH, GARCH, ARCH-M, GARCH-M, EGARCH, and other ARCH GARCH variants**

We find that the classic GARCH-M performs the best for GDPC1Q data assuming normal distribution, while TARCH (1, 1) GED performs the best for GDPC1Q data assuming non-normal distribution. Overall, models with non-normal distribution have lower AIC, BIC, HQC than models with normal distribution, excluding ARCH-M and GARCH-M. The results explicitly suggest the best volatility model to policy makers and practitioners for analyzing GDP with either normal or non-normal distribution.

---

***2) Comparison of Empirical Results***

**(i) Univariate vs. Multivariate VARMA**

*ARIMA (1, 1, 1) GDPC1Q:*

$$\Delta^{1d}Y_{1t} = 66.2629 + 0.759918\Delta^{1d}Y_{1,t-1} + \varepsilon_{1t} - 0.479305\varepsilon_{1,t-1} \tag{1.10}$$

*ARIMA (1, 1, 2) GCEC1Q:*

$$\Delta^{1d}Y_{2t} = 9.7675 + 0.811342\Delta^{1d}Y_{2,t-1} + \varepsilon_{2t} - 0.943283\varepsilon_{2,t-1} + 0.273345\varepsilon_{2,t-2} \tag{1.15}$$

*VARIMA (1, 1, 1) GDPC1Q & GCEC1Q:*

$$\begin{bmatrix}\Delta^{1d}Y_{1t}\\ \Delta^{1d}Y_{2t}\end{bmatrix} = \begin{bmatrix}82.2102\\ 8.59125\end{bmatrix} + \begin{bmatrix}0.751179 & -6.43447\\ 0.060149 & -0.261946\end{bmatrix}\begin{bmatrix}\Delta^{1d}\mathrm{Y}_{1,t-1}\\ \Delta^{1d}\mathrm{Y}_{2,t-1}\end{bmatrix} + \begin{bmatrix}\varepsilon_{1,t}\\ \varepsilon_{2,t}\end{bmatrix} + \begin{bmatrix}-0.460145 & 5.91783\\ -0.105878 & 0.279512\end{bmatrix}\begin{bmatrix}\varepsilon_{1,t-1}\\ \varepsilon_{2,t-1}\end{bmatrix} \tag{2.1'}$$

$$\Delta^{1d}Y_{1t} = 82.2102 + 0.751179\Delta^{1d}\mathrm{Y}_{1,t-1} - 6.43447\Delta^{1d}\mathrm{Y}_{2,t-1} + \varepsilon_{1,t} - 0.460145\varepsilon_{1,t-1} + 5.91783\varepsilon_{2,t-1} \tag{2.2'}$$

$$\Delta^{1d}Y_{2t} = 8.59125 + 0.060149\Delta^{1d}\mathrm{Y}_{1,t-1} - 0.261946\Delta^{1d}\mathrm{Y}_{2,t-1} + \varepsilon_{2,t} - 0.105878\varepsilon_{1,t-1} + 0.279512\varepsilon_{2,t-1}$$

*VARIMA (1, 1, 2) GDPC1Q & GCEC1Q:*

$$\begin{bmatrix}\Delta^{1d}Y_{1t}\\ \Delta^{1d}Y_{2t}\end{bmatrix} = \begin{bmatrix}46.7843\\ 0.156603\end{bmatrix} + \begin{bmatrix}0.369566 & -0.219727\\ 0.00582349 & 0.950556\end{bmatrix}\begin{bmatrix}\Delta^{1d}\mathrm{Y}_{1,t-1}\\ \Delta^{1d}\mathrm{Y}_{2,t-1}\end{bmatrix} + \begin{bmatrix}\varepsilon_{1,t}\\ \varepsilon_{2,t}\end{bmatrix} + \begin{bmatrix}-0.113708 & -0.132601\\ -0.0394123 & -1.00433\end{bmatrix}\begin{bmatrix}\varepsilon_{1,t-1}\\ \varepsilon_{2,t-1}\end{bmatrix} + \begin{bmatrix}0.216516 & -0.104326\\ 0.0111816 & 0.165578\end{bmatrix}\begin{bmatrix}\varepsilon_{1,t-2}\\ \varepsilon_{2,t-2}\end{bmatrix} \tag{2.3'}$$

$$\Delta^{1d}Y_{1t} = 46.7843 + 0.369566\Delta^{1d}\text{Y}_{1,t-1} - 0.219727\Delta^{1d}\text{Y}_{2,t-1} + \varepsilon_{1,t} - 0.113708\varepsilon_{1,t-1} - 0.132601\varepsilon_{2,t-1} + 0.216516\varepsilon_{1,t-2} - 0.104326\varepsilon_{2,t-2} \quad (2.4')$$

$$\Delta^{1d}Y_{2t} = 0.156603 + 0.00582349\Delta^{1d}\text{Y}_{1,t-1} + 0.950556\Delta^{1d}\text{Y}_{2,t-1} + \varepsilon_{2,t} - 0.0394123\varepsilon_{1,t-1} - 1.00433\varepsilon_{2,t-1} + 0.0111816\varepsilon_{1,t-2} + 0.165578\varepsilon_{2,t-2}$$

*VARI (1, 1):*

$$\begin{bmatrix}\Delta^{1d}y_{1,t}\\ \Delta^{1d}y_{2,t}\end{bmatrix} = \begin{bmatrix}49.3131\\ 11.8267\end{bmatrix} + \begin{bmatrix}0.360705 & -0.589345\\ -0.0290698 & 0.000855439\end{bmatrix}\begin{bmatrix}\Delta^{1d}y_{1,t-1}\\ \Delta^{1d}y_{2,t-1}\end{bmatrix} + \begin{bmatrix}e_{1,t}\\ e_{2,t}\end{bmatrix} \quad (3.3)$$

$$\Delta^{1d}y_{1,t} = 49.3131 + 0.360705\Delta^{1d}y_{1,t-1} - 0.589345\Delta^{1d}y_{2,t-1} + e_{1,t} \quad (3.4)$$
$$\Delta^{1d}y_{2,t} = 11.8267 - 0.0290698\Delta^{1d}y_{1,t-1} + 0.000855439\Delta^{1d}y_{2,t-1} + e_{2,t}$$

The coefficient value of $\Delta^{1d}\text{Y}_{1,t-1}$ from the first equation in (2.2') is remarkably similar to its counterpart in (1.10) (0.751179 compared with 0.759918), plus both coefficients having the same positive sign. In contrast, the coefficient value of $\Delta^{1d}\text{Y}_{2,t-1}$ from the second equation in (2.4') is very close to its counterpart in (1.15) (0.950556 compared with 0.811342), plus both coefficients having the same positive sign. Additionally, the coefficient value of $\varepsilon_{1,t-1}$ in the first equation of (2.2') is close to its counterpart in (1.10) (-0.460145 compared with -0.479305), plus both coefficients having the same negative sign. In contrast, the coefficient value of $\varepsilon_{2,t-1}$ in the second equation in (2.4') is very close to its counterpart in (1.15) (-1.00433 compared with -0.943283), plus both coefficients having the same negative sign. We find the results are very interesting, because the first equation of multivariate VARIMA (1, 1, 1) is much better than the second equation in the system as to the overall parameter's accuracy and sign from the first equation close to that of univariate ARIMA (1, 1, 1), in comparison the second equation of multivariate VARIMA (1, 1, 2) is much better than the first equation in the system as to the overall parameter's accuracy and sign from the second equation close to that of univariate ARIMA (1, 1, 2). We also see much difference in both value and sign of the parameters between first and second equation in the system (2.2') and (2.4').

**(ii) ARMAX vs. VAR**

ARMAX analysis shows us that there are positive relationships between GDP and Government Spending and between Government Spending and GDP, both concurrently in the short-run. In comparison, VAR results indicate that current GDP responses negatively to a shock in past

Government Spending permanently in the long-run, while current Government Spending not necessarily yet temporarily reacts negatively to a shock in past GDP. The combined findings suggest that policy makers should not adjust current Government Expenditure based merely on the condition and performance of past economy (GDP), because the negative relationship between current Government Spending and past GDP is statistically insignificant, thus such relationship is only temporary. Instead, policy makers should focus on both the short-term benefits of spending to the current economy and the long-term effects of expenditure on the future economy, thereby finding a balanced yet economically viable solution to both short-term and long-term propensity. Specifically, decisions on government expenditure should not be limited by the current economic condition, and abrupt increase in government spending alone (without monetary policy—interest rate—adjustment) in the short-term might not be good to the long-term health of the economy. Fundamentally, for a more balanced approach, we theorize and predict that the short-term benefits to the current economy from increasing government expenditure often largely secured by the long-term loan should outweigh the negative effects to the future economy from the long-term debt incurred by the loan, or at least equal to.

**(iii) VARMA vs. VAR**

We are unable to obtain meaningful information from VARMA analysis on GDP and Government Expenditure, because the associated parameters are statistically insignificant due to the technical difficulties mentioned on the bottom of page 34. In comparison, VAR does give us a lot insights as elaborated in the above title (ii).

**(iv) ARCH, GARCH, ARCH-M, GARCH-M, TS-GARCH, GJR, TARCH, NARCH, APARCH, EGARCH; Normal vs. Non-normal distribution**

We find that models with non-normal distribution have overall smaller AIC, BIC, HQC, compared with models with normal distribution, excluding ARCH-M and GARCH-M, suggesting that non-normally distributed volatility models generally perform better than those normally distributed. Looking into each group of normally and non-normally distributed, we find that TARCH (1, 1) GED performs the best in the group of non-normally distributed, while GARCH-M does the best in the group of normally distributed.

***3) Contributions to Literature***

The contribution of this research to the current economic literature is profound in the following ways: 1) the combined analysis of ARMAX and VAR offers better insights on the cause and effect between GDP and Government Expenditure not only in the short-run but in the long-run as well. VAR also builds a bridge enabling us to study how short-term effects long-term, interchangeably between GDP and Government Spending and between Government Spending and GDP. Thereby, the findings are practically significant to policy makers as shown in section 9; 2) theories and predictions derived from the ARMAX and VAR analysis and results are not only theoretically novel but also empirically significant, as shown in section 10, regarding the current economic debates on the topic between short-term stimulus spending and long-term debt; 3) we have empirically demonstrated that volatility models with non-normal distribution overall perform better than models distributed normally; 4) Kalman filter is both theoretically and empirically demonstrated to generate better estimations; 5) we are able to show similarity and difference between univariate ARMA and multivariate VARMA both theoretically and empirically.

***4) Limitations and Future Research***

**Model limitations**

(i) Unit root test: arbitrary lag order.

(ii) VAR: point estimate.

(iii) Multivariate VARMA/VAR: not including heteroskedasticity component like GARCH.

(iv) GARCH: univariate. (Multivariate GARCH has been discussed and used in the literature.)

**Data limitations**

Economic analysis is only limited to the U.S. data. Extension to include the foreign country data along with their comparison to the U.S. might provide further insights to the effectiveness and influence of fiscal policy on GDP.

**Future research**

*Theoretical extension1:* ARMAX extends to Multivariate VARMAX.

*Theoretical extension2:* VAR extends to VARCH or VGARCH.

*Theoretical extension3:* Consider combining GARCH with Multivariate ARMA. For example, GARCH + Multivariate VARMA = Multivariate VGAR(MA)CH. The idea is to incorporate the component of moving average to the Multivariate GARCH model.

*Empirical extension:* Use VAR and ARMAX to first study the individual effect of each monetary policy (interest rate) and fiscal policy (government spending) on GDP; then study the simultaneous effects of monetary and fiscal policy on GDP; and finally compare the results of step one and two. The goal is to not only examine application of the classic Keynesian theory in the short-term, but more importantly explore the long-term effects of both monetary policy and fiscal policy on GDP as well.

### *5) Gretl Scripts for System Equation (VAR, VARMA, etc.)*

**VAR (1)**

```
series x = diff(GDPC1Q)
series y = diff(GCEC1Q)
system
   equation x const x(-1) y(-1)
   equation y const x(-1) y(-1)
end system
estimate $system method=ols
```

Notes: The developed VAR gretl system procedure can also be used similarly in other multivariate AR analysis (such as VARX), with slight modification.

**VARMA (1, 1)**

```
series x = diff(GDPC1Q)
series y = diff(GCEC1Q)
series a = uhat1
series b = uhat2
system
   equation x const x(-1) y(-1) a(-1) b(-1)
   equation y const x(-1) y(-1) a(-1) b(-1)
end system
estimate $system method=ols
```

Notes for the VARMA gretl procedures:

Step 1: Run ARIMA (1, 1, 1) for GDPC1Q once and run ARIMA (1, 1, 1) for GCEC1Q once

Step 2: Save to the dataset each of the residuals (uhat1 and uhat2) from each univariate ARIMA (1, 1, 1) of GDPC1Q and GCEC1Q respectively.

Step 3: Input the above VARMA (1, 1) gretl scripts and get the results.

Notes: The developed VARMA procedures can be used similarly in other multivariate ARMA analysis (such as VARMAX) as well, with slight modification.

**VARMA (1, 2)**

```
series x = diff(GDPC1Q)
series y = diff(GCEC1Q)
series a = uhat1
series b = uhat2
```

```
system
   equation x const x(-1) y(-1) a(-1) b(-1) a(-2) b(-2)
   equation y const x(-1) y(-1) a(-1) b(-1) a(-2) b(-2)
end system
estimate $system method=ols
```